\documentclass[letterpaper,12pt]{article}
\usepackage{multirow}
\usepackage{enumitem}
\usepackage{bigstrut,rotating}
\usepackage[top=1in, bottom=1in, left=1in, right=1in]{geometry}
\usepackage{graphicx,rotating,amsthm,amsbsy,amssymb,amsmath, caption, natbib, subcaption, bbm, color, thmtools}
\usepackage{setspace}
\usepackage{bm}
\usepackage{algorithm}
\usepackage[noend]{algpseudocode}

\usepackage{makecell}
\usepackage{float}  
\usepackage{lipsum}
\usepackage{setspace}
\usepackage{soul}
\usepackage[T2A,T1]{fontenc}
\usepackage[russian,english]{babel}
\usepackage{amsfonts, latexsym}
\usepackage[utf8]{inputenc}

\usepackage[top=1in, bottom=1in, left=1in, right=1in]{geometry}
\usepackage{mathrsfs}
\usepackage{physics}
\usepackage{blindtext}
\usepackage{floatflt}
\usepackage[normalem]{ulem}
\usepackage{algorithm,algpseudocode,float}
\usepackage{color}
\usepackage{hyperref}

\usepackage{pifont}
\usepackage{stackengine}
\usepackage{verbatimbox}
\usepackage{mathtools}
\usepackage{csvsimple}
\usepackage{bm}

\usepackage{comment}

\usepackage{diagbox}
\usepackage{url}
\usepackage[group-separator={,},
				group-minimum-digits=4,
				table-format = 7]{siunitx}

\usepackage{xcolor}



\definecolor{darkgreen}{RGB}{0,100,0}
\definecolor{teal}{RGB}{0,128,128}

\newtheorem{theorem}{Theorem}[section]

\newtheorem{lemma}[theorem]{Lemma}

\newtheorem{defn}[theorem]{Definition} 

\newcommand{\vx}{\boldsymbol{x}}
\newcommand{\bx}{\boldsymbol{x}}
\newcommand{\obs}{\boldsymbol{x}^{\rm obs}}

\newcommand{\bj}{\boldsymbol{j}}
\newcommand{\by}{\boldsymbol{y}}
\newcommand{\bz}{\mathbf{z}}
\newcommand{\prob}{\rm {Pr}}

\newcommand{\vphi}{\vec{\phi}}

\newcommand{\vpsi}{\boldsymbol{\psi}}
\newcommand{\tildeDelta}{\Delta}

\title{\bf Bayesian nonparametric modeling of mixed-type bounded data}

\author{{\sc  Rufeng Liu, Claudia Wehrhahn, Andr\'es F. Barrientos,}\\ \sc{and Alejandro Jara}}

\begin{document}

\begin{spacing}{1}  
\date{\today}
\maketitle 
\end{spacing}

\footnotetext[1]{
Rufeng Liu is PhD(c), Department of Statistics, Florida State University, USA (rl19o@fsu.edu); Claudia Wehrhahn is Senior Actuary, HDI Seguros, Chile (clau.wehrhahn@gmail.com); Andr\'es F. Barrientos is Assistant Professor, Department of Statistics, Florida State University, USA (abarrientos@fsu.edu); Alejandro Jara is Full Professor, Department of Statistics, Pontificia Universidad Cat\'olica de Chile (atjara@uc.cl) and Visiting Professor, Department of Statistics, Brigham Young University, USA (ajara@stat.byu.edu).}

\begin{abstract}
We propose a Bayesian nonparametric model for mixed-type bounded data, where some variables are compositional and others are interval-bounded. Compositional variables are non-negative and sum to a given constant, such as the proportion of time an individual spends on different activities during the day or the fraction of different types of nutrients in a person's diet. Interval-bounded variables, on the other hand, are real numbers constrained by both a lower and an upper bound. Our approach relies on a novel class of random multivariate Bernstein polynomials, which induce a Dirichlet process mixture model of products of Dirichlet and beta densities. We study the theoretical properties of the model, including its topological support and posterior consistency. The model can be used for density and conditional density estimation, where both the response and predictors take values in the simplex space and/or hypercube. We illustrate the model's behavior through the analysis of simulated data and data from the 2005–2006 cycle of the U.S. National Health and Nutrition Examination Survey.
\end{abstract}

\section{Introduction}\label{sec:Introduction}

Modern datasets often include mixed-type variables that present unique challenges for statistical modeling. Among these, compositional variables—representing proportions or parts of a whole—are constrained to a \( d \)-dimensional simplex, \( S_d \), a non-Euclidean space that requires special consideration due to its distinct geometric structure:  
\begin{equation}\label{eq:simplex_space}  
S_d = \left\{(s_1, \ldots, s_d) \in [0,1]^{d} : \sum_{l=1}^{d} s_l \leq 1 \right\},  
\end{equation}  
where \( d+1 \) is the number of components defining the compositional variable. Additionally, interval-bounded variables—restricted to specific ranges (e.g., percentages or physical measurements with known upper and lower limits)—further complicate the modeling process.

Traditional statistical models often fail to properly capture the inherent dependencies and constraints of these variable types, leading to incorrect conclusions. This limitation underscores the need for innovative statistical methodologies that respect the geometric characteristics of these spaces. For instance, compositional data benefit from models that operate within the simplex geometry, leveraging techniques such as log-ratio transformations or models grounded in information geometry. Similarly, interval-bounded variables require approaches that explicitly account for their boundary constraints to ensure valid inferences.  

Significant attention has been devoted to the study of compositional and interval-bounded data, particularly from a parametric perspective. In the frequentist context, for example, \citet{aitchison1982statistical} laid the foundation for the statistical analysis of compositional data. However, parametric models often lack the flexibility needed to accommodate the complexities of real-world data, leading researchers to explore nonparametric approaches. In this direction, \citet{ouimet2020density,ouimet2021asymptotic,ouimet2022boundary} proposed nonparametric methods that rely on Dirichlet kernels and Bernstein polynomials. 

In the Bayesian context, early work by \citet{iyengar1996bayesian,iyengar1998box} introduced parametric models for compositional variables. \citet{petrone1999random} and \citet{petrone2002consistency} developed Bayesian nonparametric approaches for interval-bounded variables, while \citet{barrientos2015bayesian,barrientos2017fully} focused on compositional features. Additionally, \citet{ross2016bayesian} introduced a Bayesian nonparametric model leveraging compositional data for disease subtype identification, demonstrating its applicability in medical imaging and other fields.  

Despite these advances, to the best of our knowledge, no existing approach simultaneously models multiple compositional and interval-bounded variables. Developing robust statistical models for such data is not only a theoretical challenge but also of practical importance, as datasets with mixed-type variables are increasingly encountered in diverse fields, including environmental science, genomics, and social sciences. 
Our contribution addresses this need by proposing a Bayesian nonparametric approach based on a novel class of multivariate Bernstein polynomials (MBP). The proposed method can be employed for joint density estimation as well as conditional density estimation. For example, our approach can be used to estimate the conditional density of a compositional variable given interval-bounded and compositional variables. This setting can be interpreted as a regression problem, where the response variable is compositional. The advantage of our approach is that it accounts for the specific structure of the response on the simplex and the predictors on the product of a simplex and a two-dimensional unit cube.

The remainder of the article proceeds as follows. Section \ref{sec:background} reviews  Bernstein polynomials and their applications in statistical contexts. Section \ref{sec:proposedMBP} presents novel class of MBP on the product of bounded spaces, and their main properties. Section \ref{sec:randomMBP} introduces a random Bernstein polynomial process and investigates its prior and posterior properties. Section  \ref{sec:Illustration} illustrates the behavior of the model through simulations and an application to the  U.S. National Health and Nutrition Examination Survey (NHANES) dataset. Finally, Section \ref{sec:Conclusion} summarizes contributions and outlines future research directions.

\section{Bernstein polynomials}\label{sec:background}

Bernstein polynomials, introduced by \citet{bernstein1912demo}, were originally used to provide a constructive proof of the Weierstrass approximation theorem on $[0,1]$. They have since been extended to the $d$-dimensional hypercube and the $d$-dimensional simplex $S_d$, as defined in expression (\ref{eq:simplex_space}). For $F: [0,1]^d \to \mathbb{R}$, the multivariate Bernstein polynomial of degree $k$  on $[0,1]^d$ is given by
\begin{equation}\label{eq:hypercube_mbp}
\widetilde{B}_{k}^F(\vx) = \sum_{\bj \in \{0,\ldots,k\}^d} F\left(\frac{\bj}{k}\right) \prod_{l = 1}^d {\rm bin}(j_l \mid k, x_l), 
\end{equation}
where $\vx \in [0,1]^d$, $\vx = (x_1,\ldots,x_d)$, $\bj = (j_1,\ldots,j_d)$, and ${\rm bin}(\cdot \mid k, x)$ refers to the probability mass function of a binomial distribution with parameters $(k, x)$. $\widetilde{B}_{k}^F$ converges to $F$ at any point of continuity of $F$ as the degree $k \to \infty$, and the convergence is uniform if $F$ is continuous. If $F$ is the restriction of the cumulative distribution function (CDF) of a probability measure defined on $[0,1]^d$, then $\widetilde{B}_{k}^F$ is also the restriction of the CDF of a probability measured defined on $[0,1]^d$. 

Bernstein polynomials have been extensively studied from a frequentist perspective, with key asymptotic properties derived. Their application in density estimation on $[0,1]$ was introduced by \cite{vitale1975bernstein}, who analyzed bias and variance. Extensions to multivariate settings, including $[0,1]^2$ and $S_2$, were provided by \cite{tenbusch1994two}. Bernstein polynomials have also been employed for nonparametric Bayesian density estimation. \cite{petrone1999random} employed Bernstein polynomials for density estimation on $[0,1]$, demonstrating their approximation properties, and defining a prior with full support on the space of continuous densities by assuming $F$ follows a Dirichlet process (DP) and $k$ is random. Later, \cite{petrone2002consistency} established weak and Hellinger consistency of posterior distributions of this DP mixture (DPM) model. 

\cite{barrientos2015bayesian} extended the work of \cite{petrone1999random} for distributions defined on $S_d$. Their work provides the Bayesian counterpart to \cite{tenbusch1994two}'s frequentist approach, with a slight modification. This modification was required because, although the derivative of Tenbusch’s Bernstein polynomial remains consistent at the interior points of the simplex, it does not serve as a valid density function for finite $k$ and finite sample sizes. The multivariate Bernstein polynomial of degree $k$ on $S_d$ for a function $F:\mathbb{R}^d\to\mathbb{R}$
proposed by \cite{barrientos2015bayesian} is given by\begin{equation}\label{eq:simplex_mbp}
\widetilde{B}_{k}^F(\vx)=\sum_{\boldsymbol{j}\in \mathcal{G}_{d}^{k}}F\left(\frac{\boldsymbol{j}}{k}\right) {\rm mult}(\boldsymbol{j}\mid k+d-1,\vx),
\end{equation}
where $\vx \in S_d$, $\mathcal{G}_{d}^{k}=\{(j_1,\ldots,j_d)\in \{0,\ldots,k\}^d:\sum_{l=1}^d j_l\leq k+d-1\}$, and ${\rm mult}(\cdot\mid k+d-1,\vx)$ refers to the probability mass function of a multinomial distribution with parameters $(k+d-1,\vx)$. This modified class of multivariate Bernstein polynomials retains the appealing approximation properties of univariate Bernstein polynomials. Similar to Petrone's work, \cite{barrientos2015bayesian} defined a prior and posterior distribution on $S_d$. They demonstrated that this prior has full support and that the posterior distribution induced by the model is  both weak and strong consistent.

Bernstein polynomials have been successfully employed in settings beyond density \break estimation, including density regression \citep{mckay2011variable,barrientos2017fully,wehrhahn2022dependent}, time series \citep{edwards2019bayesian}, and spatio-temporal models \citep{richardson2020spatiotemporal}.




\section{The novel class of multivariate Bernstein polynomials}\label{sec:proposedMBP}


\subsection{Definition}

We define a class of MBP on the product space $\Delta_{d}$, consisting of $M$ simplex spaces and an hypercube, $\Delta_{d} \equiv S_{d_1} \times \cdots \times S_{d_M} \times [0,1]^{d_{M+1}}$, where $d=\sum_{i=1}^{M+1} d_i$ is the dimension of the space. The definition uses the product of $M$ multinomial distributions (for the product of simplices) and $d_{M+1}$ binomial distributions (for the hypercube). 
\begin{defn}{\bf{(MBP on $\Delta_{d}$)}}\label{df:MBP}
For a function $F : \mathbb{R}^{d_1+\cdots+d_{M+1}} \rightarrow \mathbb{R}$, its MBP of degree $(k_1 ,\ldots , k_M , k_{M+1})$ defined on $\Delta_{d}$  is given by
\vspace{-0.5cm}
\begin{eqnarray}\label{eq:GeneralMBP}
\nonumber & & \hspace{-14mm} B_{k_1 ,\ldots , k_M , k_{M+1}}^F(\bx) \\
&=&\sum_{\boldsymbol{j}_1\in J_{d_1}^{k_1}}\cdots\sum_{\boldsymbol{j}_M\in J_{d_M}^{k_M}}\sum_{\boldsymbol{j}_{M+1}\in \{0,\ldots,k_{M+1}\}^{{d_{M+1}}}}F\left(\frac{\boldsymbol{j}_1}{k_1-d_1+1},\ldots,\frac{\boldsymbol{j}_M}{k_M-d_M+1},\frac{\boldsymbol{j}_{M+1}}{k_{M+1}}\right) \times \nonumber\\ 
&& 
\prod_{l=1}^{M}
{\rm mult}(\boldsymbol{j}_l\mid k_l,\vx_l)\times \prod_{l=1}^{{d_{M+1}}}{\rm bin}\left(j_{(M+1),l}\mid k_{M+1},x_{(M+1),l}\right),
\end{eqnarray}
where
$\bx = (\vx_1,\ldots, \vx_{M+1}) \in \Delta_d$, with $\vx_1 \in S_{d_1}$, \ldots, $\vx_M \in S_{d_M}$, and $\vx_{M+1} \in [0,1]^{d_{M+1}}$, $\vx_l = (x_{l,1}, \ldots, x_{l,d_l})$,
$
(k_1, \dots, k_M, k_{M+1}) \in \{d_1, d_1+1 \ldots,\} \times \cdots \times \{d_M, d_M+1 \ldots,\} \times \mathbb{N}
$, $\boldsymbol{j}_m=(j_{m,1},\ldots,j_{m,d_m})$, 
$$J_{d_m}^{k_m}=\left\{(j_{m,1},\ldots,j_{m,d_m})\in\{0,\ldots,k_m\}^{d_m}:\sum_{l=1}^{d_m} j_{m,l}\leq k_m\right\},$$ 
and $m = 1,\ldots,M$.
\end{defn} 

\subsection{Properties}

Like the Bernstein polynomials described in Section \ref{sec:background}, the novel class of MBP defined above exhibits appealing approximation properties, which are summarized in the following theorem, which proof is provided in Section \ref{proof:appro2F} of the supplementary material.
\begin{theorem}\label{thm:appro2F}
    For any given function $F : \mathbb{R}^{d_1+\cdots+d_{M+1}} \rightarrow \mathbb{R}$, its MBP, $B_{k_1 ,\ldots , k_M , k_{M+1}}^F(\bx)$, converges to $F(\bx)$ at any point of continuity $\bx\in \Delta_{d}$ of $F$, as $\min\{k_1,\ldots, k_{M+1}\}\to\infty$. Moreover, this convergence is uniform when $F$ is continuous.
\end{theorem}

When $F$ is the CDF of a probability measure defined on $\Delta_{d}$, then $B_{k_1 ,\ldots , k_M , k_{M+1}}^F$ is the restriction  of the CDF of a probability measure defined on $\Delta_{d}$, provided that $F$ satisfies certain conditions on the  boundary of $\Delta_{d}$. Additionally, if $F$ admits a continuous density function $f$, 
the $(d_1+\ldots+d_{M+1})$th-order partial derivative of its MBP  uniformly converges to $f$. The proof of the following theorem is provided in Section \ref{proof:mbp_cdf} of the 
supplementary material. 

\begin{theorem}\label{prop:mbp_cdf}
    Let $F:\mathbb{R}^{d_1+\cdots+d_{M+1}} \rightarrow \mathbb{R}$ be the CDF of a probability measure defined on  $\Delta_{d}$, such that $F(\bx)=0$, for every $\bx\in B_{\Delta_{d}}$, where 
\begin{eqnarray}\label{eq:boundary} \nonumber
    {\rm B}_{\Delta_{d}}&=&{\Delta_{d}}\setminus\Bigg(\left\{\sum_{l=1}^{d_1} x_{1,l}\leq 1, x_{1,l}>0, \forall l\right\}\cap\cdots\cap\left\{\sum_{l=1}^{d_M} x_{M,l}\leq 1, x_{M,l}>0, \forall l\right\}\cap\Bigg.\\\nonumber &&\Bigg.\Big\{0<x_{(M+1),l}\leq 1, l\in\{1,\ldots,d_{M+1}\} \Big\}\Bigg).    
    \end{eqnarray}
    Then, $B_{k_1 ,\ldots , k_M , k_{M+1}}^F$ is the restriction of the CDF of an absolutely continuous probability measure, w.r.t. Lebesgue measure, defined on $\Delta_{d}$, with density function given by
\begin{eqnarray}
   \nonumber b_{k_1, \ldots, k_M, k_{M+1}}^F(\bx) &=& \sum_{\boldsymbol{j}_1\in 
    I_{d_1}^{k_1}}\cdots\sum_{\boldsymbol{j}_M\in I_{d_M}^{k_M}}\sum_{\boldsymbol{j}_{M+1}\in \{1,\ldots,k_{M+1}\}^{d_{M+1}}}
    \Bigg\{
       F\left(A_{\boldsymbol{j}_1,\ldots, \boldsymbol{j}_{M+1} } \right)\times \Bigg.
    \\
    & &\nonumber 
\prod_{l=1}^{M}
        {\rm dir}\left(\vx_l\mid \boldsymbol{\alpha}(k_l-d_l+1,\boldsymbol{j}_l)\right)\times\\
    & & \Bigg. \prod_{l=1}^{d_{M+1}}{\rm beta}(x_{(M+1),l}\mid j_{(M+1),l},k_{M+1}-j_{(M+1),l}+1) \Bigg \},
    \end{eqnarray}\label{eq:MBPdens}where
$\bx = (\vx_1,\ldots, \vx_{M+1}) \in \Delta_d$, with $\vx_1 \in S_{d_1}, \ldots, \vx_M \in S_{d_M}$, and $\vx_{M+1} \in [0,1]^{d_{M+1}}$, $\vx_l = (x_{l,1}, \ldots, x_{l,d_l})$,
$
(k_1, \dots, k_M, k_{M+1}) \in \{d_1, d_1+1 \ldots,\} \times \cdots \times \{d_M, d_M+1 \ldots,\} \times \mathbb{N}
$,  \break $I_d^k=\left\{\boldsymbol{j}\in J_d^k: j_l>0, l\in\{1,\ldots,d\}\right\}$,
$A_{\boldsymbol{j}_1,\ldots, \boldsymbol{j}_{M+1}} = \rho_{d_1,\boldsymbol{j}_1}^{k_1-d_1+1}\times\cdots\times\rho_{d_M,\boldsymbol{j}_M}^{k_M-d_M+1}\times\rho_{d_{M+1},\boldsymbol{j}_{M+1}}^{k_{M+1}}$ 
with 
$\rho_{d,\boldsymbol{j}}^k=\left(\frac{j_1-1}{k},\frac{j_1}{k}\right]\times\cdots\times\left(\frac{j_d-1}{k},\frac{j_d}{k}\right]$, ${\rm dir}\left(\cdot\mid \boldsymbol{\alpha}\right)$ refers to the density of a Dirichlet distribution with parameter $\boldsymbol{\alpha}$,
$\boldsymbol{\alpha}(k,\boldsymbol{j})=\left(\boldsymbol{j},k+d-\sum_{l=1}^d j_l\right)$, and 
${\rm beta}(\cdot \mid a,b)$ refers to the density of a Beta distribution with parameters $a$ and $b$.
\end{theorem}

The resulting density $b_{k_1, \cdots, k_M, k_{M+1}}^F$ has also appealing approximation properties. The proof of the following theorem is provided in Section \ref{proof:appro2f} of the online supplementary material.

\begin{theorem}\label{thm:appro2f} Let $F:\mathbb{R}^{d_1+\cdots+d_{M+1}} \rightarrow \mathbb{R}$ be the CDF of an absolutely continuous, w.r.t. Lebesgue measure, probability measure defined on $\Delta_{d}$. Then, $b_{k_1 ,\ldots , k_M , k_{M+1}}^F$ converges uniformly to $f$, a density of $F$, as $\min\{k_1,\ldots,k_M,k_{M+1}\}\to \infty$.
\end{theorem}

Under the Bernstein density given by expression (\ref{eq:MBPdens}), the marginal density for 
$\bx^{[-i]}=(\vx_1,\ldots,\vx_{i-1}, \vx_{i+1},\ldots, \vx_{M+1})$, $ b_{k_1, \ldots, k_M, k_{M+1}}^{[-i],F}$, is given by     
\begin{eqnarray}\nonumber
 b_{k_1, \ldots, k_M, k_{M+1}}^{[-i],F}\left(\bx^{[-i]}\right) &= & \sum_{\boldsymbol{j}_1\in 
    I_{d_1}^{k_1}}\cdots\sum_{\boldsymbol{j}_M\in I_{d_M}^{k_M}}\sum_{\boldsymbol{j}_{M+1}\in \{1,\ldots,k_{M+1}\}^{d_{M+1}}}
    \Bigg\{
       F\left(A_{\boldsymbol{j}_1,\ldots, \boldsymbol{j}_{M+1} } \right)\times \Bigg.
    \\
    &&\nonumber 
    \prod_{l\in \{1,\ldots,i-1,i+1,\dots, M\}}
        {\rm dir}\left(\vx_l\mid \boldsymbol{\alpha}(k_l-d_l+1,\boldsymbol{j}_l)\right)\times
        \\\nonumber
    &&\Bigg. \prod_{l=1}^{d_{M+1}}{\rm beta}(x_{(M+1),l}\mid j_{(M+1),l},k_{M+1}-j_{(M+1),l}+1) \Bigg \},
        \end{eqnarray}
and
\begin{eqnarray}\nonumber
  b_{k_1, \ldots, k_M, k_{M+1}}^{[-i],F}\left(\bx^{[-i]}\right) &= & \sum_{\boldsymbol{j}_1\in 
    I_{d_1}^{k_1}}\cdots\sum_{\boldsymbol{j}_M\in I_{d_M}^{k_M}}\sum_{\boldsymbol{j}_{M+1}\in \{1,\ldots,k_{M+1}\}^{d_{M+1}}}
    \Bigg\{
       F\left(A_{\boldsymbol{j}_1,\ldots, \boldsymbol{j}_{M+1} } \right)\times \Bigg.
    \\
    &&\nonumber \Bigg.
    \prod_{l=1}^{M}
        {\rm dir}\left(\vx_l\mid \boldsymbol{\alpha}(k_l-d_l+1,\boldsymbol{j}_l)\right) \Bigg \},
        \end{eqnarray}
if $i=1,\ldots,M$ and $i=M+1$, respectively. Furthermore, the conditional density for
$\vx_i$, given $\bx^{[-i]}$, is given by
\begin{eqnarray}\nonumber
   b_{k_1, \ldots, k_M, k_{M+1}}^{i,F}\left(\vx_i \mid \bx^{[-i]}\right) &= & \sum_{\boldsymbol{j}_1\in 
    I_{d_1}^{k_1}}\cdots\sum_{\boldsymbol{j}_M\in I_{d_M}^{k_M}}\sum_{\boldsymbol{j}_{M+1}\in \{1,\ldots,k_{M+1}\}^{d_{M+1}}}
    \Bigg\{
      W_{\boldsymbol{j}_1,\ldots, \boldsymbol{j}_{M+1} }\left(\bx^{[-i]}\right)       \times \Bigg.
    \\\nonumber &&
\Bigg.
   {\rm dir}\left(\vx_i\mid \boldsymbol{\alpha}(k_i-d_i+1,\boldsymbol{j}_i)\right)\Bigg \},
        \end{eqnarray}

and
\begin{eqnarray}\nonumber
 b_{k_1, \ldots, k_M, k_{M+1}}^{i,F}\left(\vx_i \mid \bx^{[-i]}\right) &= & \sum_{\boldsymbol{j}_1\in 
    I_{d_1}^{k_1}}\cdots\sum_{\boldsymbol{j}_M\in I_{d_M}^{k_M}}\sum_{\boldsymbol{j}_{M+1}\in \{1,\ldots,k_{M+1}\}^{d_{M+1}}}
    \Bigg\{
       W_{\boldsymbol{j}_1,\ldots, \boldsymbol{j}_{M+1} }\left(\bx^{[-i]}\right)\times \Bigg.
    \\
    &&\Bigg. \prod_{l=1}^{d_{M+1}}{\rm beta}(x_{(M+1),l}\mid j_{(M+1),l},k_{M+1}-j_{(M+1),l}+1) \Bigg \},
        \end{eqnarray}
if $i=1,\ldots,M$ and $i=M+1$, respectively, where
\begin{eqnarray}\hspace{-12mm}\nonumber
 W_{\boldsymbol{j}_1,\ldots, \boldsymbol{j}_{M+1}} \left(\bx^{[-i]}\right) &=&
\frac{F\left(A_{\boldsymbol{j}_1,\ldots, \boldsymbol{j}_{M+1} } \right)}{ b_{k_1, \ldots, k_M, k_{M+1}}^{[-i],F}\left(\bx^{[-i]}\right) }
\times\\\nonumber
&&\prod_{l\in \{1,\ldots,i-1,i+1,\dots, M\}}
        {\rm dir}\left(\vx_l\mid \boldsymbol{\alpha}(k_l-d_l+1,\boldsymbol{j}_l)\right)\times  \\\nonumber
&&\prod_{l=1}^{d_{M+1}}{\rm beta}(x_{(M+1),l}\mid j_{(M+1),l},k_{M+1}-j_{(M+1),l}+1),
  \end{eqnarray}
and
\begin{eqnarray}\hspace{-9cm}\nonumber
 W_{\boldsymbol{j}_1,\ldots, \boldsymbol{j}_{M+1}} \left(\bx^{[-i]}\right) &=&
\frac{F\left(A_{\boldsymbol{j}_1,\ldots, \boldsymbol{j}_{M+1} } \right)}{ b_{k_1, \ldots, k_M, k_{M+1}}^{[-i],F}\left(\bx^{[-i]}\right) }
\times\\\nonumber
&&\prod_{l\in \{1,\dots, M\}}
        {\rm dir}\left(\vx_l\mid \boldsymbol{\alpha}(k_l-d_l+1,\boldsymbol{j}_l)\right),
  \end{eqnarray}
if $i=1,\ldots,M$, and $i=M+1$, respectively.

\section{The class of random MBP on \texorpdfstring{$\Delta_{d}$}{Delta extunderscore d}}\label{sec:randomMBP}

Suppose we observed the realization of $n$  independent and identically distributed (i.i.d.) random vectors  $\obs_1, \ldots, \obs_n$, 
$$\obs_1, \ldots, \obs_n \mid G \overset{i.i.d.}{\sim} G,$$
where $G$ is  a probability measure on $\Delta_{d}$, absolutely continuous with respect to Lebesgue measure, and that admit a continuous density function. We induced a prior distribution $\Pi$ on $G$ by modeling the density function of $G$ using $b_{k_1, \cdots, k_M, k_{M+1}}^F$, 
$g \equiv b_{k_1 ,\ldots , k_M , k_{M+1}}^F$, where
$b_{k_1, \cdots, k_M, k_{M+1}}^F$ given by expression (\ref{eq:MBPdens}), and where $F$ is a discrete random probability measure defined on $\Delta_{d}$, and 
$
(k_1, \dots, k_M, k_{M+1}) \in \{d_1, d_1+1 \ldots,\} \times \cdots \times \{d_M, d_M+1 \ldots,\} \times \mathbb{N}
$ are random polynomial degrees.

\subsection{The prior model}\label{sec:PBPP}

Let $\mathscr{F}$ be the space of all probability measures defined on $\Delta_{d}$ that are absolutely continuous with respect to Lebesgue measure and have continuous density function on $\Delta_{d}$. We equip $\mathscr{F}$ with a $\sigma$-algebra, $\mathscr{B}(\mathscr{F})$, which is generated by a specified topology on $\mathscr{F}$. The choice of topology will be determined according to the requirements of the analysis and the properties of the statistical model under consideration. 
We define a random MBP process on $\Delta_{d}$ by considering random polynomial degrees and where  $F$ is a DP with precision parameter $M_0 \ge 0$ and centering measure $F_0$ on $\Delta_{d}$, denoted as $F \mid M_0, F_0\sim \text{DP}(M_0, F_0)$. 

\begin{defn}\label{df:PBPP}
Let $G$ be an $\mathscr{F}$-valued stochastic process, with probability law $\Pi$ defined on $\mathscr{F}$, such that the density function $g$ with respect to Lebesgue measure is given by  
\begin{eqnarray}\nonumber
g(\bx) &=& \sum_{\boldsymbol{j}_1\in 
    I_{d_1}^{k_1}}\cdots\sum_{\boldsymbol{j}_M\in I_{d_M}^{k_M}}\sum_{\boldsymbol{j}_{M+1}\in \{1,\ldots,k_{M+1}\}^{d_{M+1}}}
    \Bigg\{
       F\left(A_{\boldsymbol{j}_1,\ldots, \boldsymbol{j}_{M+1} } \right)\times \Bigg.
    \\
    & &\nonumber 
\prod_{l=1}^{M}
        {\rm dir}\left(\vx_l\mid \boldsymbol{\alpha}(k_l-d_l+1,\boldsymbol{j}_l)\right)\times\\\nonumber
    & & \Bigg . \prod_{l=1}^{d_{M+1}}{\rm beta}(x_{(M+1),l}\mid j_{(M+1),l},k_{M+1}-j_{(M+1),l}+1) \Bigg \},
\end{eqnarray}
for  $l=1,\ldots,M+1$,
$$
k_l \mid \lambda_l \sim  p_l(\cdot\mid \lambda_l),$$
and
$$
 F\mid M_0, F_0 \sim{\rm DP}(F_0,M_0),
$$
where $p_l(\cdot|\lambda_l)$ is probability mass function on $\mathbb{N}$ parameterized by $\lambda_l$, $l=1,\ldots,M+1$, $M_0 \in \mathbb{R}_+$, and $F_0$ is a probability measure defined on $\Delta_{d}$, which put $0$ mass on boundary ${\rm B}_{\Delta_{d}}$. The process $G$ will be referred to as the Dirichlet multivariate Bernstein polynomial \break process (DMBPP) with parameters $(\lambda_1, \ldots, \lambda_M, \lambda_{M+1}, M_0, F_0)$, and denoted by \break ${\rm DMBPP}(\lambda_1, \ldots, \lambda_M, \lambda_{M+1}, M_0, F_0)$.
\end{defn}

\subsection{Properties of the DMBPP prior}

\subsubsection{Support}\label{sec:PBPPprior}

The flexibility of a Bayesian nonparametric model for density estimation refers to its ability to capture any density within the space of possible densities. This flexibility is determined by the support of the prior distribution $\Pi$. A larger support indicates a more flexible model. In the space $\mathscr{F}$, the support of $\Pi$ is defined as the smallest closed set of probability one. Full support implies that the support of the prior is $\mathscr{F}$ itself. We show that under mild conditions $\Pi$ has full support under the topology induced by the $L_\infty$-norm, which is induced by neighborhoods of $G_0 \in \mathscr{F}$ of the form
\begin{eqnarray*}
Q_{\epsilon}({{G_0}}) & = & \left\{G\in\mathscr{F}:\|g-g_0\|_{\infty}<\epsilon\right\},
\end{eqnarray*}
where $g$ and $g_0$ are densities of $G$ and $G_0$, respectively. The proof of the following theorem is provided in Section \ref{proof:support} of the online supplementary material. 

\begin{theorem}\label{thm:support}
     Let $G$ be a DMBPP$(\lambda_1, \ldots, \lambda_M, \lambda_{M+1}, M_0, F_0)$. Suppose that DBMPP is specified such that the following conditions hold:
\begin{itemize}

   \item[(i)]  For every  $l=1,\ldots,M$,   
   $p_l(\cdot\mid \lambda_l)$ has full support on $\{d_l, d_l+1,\ldots\}$, and $p_{M+1}(\cdot \mid \lambda_{M+1})$ has full support on $\mathbb{N}$.

   \item[(ii)] $M_0>0$, and
   
\item[(iii)] $F_0$ has positive density on $\Delta_{d}$.
\end{itemize}
Then, $\Pi$, the distribution of $G$, has full $L_{\infty}$-support on $\mathscr{F}$.
\end{theorem}
It is important to emphasize that the previous theorem implies that $\Pi$ has also full support under any $L_{p}$ topology, for any $0< p < +\infty$.

\subsubsection{Posterior consistency}\label{sec:PBPPposterior}

We show that the posterior distribution induced by the DMBPP prior under i.i.d. sampling has appealing large-sample properties. Let $ \Pi\left(\cdot \mid \obs_1, \ldots, \obs_n\right)$ be a version of the posterior distribution given the data $\obs_1, \ldots, \obs_n$. We consider two different topologies.  Specifically, we consider weak and $L_1$ topologies, induced by neighborhoods of the form
\begin{eqnarray}
\nonumber
W_{\epsilon}({{G_0}}) & = & \left\{G\in \mathscr{F}:\left| \int_{\Delta_{d}} \phi(\bx) g(\bx)d\bx -\int \phi(\bx) g_0(\bx)d\bx\right| <\epsilon \right\},
\end{eqnarray}
for any bounded and continuous function $\phi$, and
\begin{eqnarray}
\nonumber
T_{\epsilon}({{G_0}}) & = &\left\{G\in \mathscr{F}:\|g-g_0\|_1=\int_{\Delta_{d}} |g(\bx)-g_0(\bx)| d\bx<\epsilon\right\},
\end{eqnarray}
respectively, with $g$ and $g_0$ being densities of $G$ and $G_0$, respectively. It is important to note that the coarsest topology is the one induced by the weak neighborhoods, followed by that induced by the $L_1$ norm.

The posterior distribution $\Pi\left(\cdot \mid \obs_1, \ldots, \obs_n\right)$ is said to be 
weak-consistent at $G_0$ if for any $\epsilon>0$,
\begin{eqnarray*}
             \Pi\left({W_{\epsilon}({G_0})}\mid \obs_1, \ldots, \obs_n\right) \underset{n \to \infty}{\longrightarrow} 1,\quad G_0-\textit{a.s.}.
\end{eqnarray*}
We show that, under mild conditions, the posterior distribution induced by the DMBPP prior is weakly consistent under i.i.d. sampling. The proof of the following  theorem is provided in Section \ref{proof:postweak} of the online supplementary material.

\begin{theorem}\label{thm:postweak}
    Assume that  $G_0\in \mathscr{F}$ admits a continuous density function $g_0$ and that $\obs_1, \ldots, \obs_n \mid G_0 \overset{i.i.d.}{\sim} G_0$ is the true data generating mechanism. Assume that the prior $\Pi$ is the law of the ${\rm DMBPP}(\lambda_1, \ldots, \lambda_M, \lambda_{M+1}, M_0, F_0)$, such that the conditions of Theorem \ref{thm:support} are met. 
    Then,  $\Pi\left(\cdot \mid \obs_1, \ldots, \obs_n\right)$ is weakly consistent at $G_0$.
\end{theorem}

The posterior $\Pi\left(\cdot \mid \obs_1, \ldots, \obs_n\right)$ is said to be $L_1$-consistent at $G_0$ (also known as strong consistency) if for any $\epsilon > 0$,
\begin{eqnarray*}
             \Pi\left(T_{\epsilon}({G_0})\mid \obs_1, \ldots, \obs_n\right) \underset{n \to \infty}{\longrightarrow} 1,\quad G_0- \textit{a.s.}.
\end{eqnarray*}
We show that under slightly stronger conditions on the prior distribution of the polynomial degrees $p(k_l)$, $l=1,\ldots, M+1$,  the posterior distribution induced by the DMBPP prior is consistent in $L_1$  under i.i.d. 
The proof of the following theorem, which is based on Theorem 2 in \cite{ghosal1999posterior}, is provided in Section \ref{proof:poststrong} of the online supplementary material.

\begin{theorem}\label{thm:poststrong}
Assume that  $G_0\in \mathscr{F}$ admits a bounded continuous density function $g_0$ and that $\obs_1, \ldots, \obs_n \mid G_0 \overset{i.i.d.}{\sim} G_0$ is the true data generating mechanism. Assume that the prior $\Pi$ is the law of the ${\rm DMBPP}(\lambda_1, \ldots, \lambda_M, \lambda_{M+1}, M_0, F_0)$, such that the conditions of Theorem \ref{thm:support} are met. Also assume that 
 there exists a sequence $(k_n^{(1)}, \ldots, k_n^{(M)}, k_n^{(M+1)})$, with $k_n^{(l)} \to \infty$, such that
\begin{itemize}
\item[(i)] $k_n^{(l)}=o(n^{\frac{1}{(M+1)d_{l}}}), l=1,\ldots, M, M+1$,
\item[(ii)] $\sum_{k_1 > k_n^{(1)}}\cdots\sum_{k_M > k_n^{(M)}}\sum_{k_{M+1} > k_n^{(M+1)}}
p_1(k_1 \mid \lambda_1)
\cdots p_{M+1}(k_{M+1} \mid \lambda_{M+1}) < ce^{-nr}$ for some $c,r>0$.
\end{itemize}    
    Then,  $\Pi\left(\cdot \mid \obs_1, \ldots, \obs_n\right)$ is $L_1$ consistent at $G_0$.
\end{theorem}


It is important to emphasize that Theorems \ref{thm:support}, \ref{thm:postweak}, and \ref{thm:poststrong} can be easily extended for a definition of the random MBP process on  $\Delta_{d}$ is based on any stick-breaking processes \citep[see, e.g.,][]{ishwaran2001gibbs}. As long as the base measure has positive density and the weights sum to one almost surely, the corresponding stick-breaking process would satisfy the required conditions. 

It is also important to emphasize that the product of truncated Poisson or truncated negative binomial distributions as a prior for the polynomial degree $(k_1, \ldots, k_{M}, k_{M+1})$, satisfy the conditions outlined in Theorems \ref{thm:support} and  \ref{thm:postweak}.
Recall that the truncation is necessary to remove the lower natural values for which the MBP is not defined. Specifically, the prior should assign 0 mass to $k_l < d_l$, for $l = 1, \ldots, M$, and $k_{M+1} < 1$. To achieve the conditions of Theorem \ref{thm:poststrong}, extra care needs to be taken. For this, we can construct a prior satisfying conditions i) and ii) using any discrete distribution to define the probabilities for the initial values and modify its tail to meet the requested conditions. Let $p_0$ denote the probability mass function of, for example, a Poisson or negative binomial distribution. Define $p_{l}(k_1 \mid \lambda_{l}) := p_0(k)$ for $d_l \leq k_l < \tilde{k}$, for $l = 1, \ldots, M$, and $1 \leq k_{M+1} < \tilde{k}$,  and define $C_{\tilde{k}} := \sum_{k=1}^{\tilde{k}-1} p_{l}(k \mid \lambda_{l})$. Then, for $j \geq 1$, define
\begin{eqnarray*}
& &
\hspace{-4mm} p_{l}(\tilde{k} + j \mid \lambda_l) 
\\ & & := 
\begin{cases} 
(1 - C_{\tilde{k}}) \exp\left( -\lambda_{l} \left( j + \tilde{k} \right)^{(M+1)d} \right), & \text{for } j = 1, \\
(1 - C_{\tilde{k}}) \left[ \exp\left( -\lambda_{l} \left( j - 1 + \tilde{k} \right)^{(M+1)d} \right) - \exp\left( -\lambda_{l} \left( j + \tilde{k} \right)^{(M+1)d} \right) \right], & \text{for } j > 1.
\end{cases}
\end{eqnarray*}
Notice that under this prior, it follows that
$$
p_{l}(k_l>\tilde{k}+j\mid\lambda_{l}) = (1-C_{\tilde{k}})\exp\left\{ -\lambda_{l}(j+\tilde{k})^{(M+1)d}\right\}.
$$
Therefore, if we choose $k_n^{(l)} = \lceil n^{c_1/(M+1)d_l} \rceil$ for some $c_1 \in (0,1)$, then condition i) is met, and 
$$
p_{l}(k_l > k_n^{(l)} \mid \lambda_{l}) = p_{l}(k > \lceil n^{c_1/(M+1)d_l} \rceil \mid \lambda_{l}) < \exp\left\{ -\lambda_{l}n^{c} \right\} \leq \exp\left\{ -\lambda_{l}n \right\}, 
$$
which implies condition ii).

\subsection{Properties of the implied prior distribution on the conditional and marginal distributions}

We now turn our attention to the estimation of the marginal and conditional distributions of $G_0$. At first glance, if we achieve consistency for the joint distribution $G_0$, it is reasonable to expect consistency for its marginals and conditionals as well. However, it is necessary to formalize this idea to determine the specific topologies under which the consistency is achieved. Set $\bx = (\vx_1,\ldots, \vx_{M+1}) \in \Delta_d$, such that $\vx_1 \in S_{d_1}$, \ldots, $\vx_M \in S_{d_M}$, and $\vx_{M+1} \in [0,1]^{d_{M+1}}$. We focus on the marginal distribution of  $\bx^{[-i]} =
(\vx_1,\ldots, \vx_{i-1},\vx_{i+1}, \ldots, \vx_{M+1}) \in \Delta_d^{[-i]}$, $i=1,\ldots,M+1$ and on the conditional distribution of $\vx_i \mid \bx^{[-i]}$. Let $d_{(i,G_0,L_1)}$ be the \break $G_0$-integrated $L_1$ distance for conditional distributions, which is defined as follows  
\begin{eqnarray*}
d_{(i,G_0, L_1)}\left(G\right)  = \int_{\Delta_d^{[-i]}} \left\{\int_{S_{d_i}}  |g\left(\vx_i\mid \bx^{[-i]}\right)-g_0\left(\vx_i\mid \bx^{[-i]}\right)|d\vx_i \right\} g_0\left(\bx^{[-i]}\right)d\bx^{[-i]},
\end{eqnarray*} 
where $g\left(\cdot\mid \bx^{[-i]}\right)$ and 
$g_0\left(\cdot\mid \bx^{[-i]}\right)$ are versions of the conditional distribution of \break $\vx_i \mid \bx^{[-i]}$ under $G$ and $G_0$, respectively, and $g_0\left(\bx^{[-i]}\right)$ denotes the marginal distribution of $\bx^{[-i]}$ under $G_0$.


We show that under the conditions of Theorem \ref{thm:poststrong}, the posterior distribution induced by the DMBPP prior is strongly consistent at the marginals and conditionals distributions. The proof of the following theorem is provided in Section \ref{proof:postmarcon} of the online supplementary material.

\begin{theorem}\label{thm:postmarcon}
Assume that the conditions of Theorem \ref{thm:poststrong} are met. Then, for every $i \in \{1,\ldots, M+1\}$, the posterior distribution $\Pi\left(\cdot \mid \obs_1, \ldots, \obs_n\right)$ is $L_1$-marginal consistent at the marginal distribution of $\bx^{[-i]}$ under $G_0$ and is $G_0$-integrated $L_1$ consistent at the conditional distribution of $\vx_i\mid\bx^{[-i]}$ under $G_0$.
\end{theorem}


\section{Illustrations}\label{sec:Illustration}
In this section, we illustrate the performance of the model with a simulation study and an application to data from the 2005 – 2006 cycle of the U.S. National Health and Nutrition Examination Survey (NHANES). The simulation scenarios illustrate the model's capability to accurately estimate the true joint density, as well as its marginal and conditional distributions. In the application, we utilize the model to estimate the marginal distributions of each data component and characterize the conditional densities. To implement the model, we use a finite-dimensional approximation of the Dirichlet process based on a truncated version of \cite{sethuraman1994constructive}'s stick-breaking representation.
For more details, we refer the reader to  Section~\ref{sec:Model_implement}, in the online supplementary materials.

\subsection{Simulated data analyses}\label{sec:Simulation}
We consider two simulation scenarios that represent different degrees of complexity and shapes. These scenarios are based on mixtures of products of Dirichlet and beta distributions, which are not particular cases of the proposed model. For Scenario I, the true joint density is a 3-component mixture of the product of a two-dimensional Dirichlet distribution and a beta distribution. For Scenario II we consider a 7-component mixture of the product of a three-dimensional Dirichlet  distribution, a two-dimensional Dirichlet distribution, and two beta distributions. Notice that Scenario II considers the same sample space of the NHANES dataset. The specifications of true joint densities for Scenarios I and II are given in Section \ref{sec:Simulation_fig} of the online supplementary material.

    

For each scenario, we consider three sample sizes: $n = 250$, $500$, and $1000$. For each scenario and sample size, we simulate 100 datasets. For each simulated dataset, 25 Markov chains of length 11,000 were generated. In each chain, a reduced chain of 40 samples was obtained after a burn-in period of 10,000, keeping 1 out of 25 samples. Thus, for each dataset, the posterior inference is based on the 1,000 samples.

Here we focus on marginal density estimation and conditional density estimation in Scenario I and II, respectively.  The results for conditional density estimation in Scenario I and marginal density estimation in Scenario II are provided in Section \ref{sec:Simulation_fig} of the online supplementary material. To assess the performance of the proposed model in estimating marginal and joint densities, we compute the mean, across Monte Carlo replicates, of posterior expected 
$L_1$ distance to the truth $g_0$, denoted by 
${\rm MPEL}_1$ and given by
    \vspace{-0.5cm}\begin{eqnarray*}
    && {\rm MPEL}_1 = \frac{1}{N}\sum_{l=1}^{N} E\left[\Vert g - g_0\Vert_1 \mid \obs_{(l,1)}, \ldots, \obs_{(l,n)}\right].
\end{eqnarray*}

\begin{figure}[H] 
\centering
\captionsetup[subfloat]{labelformat=empty, aboveskip=1.5pt, belowskip=1.5pt}
\subfloat[]{
\includegraphics[page=2]{rescaled/simu_fig/simu_plots_marginal_S1_Univariates.pdf}}
\subfloat[]{
\includegraphics[page=3]{rescaled/simu_fig/simu_plots_marginal_S1_Univariates.pdf}}
\subfloat[]{
\includegraphics[page=4] {rescaled/simu_fig/simu_plots_marginal_S1_Univariates.pdf}}
\subfloat[]{
\includegraphics[page=1]{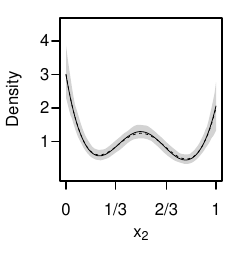}}\\[-10pt]
\subfloat[]{
\includegraphics[page=6]{rescaled/simu_fig/simu_plots_marginal_S1_Univariates.pdf}}
\subfloat[]{
\includegraphics[page=7]{rescaled/simu_fig/simu_plots_marginal_S1_Univariates.pdf}}
\subfloat[]{
\includegraphics[page=8]{rescaled/simu_fig/simu_plots_marginal_S1_Univariates.pdf}}
\subfloat[]{
\includegraphics[page=5]{rescaled/simu_fig/simu_plots_marginal_S1_Univariates.pdf}}\\[-10pt]
\subfloat[]{
\includegraphics[page=10]{rescaled/simu_fig/simu_plots_marginal_S1_Univariates.pdf}}
\subfloat[]{
\includegraphics[page=11]{rescaled/simu_fig/simu_plots_marginal_S1_Univariates.pdf}}
\subfloat[]{
\includegraphics[page=12]{rescaled/simu_fig/simu_plots_marginal_S1_Univariates.pdf}}
\subfloat[]{
\includegraphics[page=9]{rescaled/simu_fig/simu_plots_marginal_S1_Univariates.pdf}}
\caption{Simulated data -- Scenario I. Posterior inference for the marginal distributions. The dotted line represents the true marginal density, the continuous line represents the mean, across Monte Carlo simulations, of the posterior mean of the marginal density, and the gray area represents the point-wise 95\% confidence band. The first, second, and third row show the results for $n = 250$, $500$, and $1000$, respectively. The first, second, third, and fourth column show the results for the elements of the compositional variable $x_{1,1}$, $x_{1,2}$, and $x_{1,3}$, and the bounded variable $x_2$, respectively.}
\label{fig:s1mar1}
\end{figure}

Figure \ref{fig:s1mar1} displays the mean (across the sample) of the posterior mean of the marginal distributions for each variable under Scenario I. In this figure, the dotted line represents the true marginal density, the continuous line represents the mean, across Monte Carlo simulations, of the posterior mean of the marginal density, and the gray area represents the point-wise 95\% confidence band. 

\begin{figure}[H] 
\centering
\captionsetup[subfloat]{labelformat=empty, aboveskip=0pt, belowskip=0pt}
\subfloat[]{
\includegraphics[page=1]{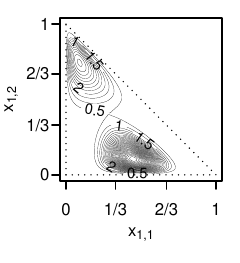}}
\subfloat[]{
\includegraphics[page=4]{rescaled/simu_fig/simu_plots_marginal_S1_Simplex.pdf}}
\subfloat[]{
\includegraphics[page=7]{rescaled/simu_fig/simu_plots_marginal_S1_Simplex.pdf}}
\subfloat[]{
\includegraphics[page=10]{rescaled/simu_fig/simu_plots_marginal_S1_Simplex.pdf}}\\[-10pt]
\subfloat[]{
\includegraphics[page=2]{rescaled/simu_fig/simu_plots_marginal_S1_Simplex.pdf}}
\subfloat[]{
\includegraphics[page=5]{rescaled/simu_fig/simu_plots_marginal_S1_Simplex.pdf}}
\subfloat[]{
\includegraphics[page=8]{rescaled/simu_fig/simu_plots_marginal_S1_Simplex.pdf}}
\subfloat[]{
\includegraphics[page=11]{rescaled/simu_fig/simu_plots_marginal_S1_Simplex.pdf}}\\[-10pt]
\subfloat[]{
\includegraphics[page=3]{rescaled/simu_fig/simu_plots_marginal_S1_Simplex.pdf}}
\subfloat[]{
\includegraphics[page=6]{rescaled/simu_fig/simu_plots_marginal_S1_Simplex.pdf}}
\subfloat[]{
\includegraphics[page=9]{rescaled/simu_fig/simu_plots_marginal_S1_Simplex.pdf}}
\subfloat[]{
\includegraphics[page=12]{rescaled/simu_fig/simu_plots_marginal_S1_Simplex.pdf}}
\caption{Simulated data -- Scenario I. Posterior inference for the bi-variate marginal distributions of the compositional variables $(x_{1,1}, x_{1,2}, x_{1,3})$. The first column display the contour plots of the true bi-variate marginal densities, while the second, third, and fourth columns show the contour plots of the mean, across Monte Carlo simulations, of the posterior mean of the bi-variate density for $n=250$, $500$, and $1000$, respectively. The first, second, and third  row show the results for $(x_{1,1}, x_{1,2})$,  $(x_{1,1}, x_{1,3})$, and  $(x_{1,2}, x_{1,3})$, respectively.}
\label{fig:s1mar2}
\end{figure}

\begin{figure}[H] 
\centering
\captionsetup[subfloat]{labelformat=empty, aboveskip=0pt, belowskip=0pt}
\subfloat[]{
\includegraphics[page=1]{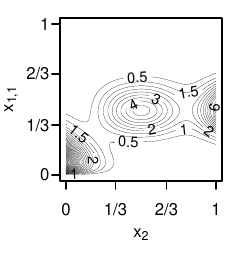}}
\subfloat[]{
\includegraphics[page=4]{rescaled/simu_fig/simu_plots_marginal_S1_Bivariates.pdf}}
\subfloat[]{
\includegraphics[page=7]{rescaled/simu_fig/simu_plots_marginal_S1_Bivariates.pdf}}
\subfloat[]{
\includegraphics[page=10]{rescaled/simu_fig/simu_plots_marginal_S1_Bivariates.pdf}}\\[-10pt]
\subfloat[]{
\includegraphics[page=2]{rescaled/simu_fig/simu_plots_marginal_S1_Bivariates.pdf}}
\subfloat[]{
\includegraphics[page=5]{rescaled/simu_fig/simu_plots_marginal_S1_Bivariates.pdf}}
\subfloat[]{
\includegraphics[page=8]{rescaled/simu_fig/simu_plots_marginal_S1_Bivariates.pdf}}
\subfloat[]{
\includegraphics[page=11]{rescaled/simu_fig/simu_plots_marginal_S1_Bivariates.pdf}}\\[-10pt]
\subfloat[]{
\includegraphics[page=3]{rescaled/simu_fig/simu_plots_marginal_S1_Bivariates.pdf}}
\subfloat[]{
\includegraphics[page=6]{rescaled/simu_fig/simu_plots_marginal_S1_Bivariates.pdf}}
\subfloat[]{
\includegraphics[page=9]{rescaled/simu_fig/simu_plots_marginal_S1_Bivariates.pdf}}
\subfloat[]{
\includegraphics[page=12]{rescaled/simu_fig/simu_plots_marginal_S1_Bivariates.pdf}}
\caption{Simulated data -- Scenario I. Posterior inference for the bi-variate marginal distributions of the bounded variable $x_2$ and the coordinates of compositional variable $(x_{1,1}, x_{1,2}, x_{1,3})$. The first column display the contour plots of the true bi-variate marginal densities, while the second, third, and fourth columns show the contour plots of the mean, across Monte Carlo simulations, of the posterior mean of the bi-variate density for $n=250$, $500$, and $1000$, respectively. The first, second, and third  row show the results for $(x_2, x_{1,1})$,  $(x_{2}, x_{1,2})$, and  $(x_{2}, x_{1,3})$, respectively.}
\label{fig:s1mar3}
\end{figure}

Figures \ref{fig:s1mar2} and \ref{fig:s1mar3} display the results for the bi-variate marginal distributions under Scenario I. Figure \ref{fig:s1mar2} display the results for the bi-variate distributions for the variables involved in the compositional compositional component $(x_{1,1}, x_{1,2}, x_{1,3})$. Figure \ref{fig:s1mar3} display the results for the bi-variate distributions of the bounded component $x_2$ and the variables involved in the compositional compositional component $(x_{1,1}, x_{1,2}, x_{1,3})$.

Table \ref{tab:MEPL1_1} presents the ${\rm MPEL}_1$ for the uni-variate marginal distributions and for the joint distribution under Scenario I, for the different samples sizes. 
\begin{table}[H]
 \centering
 \caption{Simulated Data -- Scenario I. MPEL$_1$ for the uni-variate marginals and  joint distribution, for each sample size.}
 \label{tab:MEPL1_1}
\begin{tabular}{cccc}\hline \hline
  & \multicolumn{3}{c}{Sample size} \\\cline{2-4}
 Distribution & $250$ & $500$ & $1000$\\ 
 \hline
 $x_{1,1}$   &  0.1199    & 0.0846 &  0.0612  \\
 $x_{1,2}$   &  0.1299    & 0.0970 &  0.0785  \\
 $x_{1,3}$   &  0.1128    & 0.0785 &  0.0559  \\
 $x_2$   &  0.1406    & 0.0995 &  0.0756  \\
 $(x_{1,1}, x_{1,2}, x_{1,3}, x_2)$ &  0.8408   &  0.8239   & 0.8157\\
 \hline \hline
\end{tabular}
\end{table}
As expected from the theoretical properties of the proposed model, the posterior estimates of uni-variate and bi-variate marginal densities, and the joint density  close the true ones for all sample sizes. The nonstandard true model is always covered by the confidence region.  Furthermore, the posterior estimates get closer to the true density and its sampling variability reduces as the sample size increases. 

\begin{figure}
\centering
\captionsetup[subfloat]{labelformat=empty}
\subfloat[]{
\includegraphics[page=2]{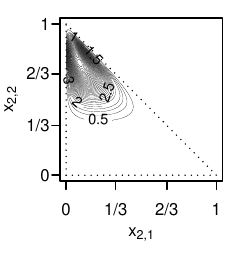}}
\subfloat[]{
\includegraphics[page=10]{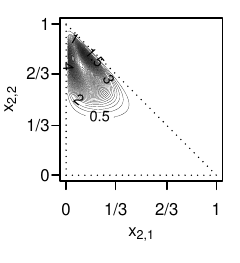}}
\subfloat[]{
\includegraphics[page=18]{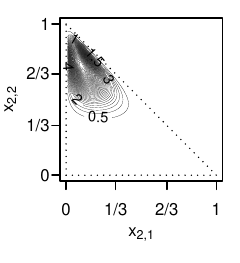}}
\subfloat[]{
\includegraphics[page=26]{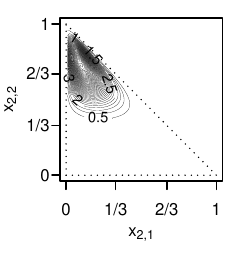}}\\[-10pt]
\subfloat[]{
\includegraphics[page=4]{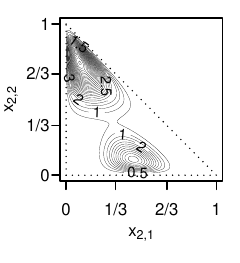}}
\subfloat[]{
\includegraphics[page=12]{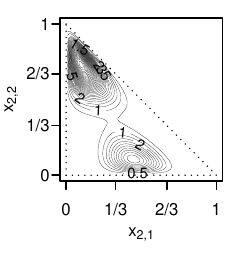}}
\subfloat[]{
\includegraphics[page=20]{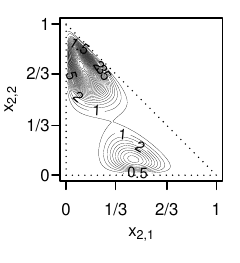}}
\subfloat[]{
\includegraphics[page=28]{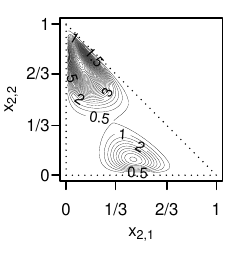}}\\[-10pt]
\subfloat[]{
\includegraphics[page=6]{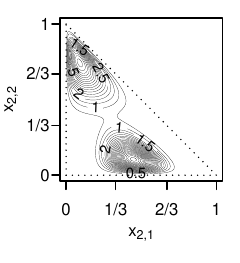}}
\subfloat[]{
\includegraphics[page=14]{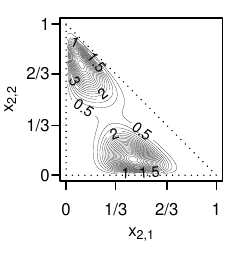}}
\subfloat[]{
\includegraphics[page=22]{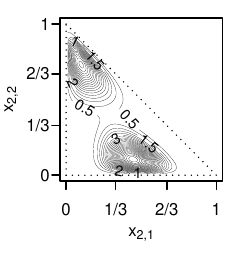}}
\subfloat[]{
\includegraphics[page=30]{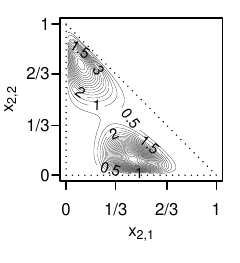}}\\[-10pt]
\subfloat[]{
\includegraphics[page=8]{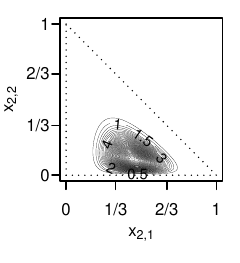}}
\subfloat[]{
\includegraphics[page=16]{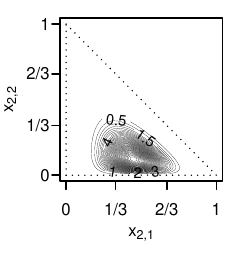}}
\subfloat[]{
\includegraphics[page=24]{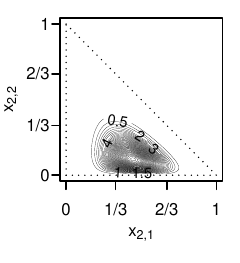}}
\subfloat[]{
\includegraphics[page=32]{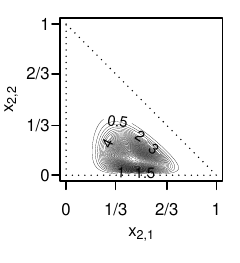}}
\caption{Simulated data -- Scenario II. Posterior inference on bi-variate conditional density of $(x_{2,1}, x_{2,2}) \mid (x_4,x_3,\bx_1) =  \vec p$. The first column display the contour plots of the true bi-variate conditional density, while the second, third, and fourth columns show the contour plots of the mean, across Monte Carlo simulations, of the posterior mean of the bi-variate conditional density for $n=250$, $500$, and $1000$, respectively. The first, second, third, and fourth row show the results for $\vec p = \vec p_1$, $\vec p_2$,  $\vec p_3$, and  $\vec p_4$, respectively.}
\label{fig:s2con}
\end{figure}

For scenario II, we consider the conditional density of $\bx_2$ given $\bx_1$, $x_3$, and $x_4$, that is,
\begin{eqnarray*}
    g(\bx_2\mid \boldsymbol{k}, F, \bx_1,x_3, x_4)&=&\frac{g(\bx_1,\bx_2,x_3, x_4 \mid \boldsymbol{k}, F)}{\int\int\int g(\bx_1,\bx_2,x_3,x_4  \mid \boldsymbol{k}, F)d x_4 d x_3 d \bx_1},\quad \bx_2 \in S_2.
\end{eqnarray*}
We illustrate the proposal by consider four in-sample data points to condition on $\vec p_1$, $\vec p_2$, $\vec p_3$, and $\vec p_4$, which values are given in Table \ref{tab:S2data} in Section \ref{sec:Simulation_fig} of the online supplementary material. Figure \ref{fig:s2con} shows the mean, across Monte Carlo simulation, of the posterior mean for the conditional densities for the different sample sizes and in-sample data points.  Table \ref{tab:MEPL1_2} presents the ${\rm MPEL}_1$ for the uni-variate marginal distributions and for the joint distribution under Scenario II, for the different samples sizes.

\begin{table}[H]
 \centering
 \caption{Simulated Data -- Scenario II. MPEL$_1$ for the uni-variate marginals and  joint distribution, for each sample size.}
 \label{tab:MEPL1_2}
\begin{tabular}{cccc}\hline \hline
  & \multicolumn{3}{c}{Sample size} \\\cline{2-4}
 Variable & $250$ & $500$ & $1000$\\ 
 \hline
 $x_{1,1}$   &  0.1386    & 0.1012 &  0.0739  \\
 $x_{1,2}$   &  0.1407    & 0.1067 &  0.0826  \\
 $x_{1,3}$   &  0.1038    & 0.0742 &  0.0535  \\
 $x_{1,4}$   &  0.0988    & 0.0702 &  0.0567  \\
 $x_{2,1}$   &  0.1285    & 0.0949 &  0.0665  \\
 $x_{2,2}$   &  0.1368    & 0.1001 &  0.0800  \\
 $x_{2,3}$   &  0.1241    & 0.0926 &  0.0646  \\
 $x_3$   &  0.2137    & 0.1779 &  0.1528  \\
 $x_4$   &  0.1288    & 0.0932 &  0.0755  \\
 $(\bx_1, \bx_2, x_3, x_4)$ &  0.9274  &  0.9130   & 0.9083\\
 \hline \hline
\end{tabular}
\end{table}

The results lead to a conclusion similar to that of Scenario I. The posterior estimates of the conditional are close to the true nonstandard conditional distributions across sample sizes. The posterior estimates get closer to the true conditional density and its sampling variability reduces as the sample size increases.   

\subsection{Application to NHANES data}\label{sec:NHANES}

\subsubsection{Data and research questions}

We consider data from the 2005–2006 cycle of NHANES by \cite{cdc_nhanes}. 
NHANES is a series of surveys that combine interviews, physical exams, and fitness and physical activity tests to assess the health and nutritional status of the American population. We  are interested in the association structure among movement behaviors, nutrient intake, body mass index (BMI), and age. Understanding these relationships offers valuable insights into health outcomes, inform prevention strategies, and guide public health interventions. For instance, identifying dietary patterns associated with varying levels of physical activity aids in designing lifestyle interventions that promote both healthy eating and physical activity. Similarly, exploring how age, movement behaviors, and nutrient intake intersect highlights critical periods for encouraging healthy habits.

Studying these associations requires a model able to account jointly for compositional and bounded data. In particular, we illustrate our approach by estimating the conditional distribution of movement behavior, a compositional attribute, given BMI, 
age, 
total nutrient intake, and nutrient intake, also a compositional attribute. 
Although age, BMI, and total nutrient intake can be assumed as not upper bounded, in practice they are constrained by biological, physiological, and behavioral factors, such as the natural limits of human lifespan, the body's ability to store and metabolize energy, and the practical constraints imposed by dietary habits, health conditions, and lifestyle choices. Thus, they were linearly map to the unit interval with any loss of generality. 
Nutrient intake is a 4-dimensional compositional feature that consists of four components representing the proportion of water, healthy nutrients (protein, vitamins, and minerals), unhealthy nutrients (carbohydrates, sugars, and fat), and fiber a person consumes on a daily basis. 

Movement behavior is a 3-dimensional compositional feature that captures the percentage of the time over seven consecutive days respondents spent in sedentary behavior (Sedentary), active behavior (Active), and periods of not wearing an activity monitoring device (Sleep). To monitor physical activity, participants were instructed to wear a uniaxial accelerometer on their right hip for 7 consecutive days. The accelerometers recorded movement intensity (magnitude of acceleration) over 10,080 consecutive 1-min intervals – one interval for each minute of the week. Sedentary behavior corresponds to low (< 100) counts per minute, as marked by the accelerometer. Active behavior corresponds to counts per minute above 100. Finally, non-wear time is defined as periods of time greater or equal to 90 consecutive minutes with movement intensity equal to zero. It is important to note that participants were asked to remove the accelerometers at bedtime and to keep them dry (e.g., when swimming or bathing). 

In our notation, for the each subject we observe $\bx = (\bx_{1}, \bx_{2}, \bx_{3})$, 
where $\bx_{1} \in S_3$ represents participants’ daily nutrient intake, $\bx_{2} \in S_2$ represents participants’ movement behavior, and $\bx_{3} \in [0,1]^3$ corresponds to participants’ rescaled total intake, age, and BMI. For this analysis, survey sampling weights are ignored and missing data were removed. Thus, the employed dataset consisted of $n = 5565$ observations. Here, we generated 25 Markov chains of length 20,000, including a burn-in of 10,000 and a thinning factor of 25. This results in 400 draws per chain, leading to a total of 10,000 posterior samples.

\subsubsection{Results}

\begin{figure}[H]
\centering
\includegraphics[page=11,trim={0 0 0 0},clip]{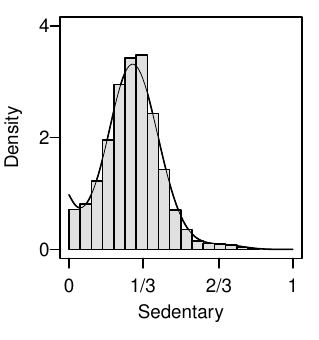}
\includegraphics[page=12,trim={0 0 0 0},clip]{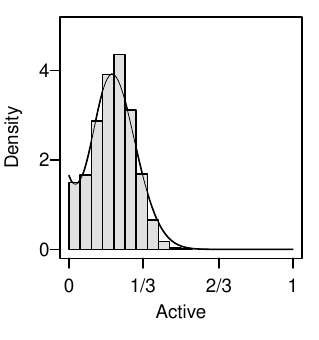}
\includegraphics[page=13,trim={0 0 0 0},clip]{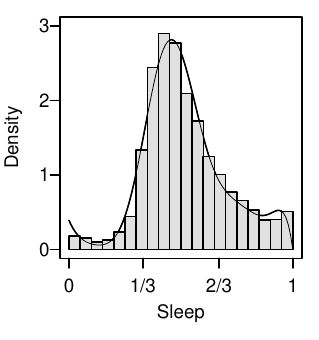}\\
\includegraphics[page=14,trim={0 0 0 0},clip]{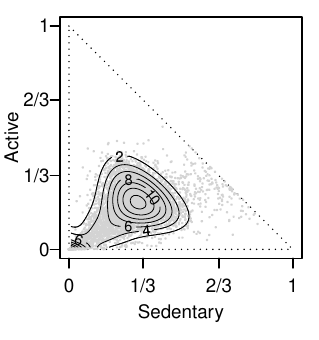}
\includegraphics[page=15,trim={0 0 0 0},clip]{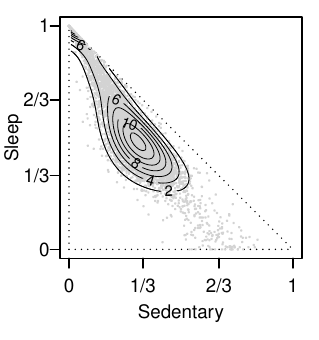}
\includegraphics[page=16,trim={0 0 0 0},clip]{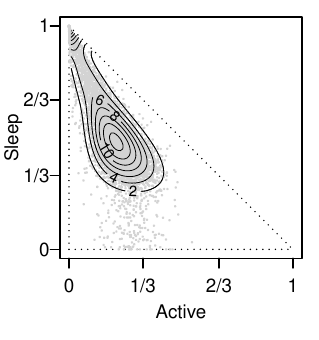}
\caption{NHANES data - Movement Behavior -- The first row display the  posterior mean of the marginal densities, along with the histogram, for the proportion of time respondents spent in sedentary behavior, active behavior, and sleep, respectively. The second row display the posterior mean of the bivariate marginal density for Active and Sedentary, Sleep and Active, and Active and Sleep, respectively. In these contour plots, the gray dots indicate the observed data points.}
\label{fig:NHANESmove}
\end{figure}

Figure \ref{fig:NHANESmove} presents the posterior mean of the marginal and bi-variate distributions for the coordinates of movement behavior. The results show a nonstandard distributional behavior, difficult to be capture by a parametric model. 
The posterior estimate for the marginal density of sedentary and active behaviors is unimodal with a positive value at zero. However, the sedentary distribution has a longer right tail, indicating that individuals tend to spend significantly more time in sedentary behavior. On the other hand, the sleep component exhibits a multimodal pattern. The bivariate distributions also display a multimodal patterns, where one of the modes corresponds to individuals whose sleep component is close to one \break (either because they are actually sleeping or not wearing the device). Additional figures for the posterior mean of the marginal and bivariate distributions for the other variables are available in Section~\ref{sec:NHANES_fig} of the supplemental material. Section~\ref{sec:NHANES_fig} also includes results illustrating that the proposed model is capable of capturing both positive and negative correlations, a commonly desired feature for models dealing with compositional data.

\begin{figure}[H]
\centering
\captionsetup[subfloat]{labelformat=empty,position=top, justification=centering}  

\subfloat[(a) \\ Low-age \\ Low-BMI]{
\includegraphics[page=1,trim={0 0 0 5.6},clip]{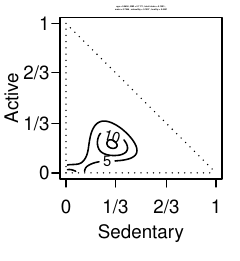}}
 \subfloat[(b) \\ Low-age \\ High-BMI]{
\includegraphics[page=2,trim={0 0 0 5.6},clip]{rescaled/real_fig/alldata_multi_with_inits_Conditional.pdf}}
 \subfloat[(c) \\ High-age \\ Low-BMI]{
\includegraphics[page=3,trim={0 0 0 5.6},clip]{rescaled/real_fig/alldata_multi_with_inits_Conditional.pdf}}
 \subfloat[(d) \\ High-age \\ High-BMI]{
\includegraphics[page=4,trim={0 0 0 5.6},clip]{rescaled/real_fig/alldata_multi_with_inits_Conditional.pdf}}

\subfloat[(e) \\ Low-total-intake \\ Low-water-intake]{
\includegraphics[page=5,trim={0 0 0 5.6},clip]{rescaled/real_fig/alldata_multi_with_inits_Conditional.pdf}}
\subfloat[(f) \\ 
Low-total-intake \\ High-wate-intake]{
\includegraphics[page=6,trim={0 0 0 5.6},clip]{rescaled/real_fig/alldata_multi_with_inits_Conditional.pdf}}
\subfloat[(g) \\ High-total-intake \\ Low-water-intake]{
\includegraphics[page=7,trim={0 0 0 5.6},clip]{rescaled/real_fig/alldata_multi_with_inits_Conditional.pdf}}
\subfloat[(h) \\ High-total-intake \\ High-water-intake]{
\includegraphics[page=8,trim={0 0 0 5.6},clip]{rescaled/real_fig/alldata_multi_with_inits_Conditional.pdf}}

\subfloat[(i) \\ Low-age \\ Low-water-intake]{
\includegraphics[page=9,trim={0 0 0 5.6},clip]{rescaled/real_fig/alldata_multi_with_inits_Conditional.pdf}}
\subfloat[(j) \\ Low-age \\ High-water-intake]{
\includegraphics[page=10,trim={0 0 0 5.6},clip]{rescaled/real_fig/alldata_multi_with_inits_Conditional.pdf}}
\subfloat[(k) \\ High-age \\ Low-water-intake]{
\includegraphics[page=11,trim={0 0 0 5.6},clip]{rescaled/real_fig/alldata_multi_with_inits_Conditional.pdf}}
\subfloat[(l) \\ High-age \\ High-water-intake]{
\includegraphics[page=12,trim={0 0 0 5.6},clip]{rescaled/real_fig/alldata_multi_with_inits_Conditional.pdf}}

\caption{MHANES data -- Conditional density for active and sedentary behaviors, given different values of age, BMI, total intake, and nutrient intake. 
Panels (a) -- (d) display the results for 
low-age and low-BMI,
low-age and high-BMI,
high-age and low-BMI,
high-age and high-BMI, respectively.
Panels (e) -- (h) display the results for 
low-total-intake and low-water-intake,
low-total-intake and high-water-intake,
high-total-intake and low-water-intake,
high-total-intake and high-water-intake, respectively.
Panels (i) -- (l) display the results for 
low-age and low-water-intake,
low-age and high-water-intake,
high-age and low-water-intake, and
high-age and high-water-intake, respectively.
Here low-age = 16.5 years, high-age = 62 years, low-BMI = 19.46, high-BMI = 35.54, low-total-intake = 1912.82 grams, high-total-intake = 4059.21 grams, low-water-intake 64\%, and high-water-intake = 82\%. For each panel, while the two corresponding variables are fixed at either low or high values, the remaining variables are set to representative central values.}
\label{fig:NHANESconcontour1}
\end{figure}

The posterior inferences for the conditional density of the proportions of time spent in active and sedentary movement behaviors, given varying values of nutrient intake, total intake, BMI, and age, are shown in Figure \ref{fig:NHANESconcontour1}, which presents the bivariate conditional distribution of active and sedentary behavior. In these conditional distributions, we use fixed and representative center values for the remaining variables. Additionally, Section~\ref{sec:NHANES_fig} of the supplemental material includes the bivariate conditional distributions of active and sleep behavior, as well as sedentary and sleep behavior.

Generally, we observe non-standard conditional densities, with bi-modal behaviors and where higher-density regions (inner contours) are closely packed, suggesting that most data points fall within a narrow range of movement behavior distributions. However, the spread of contours varies across conditions, indicating differences in movement patterns based on age, BMI, water intake, and total intake.

From panels (a) -- (d) in Figure \ref{fig:NHANESconcontour1}, we see that a greater probability mass is concentrated near the origin for lower age groups compared to higher age groups, indicating a lower proportion of sedentary and active behavior among younger individuals. However, the effect of BMI changes is less apparent. From panels (e) -- (h) in Figure \ref{fig:NHANESconcontour1}, we observe that higher water intake leads to a reduced probability mass near the origin, suggesting increased sedentary and active behaviors. Conversely, changes in total intake do not appear to substantially affect movement behavior, particularly when water consumption is low. Finally, from panels (i) -- (l) in Figure \ref{fig:NHANESconcontour1}, we observe that within the lower-age group, changes in water intake more noticeably influence movement behavior. However, these observations from the contour plots are not entirely straightforward. 

\begin{figure}
\captionsetup[subfloat]{labelformat=empty,position=top, justification=centering}  
\subfloat[\hspace{0.8cm}(a) age and BMI]{
\includegraphics[page=1,trim={0 0 0 8},clip]{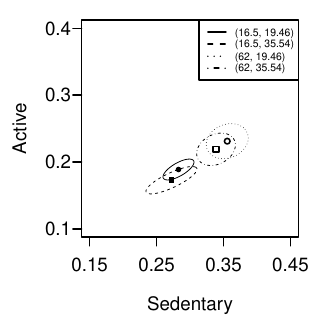}}
\subfloat[\hspace{0.8cm}(b) total and water intake]{
\includegraphics[page=2,trim={0 0 0 8},clip]{rescaled/real_fig/alldata_multi_with_inits_Ellipse.pdf}}
\subfloat[\hspace{0.6cm}(c) age and water intake]{
\includegraphics[page=3,trim={0 0 0 8},clip]{rescaled/real_fig/alldata_multi_with_inits_Ellipse.pdf}}\\[7pt]
\subfloat[\hspace{0.8cm}(d) age and BMI]{
\includegraphics[page=4,trim={0 0 0 8},clip]{rescaled/real_fig/alldata_multi_with_inits_Ellipse.pdf}}
\subfloat[\hspace{0.8cm}(e) total and  water intake]{
\includegraphics[page=5,trim={0 0 0 8},clip]{rescaled/real_fig/alldata_multi_with_inits_Ellipse.pdf}}
\subfloat[\hspace{0.6cm}(f) age and water intake]{
\includegraphics[page=6,trim={0 0 0 8},clip]{rescaled/real_fig/alldata_multi_with_inits_Ellipse.pdf}}\\[7pt]
\subfloat[\hspace{0.8cm}(g) age and BMI]{
\includegraphics[page=7,trim={0 0 0 8},clip]{rescaled/real_fig/alldata_multi_with_inits_Ellipse.pdf}}
\subfloat[\hspace{0.8cm}(h) total and water intake]{
\includegraphics[page=8,trim={0 0 0 8},clip]{rescaled/real_fig/alldata_multi_with_inits_Ellipse.pdf}}
\subfloat[\hspace{0.6cm}(i) age and water intake]{
\includegraphics[page=9,trim={0 0 0 8},clip]{rescaled/real_fig/alldata_multi_with_inits_Ellipse.pdf}}
\caption{NHANES data -- Posterior mean, and 95\% credible regions around the posterior mean, for the conditional expected movement behaviors for varying values of age, BMI, total intake, and water intake. Panels (a), (d), and (g) display the results for varying values of age and BMI. Panels (b), (e), and (h) display the results for varying values of
total intake and water intake. Panels (c), (f), and (i)
display the results for varying values of
age and water intake. In each panel, the solid ellipse represents the 95\% credible region when both components in the combination are low. The dashed ellipse represents the 95\% credible region when the first component is low and the second component is high. The dotted ellipse represents the 95\% credible region when the first component is high and the second component is low. Finally, the dotted-dashed ellipse represents the 95\% credible region when both components are high. For example, in the first panel, the solid ellipse represents the 95\% credible region for sedentary and active behaviors when both age and BMI are low. The dashed ellipse represents the credible region when age is low and BMI is high. The dotted ellipse represents the credible region when age is high and BMI is low. The dotted-dashed ellipse represents the credible region when both age and BMI are high.
}
\label{fig:NHANESconregn}
\end{figure}

To further investigate these effects, Figure \ref{fig:NHANESconregn} displays the posterior mean of the conditional mean of movement behavior with 95\% credible regions. We consider low and high observed values for rescaled variables: age (16.5 and 62 years), BMI (19.46 and 35.54), \break total intake (1912.82 and 4059.21 grams), and water intake percentage (64\% and 82\%). An interesting pattern emerges when comparing groups of individuals with the same BMI (panels (a), (d), and (g)), as represented by the solid and dotted ellipses or the dashed and dot-dashed ellipses. Older individuals tend to exhibit both more sedentary and more active behaviors simultaneously, resulting in less sleep behavior. This finding suggests that as people age, their daily behavior becomes more polarized, with distinct periods of high activity interspersed with extended sedentary time. A possible explanation for this pattern is that older individuals may engage in deliberate bouts of physical activity (such as walking, gardening, or exercise classes) while also spending longer periods being sedentary (e.g., sitting for extended periods at home or resting). This could reflect a lifestyle where older adults prioritize physical activity at specific times while compensating with increased rest due to lower overall energy levels or recovery needs. Other explanation is that  older individuals might be more diligent in wearing the accelerometer. 

When comparing groups of individuals of the same age (e.g., solid and dashed ellipses, or dotted and dotted-dashed ellipses), there is substantial overlap between the groups, making it challenging to draw definitive conclusions about the influence of BMI on movement behaviors. However, we observe a subtle trend. Ellipses representing higher BMI tend to capture a combination of both less active and less sedentary behaviors compared to those for lower BMI groups. While the overlap suggests that BMI alone may not be a strong predictor of movement patterns, this slight shift toward less sedentary and active behaviors in higher BMI groups warrants further investigation.

 When comparing groups of individuals with the same total nutrient intake (e.g., solid and dashed ellipses, or dotted and dotted-dashed ellipses; panels (b), (e), and (h)), a subtle trend emerges: ellipses representing a higher percentage of water intake tend to capture a combination of more sedentary behavior and less non-wear time. However, the significant overlap of ellipses across all movement behaviors , as well as across different levels of total and water intake, makes it challenging to identify clear differences in movement behavior across these conditions. This overlap limits our ability to isolate the effects of total and water intake, highlighting the need for further investigation.

Finally, when comparing groups with the same water intake percentage (represented by the solid and dotted ellipses or the dashed and dot-dashed ellipses; panels (c), (f), and (l)), we observe that older individuals exhibit both more sedentary and more active behaviors. This pattern is consistent with the trends seen in the first set of conditions displayed in panels (a), (d), and (g). However, individuals with a higher water intake percentage display less variation in movement behavior across different age groups, suggesting a potential stabilizing effect of water intake on movement patterns as people age. For groups of the same age (e.g., solid and dashed  ellipses or dotted and dotted-dashed  ellipses), the results reveal a nuanced pattern. Among younger individuals (solid and dashed  ellipses), a higher water intake percentage corresponds with more sedentary behavior. In contrast, among older individuals (dotted and dot-dashed  ellipses), water intake percentage seems to have a minimal impact on movement behaviors. This suggests that water intake may play a more significant role in shaping sedentary behavior in younger populations, while other factors might have a stronger influence among older groups.

\section{Concluding remarks}\label{sec:Conclusion}


We have proposed a novel Bayesian nonparametric novel for mixed-type bounded variables, where some variables are compositional and others are interval-bounded. The proposal is based on an extension of multivariate Bernstein polynomials, which induce a particular Dirichlet process mixture model of products of Dirichlet and beta distributions. The proposed model class has appealing theoretical properties such as full support and consistent posterior distribution. The model can be used for density and conditional density estimation, where both the response and predictors take values in the simplex space and/or hypercube. We illustrated the model's behavior through the analysis of simulated data and real data. The generalization of the approach to incorporate unbounded variables is the subject of ongoing research.

\section*{Acknowledgments}

A. Jara’s research is supported by ANID-Fondecyt 1220907 grant.

\bibliographystyle{asa}
\typeout{}
\bibliography{myrefs}

\newpage


\clearpage  
\makeatletter 
\renewcommand{\thefigure}{S\@arabic\c@figure}
\renewcommand{\thetable}{S\@arabic\c@table}
\renewcommand{\thesection}{S\@arabic\c@section}
\renewcommand{\theHsection}{S\arabic{section}}  
\renewcommand{\theHfigure}{S\arabic{figure}}    
\renewcommand{\theHtable}{S\arabic{table}}      
\makeatother
\setcounter{figure}{0}
\setcounter{table}{0}
\setcounter{section}{0}

\begin{center}
{\Large \textbf{Supplementary Material: Bayesian nonparametric modeling of mixed-type bounded data}}
\end{center}


\section{Proofs of results from Section \ref{sec:proposedMBP}} \label{sec:MBP_proof}

In this section, we provide the proofs of the theoretical results from Section \ref{sec:proposedMBP}. To simplify the notation in the proofs and without loss of generality, we now limit Definition \ref{df:MBP} to the case of two simplices instead of $M$. Set $\tildeDelta_d = S_{d_1} \times S_{d_2} \times [0,1]^{d_3}$, where $d = d_1 + d_2 + d_3$. 

\begin{defn}{\bf{(MBP on space $\tildeDelta_d$)}}\label{df:MBP_low}
The MBP of degree $(k_1,k_2,k_{3})$ on $\tildeDelta_d$ for a given $F : \mathbb{R}^{d_1+d_2+d_3} \rightarrow \mathbb{R}$ is given by
\begin{eqnarray}\label{eq:MBP}
    \nonumber    B_{k_1,k_2,k_3}^F(\bx)&=&\sum_{\boldsymbol{j}_1\in J_{d_1}^{k_1}}\sum_{\boldsymbol{j}_2\in J_{d_2}^{k_2}}\sum_{\boldsymbol{j}_3\in \{0,\ldots,k_3\}^{d_3}} F\left(\frac{\boldsymbol{j}_1}{k_1-d_1+1},\frac{\boldsymbol{j}_2}{k_2-d_2+1},\frac{\boldsymbol{j}_3}{k_3}\right)\\&&\times {\rm mult}(\boldsymbol{j}_1\mid k_1,\vx_1)\times {\rm mult}(\boldsymbol{j}_2\mid k_2,\vx_2)\times \prod_{l=1}^{d_3}{\rm bin}(j_{3,l}\mid k_3,x_{3,l}).
    \end{eqnarray}
    where
$\bx = (\vx_1,\vx_2, \vx_3) \in \Delta_d$, with $\vx_1 \in S_{d_1}$, $\vx_2 \in S_{d_2}$, and $\vx_{3} \in [0,1]^{d_{3}}$, \break
$(k_1, k_2, k_3) \in \{d_1, d_1+1 \ldots,\} \times \{d_2, d_2+1 \ldots,\} \times \mathbb{N}$, $\boldsymbol{j}_m=(j_{m,1},\ldots,j_{m,d_m})$,

$$J_{d_m}^{k_m}=\left\{(j_{m,1},\ldots,j_{m,d_m})\in\{0,\ldots,k_m\}^{d_m}:\sum_{l=1}^{d_m} j_{m,l}\leq k_m\right\},$$ 
and $m = 1,2$.
\end{defn}

\subsection{Proof of Theorem \ref{thm:appro2F}}\label{proof:appro2F}
First, we prove that the convergence holds at any point of continuity of $F$, say at $\bx=(\vx_1,\vx_2,\vx_3)$. To prove this, we begin with constructing random vectors $\frac{\boldsymbol{\Psi}_{k_1}^{d_1}}{k_1}$, $\frac{\boldsymbol{\Psi}_{k_2}^{d_2}}{k_2}$ and $\frac{\boldsymbol{\Phi}_{k_3}^{d_3}}{k_3}$ converge to $\vx_1\in S_{d_1}$, $\vx_2\in S_{d_2}$ and $\vx_3\in [0,1]^{d_3}$, respectively. 

For the simplex case, say $\vx = (x_{1},\ldots, x_{d})\in S_d$, let $\vpsi_1,\vpsi_2,\ldots, \vpsi_{k}$ be independent and identically distributed random vectors, following the categorical distribution,
\begin{equation*}
\begin{aligned}
  &{\prob}\left(\vpsi_t=(\overbrace{1,0,0,\ldots,0}^{d})\right)=x_{1},\quad {\prob}\left(\vpsi_t=(\overbrace{0,1,0,\ldots,0}^{d})\right)=x_{2},\quad\ldots\quad,
  \\&{\prob}\left(\vpsi_t=(\overbrace{0,0,\ldots,0,1}^{d})\right)=x_{d},\quad {\prob}\left(\vpsi_t=(\overbrace{0,0,\ldots,0}^{d})\right)=1-\sum_{l=1}^{d} x_{l},\quad t=1,\ldots,k,
\end{aligned}
\end{equation*}
and define $\boldsymbol{\Psi}_{k}^{d}=\sum_{t=1}^{k}\vpsi_t$. Notice that $\boldsymbol{\Psi}_{k}^{d}$ has a multinomial distribution with number of trials $k$ and $d$-dimensional success probability vector $\boldsymbol{p}=\vx=(x_{1},x_{2},\ldots,x_{d})$. Therefore,
\begin{equation*}
\begin{aligned}
 {\rm E}\left[\boldsymbol{\Psi}_{k}^{d}\right]=(k x_{1}, \ldots, k x_{d})=k\vx,\quad {\rm Var}\left[\Psi_{k,l}^d\right]=k x_{l}(1-x_{l}),\quad l = 1,\ldots,d,
\end{aligned}
\end{equation*}
implying that,
\begin{equation*}
\begin{aligned}
 {\rm E}\left[\frac{\boldsymbol{\Psi}_{k}^{d}}{k}\right]=\vx,\quad {\rm Var}\left[\frac{\Psi_{k,l}^d}{k}\right]=\frac{x_{l}(1-x_{l})}{k}\leq \frac{1}{4k},\quad l = 1,\ldots,d,
\end{aligned}
\end{equation*}
and from the weak law of large numbers, we have $\frac{\boldsymbol{\Psi}_{k}^{d}}{k}\stackrel{P}{\to}\vx$ for $\vx\in S_d$.

For the unit cube case, say $\vx = (x_{1}, \ldots, x_{d})\in [0,1]^{d}$, let $\vphi_1,\vphi_2,\ldots,\vphi_{k}$ be independent and identically distributed, where $\vphi_t=(\phi_{t,1},\ldots,\phi_{t,d}),\ t = 1,\ldots,k$, and $\phi_{t,1},\ldots,\phi_{t,d}$ independently follow Bernoulli distributions,
\begin{equation*}
{\prob}(\phi_{t,l}=1)=x_{l}=1-{\prob}(\phi_{t,l}=0),\quad t=1,\ldots,k, \quad l=1,\ldots,d,
\end{equation*}
and define $\boldsymbol{\Phi}_{k}^{d}=\sum_{t=1}^{k} \vphi_t$. Notice that in $\boldsymbol{\Phi}_{k}^{d}$, each $\Phi_{k,l}$ has a binomial distribution with probability $x_{l}$ and 
\begin{equation*}
\begin{aligned}
  {\rm E}\left[\boldsymbol{\Phi}_{k}^{d}\right]=(k x_{1}, \ldots, k x_{d})=k\vx,\quad {\rm Var}\left[\Phi_{k,l}^{d}\right]=k x_{l}(1-x_{l}),\quad l = 1,\ldots,d.
\end{aligned}
\end{equation*}
Hence,
\begin{equation*}
\begin{aligned}
 {\rm E}\left[\frac{\boldsymbol{\Phi}_{k}^{d}}{k}\right]=\vx,\quad {\rm Var}\left[\frac{\Phi_{k,l}^{d}}{k}\right]=\frac{x_{l}(1-x_{l})}{k}\leq \frac{1}{4k},\quad l = 1,\ldots,d,
\end{aligned}
\end{equation*}
and from the weak law of large numbers, $\frac{\boldsymbol{\Phi}_{k}^{d}}{k}\stackrel{P}{\to}\vx$ for $\vx\in [0,1]^{d}$.\\

As $\frac{k_1}{k_1-d_1+1}{\to}1$ and $\frac{k_2}{k_2-d_2+1}{\to}1$, by continuous mapping theorem, we have \break $\frac{\boldsymbol{\Psi}_{k_1}^{d_1}}{k_1-d_1+1}=\frac{\boldsymbol{\Psi}_{k_1}^{d_1}}{k_1}\frac{k_1}{k_1-d_1+1}\stackrel{P}{\to}\vx_1$ and $\frac{\boldsymbol{\Psi}_{k_2}^{d_2}}{k_2-d_2+1}=\frac{\boldsymbol{\Psi}_{k_2}^{d_2}}{k_2}\frac{k_2}{k_2-d_2+1}\stackrel{P}{\to}\vx_2$, respectively.
Given that \break $\min\{k_1,k_2,k_3\}\to\infty$, we have $\left(\frac{\boldsymbol{\Psi}_{k_1}^{d_1}}{k_1-d_1+1},\frac{\boldsymbol{\Psi}_{k_2}^{d_2}}{k_2-d_2+1},\frac{\boldsymbol{\Phi}_{k_3}^{d_3}}{k_3}\right)\stackrel{P}{\to}\bx=(\vx_1,\vx_2,\vx_3)$.
By continuous mapping theorem, for the continuous $F$, $F\left(\frac{\boldsymbol{\Psi}_{k_1}^{d_1}}{k_1-d_1+1},\frac{\boldsymbol{\Psi}_{k_2}^{d_2}}{k_2-d_2+1},\frac{\boldsymbol{\Phi}_{k_3}^{d_3}}{k_3}\right)\stackrel{P}{\to}F(\bx)$. Therefore, it follows that $F\left(\frac{\boldsymbol{\Psi}_{k_1}^{d_1}}{k_1-d_1+1},\frac{\boldsymbol{\Psi}_{k_2}^{d_2}}{k_2-d_2+1},\frac{\boldsymbol{\Phi}_{k_3}^{d_3}}{k_3}\right)\stackrel{D}{\to}F(\bx)$, that is to say, 
\begin{eqnarray}
{\rm E}\left[F\left(\frac{\boldsymbol{\Psi}_{k_1}^{d_1}}{k_1-d_1+1},\frac{\boldsymbol{\Psi}_{k_2}^{d_2}}{k_2-d_2+1},\frac{\boldsymbol{\Phi}_{k_3}^{d_3}}{k_3}\right)\right]\to {\rm E}\left[F(\bx)\right]=F(\bx).\label{eq:pointwise_converge}
\end{eqnarray}
Recall that the random vectors $\boldsymbol{\Psi}_{k_1}^{d_1}$ and $\boldsymbol{\Psi}_{k_2}^{d_2}$ each follow a multinomial distribution, while $\boldsymbol{\Phi}_{k_3}^{d_3}$ follows a product of Binomial distributions. More precisely, $\boldsymbol{\Psi}_{k_1}^{d_1} \sim {\rm multinomial}(k_1,\vx_1)$, $\boldsymbol{\Psi}_{k_2}^{d_2} \sim {\rm multinomial}(k_2,\vx_2)$ and in $\boldsymbol{\Phi}_{k_3}^{d_3}$, ${\Phi}_{k_3,l} \overset{ind.}{\sim} {\rm binomial}(k_3,x_l)$ for all $l\in\{1,\ldots,d_3\}$. 

Therefore,
\begin{eqnarray}
    & &\hspace{-26mm}{\rm E}\left[F\left(\frac{\boldsymbol{\Psi}_{k_1}^{d_1}}{k_1-d_1+1},\frac{\boldsymbol{\Psi}_{k_2}^{d_2}}{k_2-d_2+1},\frac{\boldsymbol{\Phi}_{k_3}^{d_3}}{k_3}\right)\right]\nonumber
    \\
    &=& \sum_{\boldsymbol{j}_1\in J_{d_1}^{k_1}}\sum_{\boldsymbol{j}_2\in J_{d_2}^{k_2}}\sum_{\boldsymbol{j}_3\in \{0,\ldots,k_3\}^{d_3}}\nonumber F\left(\frac{\boldsymbol{j}_1}{k_1-d_1+1},\frac{\boldsymbol{j}_2}{k_2-d_2+1},\frac{\boldsymbol{j}_3}{k_3}\right)\nonumber\\&&\times {\rm mult}(\boldsymbol{j}_1\mid k_1,\vx_1){\rm mult}(\boldsymbol{j}_2\mid k_2,\vx_2)\prod_{l=1}^{d_3}{\rm bin}(j_{3,l}\mid k_3,x_{3,l})\nonumber\\
    &=&B_{k_1,k_2,k_3}^F(\bx) \label{eq:pointwise_expect}.
\end{eqnarray}
From (\ref{eq:pointwise_converge}) and (\ref{eq:pointwise_expect}), we can conclude that $B_{k_1,k_2,k_3}^F(\bx)\to F(\bx)$ at any point of continuity of $F$.

We have shown that the Bernstein polynomial converges to $F(\bx)$ at any point of continuity of $F$. Now, we demonstrate that it converges to $F(\bx)$ uniformly, assuming that $F$ is continuous. Let the distance between two vectors be the $l_2$ distance, say, $d(\bx,\bz)=\left(\sum_{l=1}^{d_1+d_2+d_3} (x_l-z_l)^2\right)^{\frac{1}{2}}$, since $F$ is continuous and $\tildeDelta_d$ is compact (closed, bounded), then by Heine–Cantor theorem, $F$ is uniformly continuous. Define the modulus of continuity as \[\omega(\epsilon)=\sup_{\substack{d(\bx,\bz)<\epsilon \\ \bx,\bz\in \tildeDelta_d}} \left|F(\bx)-F(\bz)\right|.\] Therefore, uniform convergence means \[\lim_{\epsilon\downarrow 0}\omega(\epsilon)=0.\]
Define $\|F\|=\sup\left\{|F(\bx)|:\bx\in \tildeDelta_d\right\}$, and let $\by=\left(\frac{\boldsymbol{\Psi}_{k_1}^{d_1}}{k_1-d_1+1},\frac{\boldsymbol{\Psi}_{k_2}^{d_2}}{k_2-d_2+1},\frac{\boldsymbol{\Phi}_{k_3}^{d_3}}{k_3}\right)$, with (\ref{eq:pointwise_converge}) we have
\begin{eqnarray}
\sup_{\bx}\left|B_{k_1,k_2,k_3}^F(\bx)-F(\bx)\right|
& = & \sup_{\bx}\left|{\rm E}\left[F(\by)\right]-F(\bx)\right|,\nonumber\\
& \leq & \sup_{\bx}{\rm E}\left[|F(\by)-F(\bx)|\right],\nonumber\\ 
& \leq & \sup_{\bx} \bigg\{\{ {\rm E}\left[\left|F(\by)-F(\bx)\right|\mathbbm{1}_{\{d(\by,\bx)\leq \epsilon\}}\right] \bigg.\nonumber\\
& & \bigg.+\sup_{\bx}{\rm E}\left[|F(\by)-F(\bx)|\mathbbm{1}_{\{d(\by,\bx)>\epsilon\}}\right]\bigg\},\nonumber\\
& \leq & \omega(\epsilon)\sup_{\bx}{\prob}\left\{d(\by,\bx)\leq \epsilon\right\}+2\|F\|\sup_{\bx}{\prob}\left\{d(\by,\bx)> \epsilon\right\}, \nonumber\\
& \leq & \omega(\epsilon)+2\|F\|\sup_{\bx}{\prob}\left\{d(\by,\bx)> \epsilon\right\}\label{eq:unif_approx}.
\end{eqnarray}
The second term  of the summation in (\ref{eq:unif_approx}) can also be upper-bounded, as follows,
\begin{eqnarray}
2\|F\|\sup_{\bx}{\prob}\{d(\by,\bx)> \epsilon\} &=&2\|F\|\sup_{\bx}{\prob}\left\{\sum_{l=1}^{d_1} \left(\frac{\Psi_{k_{1},l}^{d_1}}{k_1-d_1+1}-x_{1,l}\right)^2\right.\nonumber\nonumber\\& & \left.+\sum_{l=1}^{d_2} \left(\frac{\Psi_{k_{2},l}^{d_2}}{k_2-d_2+1}-x_{2,l}\right)^2
+\sum_{l=1}^{d_3} \left(\frac{\Phi_{k_{3},l}^{d_3}}{k_3}-x_{3,l}\right)^2> \epsilon^2\right\},\nonumber\\
&\leq&\frac{2\|F\|}{\epsilon^2}\sup_{\bx}{\rm E}\left[\sum_{l=1}^{d_1} \left(\frac{\Psi_{k_{1},l}^{d_1}}{k_1-d_1+1}-x_{1,l}\right)^2\right.\nonumber\\& &\left.+\sum_{l=1}^{d_2} \left(\frac{\Psi_{k_{2},l}^{d_2}}{k_2-d_2+1}-x_{2,l}\right)^2
+\sum_{l=1}^{d_3} \left(\frac{\Phi_{k_{3},l}^{d_3}}{k_3}-x_{3,l}\right)^2\right],\nonumber\\
&=&\frac{2\|F\|}{\epsilon^2}\sup_{\bx}\left\{\sum_{l=1}^{d_1} {\rm E}\left[\left(\frac{\Psi_{k_{1},l}^{d_1}}{k_1-d_1+1}-x_{1,l}\right)^2\right]\right.\nonumber\\ & &\left.+\sum_{l=1}^{d_2} {\rm E}\left[\left(\frac{\Psi_{k_{2},l}^{d_2}}{k_2-d_2+1}-x_{2,l}\right)^2\right]+\sum_{l=1}^{d_3} {\rm E}\left[\left(\frac{\Phi_{k_{3},l}^{d_3}}{k_3}-x_{3,l}\right)^2\right]\right\},\nonumber\\
&=&\frac{2\|F\|}{\epsilon^2}\left\{\sup_{\vx_1}\sum_{l=1}^{d_1} {\rm Var}\left[\frac{\Psi_{k_{1},l}^{d_1}}{k_1-d_1+1}\right]\right.\nonumber\\& & \left. +\sup_{\vx_2}\sum_{l=1}^{d_2} {\rm Var}\left[\frac{\Psi_{k_{2},l}^{d_2}}{k_2-d_2+1}\right]+\sup_{\vx_3}\sum_{l=1}^{d_3} {\rm Var}\left[\frac{\Phi_{k_{3},l}^{d_3}}{k_3}\right]\right\},\nonumber\\
&\leq &\frac{\|F\|}{2\epsilon^2}\left[\frac{d_1k_1}{(k_1-d_1+1)^2}+\frac{d_2k_2}{(k_2-d_2+1)^2}+\frac{d_3}{k_3}\right]\label{eq:unif_approx2}.
\end{eqnarray}

From inequalities (\ref{eq:unif_approx}) and (\ref{eq:unif_approx2}), we have \[\sup_{\bx\in \tildeDelta_d}\left|B_{k_1,k_2,k_3}^F(\bx)-F(\bx)\right|\leq \omega(\epsilon)+\frac{\|F\|}{2\epsilon^2}\left[\frac{d_1k_1}{(k_1-d_1+1)^2}+\frac{d_2k_2}{(k_2-d_2+1)^2}+\frac{d_3}{k_3}\right].\]

Now we can show that \[\limsup_{\min\{k_1,k_2,k_3\}\to\infty}\sup_{\bx\in \tildeDelta_d}\left|B_{k_1,k_2,k_3}^F(\bx)-F(\bx)\right|\leq \omega(\epsilon).\]
Since $\omega(\epsilon)\to 0$ as $\epsilon\to 0$, the uniform convergence is proved.

\subsection{Proof of Theorem \ref{prop:mbp_cdf}}\label{proof:mbp_cdf}
Notice that for any continuous function $H=H(x_1,x_2)$ and any constants $c_1$ and $c_2$, the result
\[\frac{\partial^2 H(x_1,x_2)}{\partial x_1 \partial x_2} = \frac{\partial^2 H(x_1,x_2)+c_1x_1+c_2x_2}{\partial x_1 \partial x_2},\]
always holds. As we can hardly measure the effect on the boundary of a space $\tildeDelta_d$ by taking derivative, we now rewrite the summation into two parts, when $\boldsymbol{j}$ taking value on the boundary of the product space ($\boldsymbol{j}\in {\rm B}_{\tildeDelta_d}$) and otherwise, $\boldsymbol{j}\in \tildeDelta_d\setminus {\rm B}_{\tildeDelta_d}$, 
that is,
\begin{eqnarray*}
B_{k_1,k_2,k_3}^F(\bx)&=&\sum_{\boldsymbol{j}_1\in I_{d_1}^{k_1}}\sum_{\boldsymbol{j}_2\in I_{d_2}^{k_2}}\sum_{\boldsymbol{j}_3\in \{1,\ldots,k_3\}^{d_3}} F\left(\frac{\boldsymbol{j}_1}{k_1-d_1+1},\frac{\boldsymbol{j}_2}{k_2-d_2+1},\frac{\boldsymbol{j}_3}{k_3}\right)\\&& \times{\rm mult}(\boldsymbol{j}_1\mid k_1,\vx_1){\rm mult}(\boldsymbol{j}_2\mid k_2,\vx_2)\prod_{l=1}^{d_3}{\rm bin}(j_{3,l}\mid k_3,x_{3,l})\\
&&+\sum_{\boldsymbol{j}_1\in { J_{d_1}^{k_1} \setminus I_{d_1}^{k_1}}}\sum_{\boldsymbol{j}_2\in { J_{d_2}^{k_2} \setminus I_{d_2}^{k_2}}}\sum_{\boldsymbol{j}_3\in {\{0,\ldots,k_3\}^{d_3} \setminus \{1,\ldots,k_3\}^{d_3}} } F\left(\frac{\boldsymbol{j}_1}{k_1-d_1+1},\frac{\boldsymbol{j}_2}{k_2-d_2+1},\frac{\boldsymbol{j}_3}{k_3}\right)\\&& \times{\rm mult}(\boldsymbol{j}_1\mid k_1,\vx_1){\rm mult}(\boldsymbol{j}_2\mid k_2,\vx_2)\prod_{l=1}^{d_3}{\rm bin}(j_{3,l}\mid k_3,x_{3,l}).
\end{eqnarray*}
Given condition that $F(\bx)=0$ when $\bx\in B_{\tildeDelta_d}$, we have 
\begin{eqnarray*}
    B_{k_1,k_2,k_3}^F(\bx)
    &=&\sum_{\boldsymbol{j}_1\in I_{d_1}^{k_1}}\sum_{\boldsymbol{j}_2\in I_{d_2}^{k_2}}\sum_{\boldsymbol{j}_3\in \{1,\ldots,k_3\}^{d_3}} F\left(\frac{\boldsymbol{j}_1}{k_1-d_1+1},\frac{\boldsymbol{j}_2}{k_2-d_2+1},\frac{\boldsymbol{j}_3}{k_3}\right)\\&& \times{\rm mult}(\boldsymbol{j}_1\mid k_1,\vx_1){\rm mult}(\boldsymbol{j}_2\mid k_2,\vx_2)\prod_{l=1}^{d_3}{\rm bin}(j_{3,l}\mid k_3,x_{3,l}).      
\end{eqnarray*}
Consider the derivative of Bernstein polynomial above, we denote that 
\begin{eqnarray}
   \nonumber b_{k_1,k_2,k_3}^F(\bx)&=&\sum_{\boldsymbol{j}_1\in 
    I_{d_1}^{k_1}}\sum_{\boldsymbol{j}_2\in I_{d_2}^{k_2}}\sum_{\boldsymbol{j}_3\in \{1,\ldots,k_3\}^{d_3}}F\left(\rho_{d_1,\boldsymbol{j}_1}^{k_1-d_1+1}\times\rho_{d_2,\boldsymbol{j}_2}^{k_2-d_2+1}\times\rho_{d_3,\boldsymbol{j}_3}^{k_3}\right)\\
    &&\nonumber \times{\rm dir}\left(\vx_1\mid\alpha(k_1-d_1+1,\boldsymbol{j}_1)\right){\rm dir}(\vx_2\mid\alpha(k_2-d_2+1,\boldsymbol{j}_2))\\
    &&\times \prod_{l=1}^{d_3}{\rm beta}(x_{3,l}\mid j_{3,l},k_3-j_{3,l}+1)\nonumber.
\end{eqnarray}
Given that the mixing weights $w_{\boldsymbol{j},\boldsymbol{k}}(F)=F\left({\rho_{d_1,\boldsymbol{j}_1}^{k_1-d_1+1}}\times{\rho_{d_2,\boldsymbol{j}_2}^{k_2-d_2+1}}\times\rho_{d_3,\boldsymbol{j}_3}^{k_3}\right)$ are the probabilities of partition elements $(\boldsymbol{j}, \boldsymbol{k})$ under $F$, we have $\int_{\tildeDelta_d} b_{k_1,k_2,k_3}^F(\bx)d \bx =1$. As $b_{k_1,k_2,k_3}^F(\bx)\geq 0$ for all $\bx$, and $\int_{\tildeDelta_d} b_{k_1,k_2,k_3}^F(\bx)d \bx =1$, then we can conclude that $b_{k_1,k_2,k_3}^F(\bx)$ is a density function given by a mixture model of the product of Dirichlet distributions and beta distributions.

By fundamental theorem of calculus, the MBP can be rewritten as the integral of the corresponding density, $b_{k_1,k_2,k_3}^F(\by)$, 
\begin{eqnarray*}       B_{k_1,k_2,k_3}^F(\bx)&=&\int_{0}^{x_{1,1}}\cdots\int_{0}^{x_{1,{d_1}}}\int_{0}^{x_{2,1}}\cdots\int_{0}^{x_{2,{d_2}}}\int_{0}^{x_{3,1}}\cdots\int_{0}^{x_{3,{d_3}}}\\&&b_{k_1,k_2,k_3}^F(\by)d{y_{3,{d_3}}}\cdots d{y_{3,1}}d{y_{2,{d_2}}}\cdots d{y_{2,1}}d{y_{1,{d_1}}}\cdots d{y_{1,1}},\quad\bx\in \tildeDelta_d.
\end{eqnarray*}
As $b_{k_1,k_2,k_3}^F(\by)$ is a probability density function, then the theorem is proved. 

\subsection{Proof of Theorem \ref{thm:appro2f}}\label{proof:appro2f}
Recall that the MBP density is given by
\begin{eqnarray*}
    b_{k_1,k_2,k_3}^F(\bx)&=&\sum_{\boldsymbol{j}_1\in I_{d_1}^{k_1}}\sum_{\boldsymbol{j}_2\in I_{d_2}^{k_2}}\sum_{\boldsymbol{j}_3\in \{1,\ldots,k_3\}^{d_3}}F(\rho_{d_1,\boldsymbol{j}_1}^{k_1-d_1+1}\times\rho_{d_2,\boldsymbol{j}_2}^{k_2-d_2+1}\times\rho_{d_3,\boldsymbol{j}_3}^{k_3})\\
    &&\times {\rm dir}(\vx_1\mid \alpha(k_1-d_1+1,\boldsymbol{j}_1)){\rm dir}(\vx_2\mid \alpha(k_2-d_2+1,\boldsymbol{j}_2))\\
    &&\times \prod_{l=1}^{d_3}{\rm beta}(x_{3,l}\mid j_{3,l},k_3-j_{3,l}+1).
\end{eqnarray*}
Notice that, the MBP density can also be written as,
\begin{eqnarray*}
    b_{k_1,k_2,k_3}^F(\bx)
    &=&\sum_{\boldsymbol{j}_1\in J_{d_1}^{k_1-d_1}}\sum_{\boldsymbol{j}_2\in J_{d_2}^{k_2-d_2}}\sum_{\boldsymbol{j}_3\in \{0,\ldots,k_3-1\}^{d_3}}F({\tilde{\rho}_{d_1,\boldsymbol{j}_1}^{k_1-d_1+1}}\times{\tilde{\rho}_{d_2,\boldsymbol{j}_2}^{k_2-d_2+1}}\times{\tilde{\rho}_{d_3,\boldsymbol{j}_3}^{k_3}})\\
    &&\times \frac{k_1!}{(k_1-d_1)!}{\rm mult}(\boldsymbol{j}_1\mid k_1-d_1,\vx_1)\frac{k_2!}{(k_2-d_2)!}{\rm mult}(\boldsymbol{j}_2\mid k_2-d_2,\vx_2)\\
    &&\prod_{l=1}^{d_3}[k_3\cdot{\rm bin}(j_{3,l}\mid k_3-1,x_{3,l})],\\
    J_{d}^{k-d}&=&\left\{\boldsymbol{j}\in \{0,\ldots,k-d\}^d:\sum_{l=1}^d j_l\leq k-d\right\},\\
    {\tilde{\rho}_{d,\boldsymbol{j}}^{k-d+1}}&=&\left(\frac{j_1}{{k-d+1}},\frac{j_1+1}{{k-d+1}}\right]\times\cdots\times\left(\frac{j_d}{{k-d+1}},\frac{j_d+1}{{k-d+1}}\right].
\end{eqnarray*}
From the generalized mean value theorem in Chapter 2 of \cite{ash2008mean}, there exists $\boldsymbol{C}_{\boldsymbol{j}}\in {\tilde{\rho}_{d_1,\boldsymbol{j}_1}^{k_1-d_1+1}}\times{\tilde{\rho}_{d_2,\boldsymbol{j}_2}^{k_2-d_2+1}}\times\tilde{\rho}_{d_3,\boldsymbol{j}_3}^{k_3}$, such that,
\begin{eqnarray*}
b_{k_1,k_2,k_3}^F(\bx)
        &=&\sum_{\boldsymbol{j}_1\in J_{d_1}^{k_1-d_1}}\sum_{\boldsymbol{j}_2\in J_{d_2}^{k_2-d_2}}\sum_{\boldsymbol{j}_3\in \{0,\ldots,k_3-1\}^{d_3}} \frac{f(\boldsymbol{C}_{\boldsymbol{j}})}{(k_1-d_1+1)^{d_1}(k_2-d_2+1)^{d_2}{k_3}^{d_3}}\\
        &&\times \frac{k_1!}{(k_1-d_1)!}{\rm mult}(\boldsymbol{j}_1\mid k_1-d_1,\vx_1)\frac{k_2!}{(k_2-d_2)!}{\rm mult}(\boldsymbol{j}_2\mid k_2-d_2,\vx_2)\\
        &&\times\prod_{l=1}^{d_3}[k_3\cdot{\rm bin}(j_{3,l}\mid k_3-1,x_{3,l})],\\
        &=&\sum_{\boldsymbol{j}_1\in J_{d_1}^{k_1-d_1}}\sum_{\boldsymbol{j}_2\in J_{d_2}^{k_2-d_2}}\sum_{\boldsymbol{j}_3\in \{0,\ldots,k_3-1\}^{d_3}}\left[f(\boldsymbol{C}_{\boldsymbol{j}})\pm f\left(\frac{\boldsymbol{j}_1}{k_1-2d_1+1},\frac{\boldsymbol{j}_2}{k_2-2d_2+1},\frac{\boldsymbol{j}_3}{k_3-1}\right)\right]\\
        &&\times\frac{\prod_{l=0}^{d_1-2}(k_1-l)}{(k_1-d_1+1)^{d_1-1}}\frac{\prod_{l=0}^{d_2-2}(k_2-l)}{(k_2-d_2+1)^{d_2-1}}{\rm mult}(\boldsymbol{j}_1\mid k_1-d_1,\vx_1)\\
        &&\times{\rm mult}(\boldsymbol{j}_2\mid k_2-d_2,\vx_2)\prod_{l=1}^{d_3}{\rm bin}(j_{3,l}\mid k_3-1,x_{3,l}).
\end{eqnarray*}
Thus, the MBP density can be express as the summation of two parts,
\begin{eqnarray}        
& & \hspace{-12mm} b_{k_1,k_2,k_3}^F(\bx)\nonumber\\
        &=&\nonumber\sum_{\boldsymbol{j}_1\in J_{d_1}^{k_1-d_1}}\sum_{\boldsymbol{j}_2\in J_{d_2}^{k_2-d_2}}\sum_{\boldsymbol{j}_3\in \{0,\ldots,k_3-1\}^{d_3}}\left[f(\boldsymbol{C}_{\boldsymbol{j}})- f\left(\frac{\boldsymbol{j}_1}{k_1-2d_1+1},\frac{\boldsymbol{j}_2}{k_2-2d_2+1},\frac{\boldsymbol{j}_3}{k_3-1}\right)\right]\\
        &&\nonumber\times\frac{\prod_{l=0}^{d_1-2}(k_1-l)}{(k_1-d_1+1)^{d_1-1}}\frac{\prod_{l=0}^{d_2-2}(k_2-l)}{(k_2-d_2+1)^{d_2-1}}{\rm mult}(\boldsymbol{j}_1\mid k_1-d_1,\vx_1)\\
        &&\nonumber\times{\rm mult}(\boldsymbol{j}_2\mid k_2-d_2,\vx_2)\prod_{l=1}^{d_3}{\rm bin}(j_{3,l}\mid k_3-1,x_{3,l})\\
        &&\nonumber+\sum_{\boldsymbol{j}_1\in J_{d_1}^{k_1-d_1}}\sum_{\boldsymbol{j}_2\in J_{d_2}^{k_2-d_2}}\sum_{\boldsymbol{j}_3\in \{0,\ldots,k_3-1\}^{d_3}} f\left(\frac{\boldsymbol{j}_1}{k_1-2d_1+1},\frac{\boldsymbol{j}_2}{k_2-2d_2+1},\frac{\boldsymbol{j}_3}{k_3-1}\right)\\
        &&\nonumber\times\frac{\prod_{l=0}^{d_1-2}(k_1-l)}{(k_1-d_1+1)^{d_1-1}}\frac{\prod_{l=0}^{d_2-2}(k_2-l)}{(k_2-d_2+1)^{d_2-1}}{\rm mult}(\boldsymbol{j}_1\mid k_1-d_1,\vx_1)\\
        &&\times{\rm mult}(\boldsymbol{j}_2\mid k_2-d_2,\vx_2)\prod_{l=1}^{d_3}{\rm bin}(j_{3,l}\mid k_3-1,x_{3,l}).\label{eq:density_approx}
\end{eqnarray}
For the first term of the summation in (\ref{eq:density_approx}), it is easy to see that 
\begin{eqnarray*} 
&&\hspace{-1.8cm}f(\boldsymbol{C}_{\boldsymbol{j}})- f\left(\frac{\boldsymbol{j}_1}{k_1-2d_1+1},\frac{\boldsymbol{j}_2}{k_2-2d_2+1},\frac{\boldsymbol{j}_3}{k_3-1}\right)\\&\leq&\left|f(\boldsymbol{C}_{\boldsymbol{j}})- f\left(\frac{\boldsymbol{j}_1}{k_1-2d_1+1},\frac{\boldsymbol{j}_2}{k_2-2d_2+1},\frac{\boldsymbol{j}_3}{k_3-1}\right)\right|.
\end{eqnarray*}
For $\boldsymbol{C}_{\boldsymbol{j}}\in {\tilde{\rho}_{d_1,\boldsymbol{j}_1}^{k_1-d_1+1}}\times{\tilde{\rho}_{d_2,\boldsymbol{j}_2}^{k_2-d_2+1}}\times\tilde{\rho}_{d_3,\boldsymbol{j}_3}^{k_3}$, let $\boldsymbol{C}_{\boldsymbol{j}}=(\boldsymbol{C}_{\boldsymbol{j}_1},\boldsymbol{C}_{\boldsymbol{j}_2},\boldsymbol{C}_{\boldsymbol{j}_3})$, where $\boldsymbol{C}_{\boldsymbol{j}_1}=(C_{\boldsymbol{j}_1,1},\ldots,C_{\boldsymbol{j}_1,d_1})$, $\boldsymbol{C}_{\boldsymbol{j}_2}=(C_{\boldsymbol{j}_2,1},\ldots,C_{\boldsymbol{j}_2,d_2})$, $\boldsymbol{C}_{\boldsymbol{j}_3}=(C_{\boldsymbol{j}_3,1},\ldots,C_{\boldsymbol{j}_3,d_3})$. That is to say, $C_{\boldsymbol{j}_1,r}\in \left(\frac{j_{1,r}}{{k_1-d_1+1}},\frac{j_{1,r}+1}{{k_1-d_1+1}}\right]$, $r=1,\ldots,d_1$,
 $C_{\boldsymbol{j}_2,s}\in \left(\frac{j_{2,s}}{{k_2-d_2+1}},\frac{j_{2,s}+1}{{k_2-d_2+1}}\right]$, $s=1,\ldots,d_2$,
 $C_{\boldsymbol{j}_3,t}\in \left(\frac{j_{3,t}}{{k_3}},\frac{j_{3,t}+1}{{k_3}}\right]$, $t=1,\ldots,d_3$.
 
For any $\delta >0$, there exists $N_1>0$, $N_2>0$ and $N_3>0$, such that in $\boldsymbol{j}_1$, when $k_1 > N_1$, \begin{eqnarray*} \left|C_{\boldsymbol{j}_1,r}-\frac{j_{1,r}}{k_1-2d_1+1}\right|&\leq&\max\left\{\left|\frac{j_{1,r}}{{k_1-d_1+1}}-\frac{j_{1,r}}{k_1-2d_1+1}\right|,\left|\frac{j_{1,r}+1}{{k_1-d_1+1}}-\frac{j_{1,r}}{k_1-2d_1+1}\right|\right\}\\&<&\delta \cdot (d_1+d_2+d_3)^{-\frac{1}{2}},\end{eqnarray*}
in $\boldsymbol{j}_2$, when $k_2 > N_2$, \begin{eqnarray*} \left|C_{\boldsymbol{j}_2,s}-\frac{j_{2,s}}{k_2-2d_2+1}\right|&\leq&\max\left\{\left|\frac{j_{2,s}}{{k_2-d_2+1}}-\frac{j_{2,s}}{k_2-2d+1}\right|,\left|\frac{j_{2,s}+1}{{k_2-d_2+1}}-\frac{j_{2,s}}{k_2-2d_2+1}\right|\right\}\\&<&\delta \cdot (d_1+d_2+d_3)^{-\frac{1}{2}},\end{eqnarray*}
in $\boldsymbol{j}_3$, when $k_3 > N_3$, \begin{eqnarray*} \left|C_{\boldsymbol{j}_3,t}-\frac{j_{3,t}}{k_3-1}\right|&\leq& \max\left\{\left|\frac{j_{3,t}}{k_3-1}-\frac{j_{3,t}}{k_3}\right|,\left|\frac{j_{3,t}}{k_3-1}-\frac{j_{3,t}+1}{k_3}\right|\right\},\\&<&\delta \cdot (d_1+d_2+d_3)^{-\frac{1}{2}}.\end{eqnarray*}
Therefore, when $\min\{k_1,k_2,k_3\}>\max\{N_1,N_2,N_3\}$, we have \[d\left(\boldsymbol{C}_{\boldsymbol{j}},\left(\frac{\boldsymbol{j}_1}{k_1-2d_1+1},\frac{\boldsymbol{j}_2}{k_2-2d_2+1},\frac{\boldsymbol{j}_3}{k_3-1}\right)\right)<\delta.\]
By Heine–Cantor theorem, for continuous $f$ and any $\epsilon>0$, there exists $\delta>0$, such that \[d\left(\boldsymbol{C}_{\boldsymbol{j}},\left(\frac{\boldsymbol{j}_1}{k_1-2d_1+1},\frac{\boldsymbol{j}_2}{k_2-2d_2+1},\frac{\boldsymbol{j}_3}{k_3-1}\right)\right)<\delta,\] which implies that \[\left|f(\boldsymbol{C}_{\boldsymbol{j}})- f\left(\left(\frac{\boldsymbol{j}_1}{k_1-2d_1+1},\frac{\boldsymbol{j}_2}{k_2-2d_2+1},\frac{\boldsymbol{j}_3}{k_3-1}\right)\right)\right|<\epsilon.\]\\
Since \[\frac{\prod_{l=0}^{d-2}(k-l)}{(k-d+1)^{d-1}}\stackrel{k\to\infty}{\longrightarrow}1,\]
when $\min\{k_1,k_2,k_3\}\to \infty$, we can show that the first term of the summation in (\ref{eq:density_approx}) goes to $0$.

Notice that the second term of the summation in (\ref{eq:density_approx}) can be rewritten as 
\begin{eqnarray}
&&\frac{\prod_{l=0}^{d_1-2}(k_1-l)}{(k_1-d_1+1)^{d_1-1}}\frac{\prod_{l=0}^{d_2-2}(k_2-l)}{(k_2-d_2+1)^{d_2-1}}\nonumber\\&\times&\sum_{\boldsymbol{j}_1\in J_{d_1}^{k_1^{*}}}\sum_{\boldsymbol{j}_2\in J_{d_2}^{k_2^{*}}}\sum_{\boldsymbol{j}_3\in \{0,\ldots,k_3^{*}\}^{d_3}} f\left(\frac{\boldsymbol{j}_1}{k_1^{*}-d_1+1},\frac{\boldsymbol{j}_2}{k_2^{*}-d_2+1},\frac{\boldsymbol{j}_3}{k_3^{*}}\right)\nonumber\\
&\times&{\rm mult}(\boldsymbol{j}_1\mid k_1^{*},\vx_1){\rm mult}(\boldsymbol{j}_2\mid k_2^{*},\vx_2)\prod_{l=1}^{d_3}{\rm bin}(j_{3,l}\mid k_3^{*},x_{3,l}),\label{eq:converge2f}   
\end{eqnarray}
where $k_1^{*}=k_1-d_1$, $k_2^{*}=k_2-d_2$ and $k_3^{*}=k_3-1$. Since $\frac{\prod_{l=0}^{d_1-2}(k_1-l)}{(k_1-d_1+1)^{d_1-1}}\frac{\prod_{l=0}^{d_2-2}(k_2-l)}{(k_2-d_2+1)^{d_2-1}}\to 1$ when $\min\{k_1,k_2\}\to \infty$, (\ref{eq:converge2f}) converges to $B_{k_1^{*}k_2^{*}k_3^{*}}^f(\bx)$, which has been proved to uniformly converge to $f(\bx)$ In Theorem \ref{thm:appro2F}. Thus, $b_{k_1,k_2,k_3}^F(\bx)\xrightarrow[]{\min\{k_1,k_2,k_3\}\to \infty}f(\bx)$.
\newpage
\section{Proofs of results from Section \ref{sec:randomMBP}} \label{sec:RMBP_proof}

In this section, we provide the proofs of the theoretical results from Section \ref{sec:randomMBP}. To simplify the notation in the proofs and without loss of generality, we now limit Definition \ref{df:MBP} to the case of two simplices instead of $M$. Set $\tildeDelta_d = S_{d_1} \times S_{d_2} \times [0,1]^{d_3}$, where $d = d_1 + d_2 + d_3$, which was defined in Definition \ref{df:MBP_low}. 

We also provide the proof of supplementary lemmas supporting the proof of posterior consistency for marginal and conditional densities.

\subsection{Proof of Theorem \ref{thm:support}}\label{proof:support}
This proof employs techniques similar to those utilized in the proof of Theorem 2 in \citet{petrone1999random}. For space $\mathscr{F}$, we consider the probability space $(\mathscr{F},\mathscr{B}(\mathscr{F}),\pi)$, where $g\sim \pi$, $\mathscr{B}(\mathscr{F})$ is the Borel $\sigma$-algebra, with respect to $L_\infty$-norm.
    
To prove $G$ has full $L_{\infty}$-support on $\mathscr{F}$, it suffices to prove that, for any open set $O\subset \mathscr{F}, O\in \mathscr{B}$, $\pi(O)>0$ is satisfied. Notice that to show that the open set $O$ has positive probability, it suffices to show that any open ball in $O$ has positive mass. This is because for any set $O$, there exists an open ball $Q_{\epsilon}({{G_0}})\subseteq O$, where 
\[Q_{\epsilon}({{G_0}})=\left\{G\in\mathscr{F}:\|g-g_0\|_{\infty}<\epsilon\right\}.\] 
Since $\pi(O)\geq\pi(Q)$, to prove $\pi(O)>0$, we only need to prove $\pi(Q)>0$.
    
It has been shown in Theorem \ref{thm:appro2F} that, for any $\epsilon>0$ and $G_0\in\mathscr{F}$ with continuous density function $g_0$, there exists $k_1^0$, $k_2^0$, $k_3^0>0$, such that $\left\|b_{k_1^0k_2^0k_3^0}^{G_0}-g_0\right\|_{\infty}<\frac{\epsilon}{2}$. Therefore, it suffices to show that
\begin{eqnarray}
\pi\left(Q_{\frac{\epsilon}{2}}({{b_{k_{1}^{0}k_{2}^{0}k_{3}^{0}}^{G_{0}}}})\right) =  {\prob}\left(G\left(\cdot|k_1,k_2,k_3,F\right)\in Q_{\frac{\epsilon}{2}}({{b_{k_{1}^{0}k_{2}^{0}k_{3}^{0}}^{G_{0}}}})\right)>0,\label{eq:posi_prob_ball}
\end{eqnarray}
as $Q_{\frac{\epsilon}{2}}({{b_{k_{1}^{0}k_{2}^{0}k_{3}^{0}}^{G_{0}}}})\subset Q_{\epsilon}({{G_0}}) \subseteq O$.

Notice that,
\begin{eqnarray*}
& & \hspace{-15mm} {\prob}\left(G(\cdot|k_1,k_2,k_3,F)\in Q_{\frac{\epsilon}{2}}({{b_{k_{1}^{0}k_{2}^{0}k_{3}^{0}}^{G_{0}}}})\right)\\
& = &{\prob}\left(\left\Vert g(\cdot|k_1,k_2,k_3,F)-b_{k_{1}^{0}k_{2}^{0}k_{3}^{0}}^{G_{0}}(\cdot)\right\Vert_{\infty}<\frac{\epsilon}{2}\right),\\
& > & {\prob}\left(k_{1}=k_{1}^{0},k_{2}=k_{2}^{0},k_{3}=k_{3}^{0},\left\Vert g(\cdot|k_{1}^{0},k_{2}^{0},k_{3}^{0},F)-b_{k_{1}^{0}k_{2}^{0}k_{3}^{0}}^{G_{0}}(\cdot)\right\Vert _{\infty}<\frac{\epsilon}{2}\right),\\
&=& {\prob}\left(k_{1}=k_{1}^{0},k_{2}=k_{2}^{0},k_{3}=k_{3}^{0}\right)\times {\prob}\left(\left\Vert g(\cdot|k_{1}^{0},k_{2}^{0},k_{3}^{0},F)-b_{k_{1}^{0}k_{2}^{0}k_{3}^{0}}^{G_{0}}(\cdot)\right\Vert _{\infty}<\frac{\epsilon}{2}\right),
\end{eqnarray*}
and given condition(i), we have ${\prob}(k_{1}=k_{1}^{0},k_{2}=k_{2}^{0},k_{3}=k_{3}^{0})>0$, therefore, to show (\ref{eq:posi_prob_ball}), it suffices to show that
\begin{eqnarray}
    {\prob}\left(\left\Vert g(\cdot|k_{1}^{0},k_{2}^{0},k_{3}^{0},F)-b_{k_{1}^{0}k_{2}^{0}k_{3}^{0}}^{G_{0}}(\cdot)\right\Vert _{\infty}<\frac{\epsilon}{2}\right)>0.\label{eq:posi_prob_infty}
\end{eqnarray}

Now we will find an upper bound for $\left\Vert g(\cdot|k_{1}^{0},k_{2}^{0},k_{3}^{0},F)-b_{k_{1}^{0}k_{2}^{0}k_{3}^{0}}^{G_{0}}(\cdot)\right\Vert _{\infty}$, which will help us to show (\ref{eq:posi_prob_infty}).

Given the fact that, 
\begin{eqnarray*}
g(\vx_{1},\vx_2,\vx_3|k_{1},k_{2},k_{3},F)&=&\sum_{\boldsymbol{j}_{1}\in I_{d_1}^{k_1}}\sum_{\boldsymbol{j}_{2}\in I_{d_2}^{k_2}}\sum_{\boldsymbol{j}_{3}\in\{1,\ldots,k_{3}\}^{d_{3}}}F\left(\rho_{d_{1},\boldsymbol{j}_{1}}^{k_{1}-d_{1}+1}\times\rho_{d_{2},\boldsymbol{j}_{2}}^{k_{2}-d_{2}+1}\times\rho_{d_{3},\boldsymbol{j}_{3}}^{k_{3}}\right)\\
&&\times{{\rm dir}(\vx_{1}\mid \alpha(k_{1}-d_{1}+1,\boldsymbol{j}_{1})){\rm dir}(\vx_2\mid\alpha(k_{2}-d_{2}+1,\boldsymbol{j}_{2}))}\\
&&\times\prod_{l=1}^{d_{3}}{{\rm beta}(x_{3,l}\mid j_{3,l},k_{3}-j_{3,l}+1)},
\end{eqnarray*}
and 
\begin{eqnarray*}
b_{k_{1}^{0}k_{2}^{0}k_{3}^{0}}^{G_{0}}(\vx_{1},\vx_2,\vx_3)&=&\sum_{\boldsymbol{j}_{1}\in I_{d_1}^{k_1^0}}\sum_{\boldsymbol{j}_{2}\in I_{d_2}^{k_2^0}}\sum_{\boldsymbol{j}_{3}\in\{1,\ldots,k_{3}^{0}\}^{d_{3}}}G_{0}\left(\rho_{d_{1},\boldsymbol{j}_{1}}^{k_{1}^{0}-d_{1}+1}\times\rho_{d_{2},\boldsymbol{j}_{2}}^{k_{2}^{0}-d_{2}+1}\times\rho_{d_{3},\boldsymbol{j}_{3}}^{k_{3}^{0}}\right)\\
&&\times{\rm dir}(\vx_{1}\mid\alpha(k_{1}^{0}-d_{1}+1,\boldsymbol{j}_{1})){\rm dir}(\vx_2\mid\alpha(k_{2}^{0}-d_{2}+1,\boldsymbol{j}_{2}))\\
&&\times\prod_{l=1}^{d_{3}}{{\rm beta}(x_{3,l}\mid j_{3,l},k_{3}^{0}-j_{3,l}+1),}
\end{eqnarray*}

we have
\begin{eqnarray*}
 & & \hspace{-20mm} \left\Vert g(\cdot|k_{1}^{0},k_{2}^{0},k_{3}^{0},F)-b_{k_{1}^{0}k_{2}^{0}k_{3}^{0}}^{G_{0}}(\cdot)\right\Vert _{\infty} \\ & =  &\sup_{\vx_{1},\vx_2,\vx_3}\left|g(\vx_{1},\vx_2,\vx_3|k_{1}^{0},k_{2}^{0},k_{3}^{0},F)-b_{k_{1}^{0}k_{2}^{0}k_{3}^{0}}^{G_{0}}(\vx_{1},\vx_2,\vx_3)\right|,\\&<&\sum_{\boldsymbol{j}_{1}\in I_{d_1}^{k_1^0}}\sum_{\boldsymbol{j}_{2}\in I_{d_2}^{k_2^0}}\sum_{\boldsymbol{j}_{3}\in\{1,\ldots,k_{3}^{0}\}^{d_{3}}}\\&&\left|F\left(\rho_{d_{1},\boldsymbol{j}_{1}}^{k_{1}^{0}-d_{1}+1}\times\rho_{d_{2},\boldsymbol{j}_{2}}^{k_{2}^{0}-d_{2}+1}\times\rho_{d_{3},\boldsymbol{j}_{3}}^{k_{3}^{0}}\right)-G_{0}\left(\rho_{d_{1},\boldsymbol{j}_{1}}^{k_{1}^{0}-d_{1}+1}\times\rho_{d_{2},\boldsymbol{j}_{2}}^{k_{2}^{0}-d_{2}+1}\times\rho_{d_{3},\boldsymbol{j}_{3}}^{k_{3}^{0}}\right)\right|\\
&&\times{\sup_{\vx_{1}}{\rm dir}(\vx_{1}\mid\alpha(k_{1}^{0}-d_{1}+1,\boldsymbol{j}_{1})){\sup_{\vx_2}{\rm dir}(\vx_2\mid\alpha(k_{2}^{0}-d_{2}+1,\boldsymbol{j}_{2}))}}\\
&&\times\sup_{\vx_3}\prod_{l=1}^{d_{3}}{{\rm beta}(x_{3,l}\mid j_{3,l},k_{3}^{0}-j_{3,l}+1)},\\
&=&\sum_{\boldsymbol{j}_{1}\in I_{d_1}^{k_1^0}}\sum_{\boldsymbol{j}_{2}\in I_{d_2}^{k_2^0}}\sum_{\boldsymbol{j}_{3}\in\{1,\ldots,k_{3}^{0}\}^{d_{3}}}\\
&&\left|F\left(\rho_{d_{1},\boldsymbol{j}_{1}}^{k_{1}^{0}-d_{1}+1}\times\rho_{d_{2},\boldsymbol{j}_{2}}^{k_{2}^{0}-d_{2}+1}\times\rho_{d_{3},\boldsymbol{j}_{3}}^{k_{3}^{0}}\right)-G_{0}\left(\rho_{d_{1},\boldsymbol{j}_{1}}^{k_{1}^{0}-d_{1}+1}\times\rho_{d_{2},\boldsymbol{j}_{2}}^{k_{2}^{0}-d_{2}+1}\times\rho_{d_{3},\boldsymbol{j}_{3}}^{k_{3}^{0}}\right)\right|\\
&&\times C(k_{1}^{0},k_{2}^{0},k_{3}^{0}),
\end{eqnarray*}
where 
\begin{eqnarray}
\nonumber C(k_{1}^{0},k_{2}^{0},k_{3}^{0})&=&{ \sup_{\vx_{1}}{\rm dir}(\vx_{1}\mid\alpha(k_{1}^{0}-d_{1}+1,\boldsymbol{j}_{1})){\sup_{\vx_2}{\rm dir}(\vx_2\mid\alpha(k_{2}^{0}-d_{2}+1,\boldsymbol{j}_{2}))}}\\&&\times\sup_{\vx_3}\prod_{l=1}^{d_{3}}{ {\rm beta}(x_{3,l}\mid j_{3,l},k_{3}^{0}-j_{3,l}+1)}.\label{eq:mixture_ub_const}
\end{eqnarray}
Therefore, to prove (\ref{eq:posi_prob_infty}), it suffices to show that
\begin{eqnarray*}
{\prob}&\bigg(&\sum_{\boldsymbol{j}_{1}\in I_{d_1}^{k_1^0}}\sum_{\boldsymbol{j}_{2}\in I_{d_2}^{k_2^0}} \sum_{\boldsymbol{j}_{3}\in\{1,\ldots,k_{3}^{0}\}^{d_{3}}}C(k_{1}^{0},k_{2}^{0},k_{3}^{0})\bigg.\\&&\bigg.\times\left|F\left(\rho_{d_{1},\boldsymbol{j}_{1}}^{k_{1}^{0}-d_{1}+1}\times\rho_{d_{2},\boldsymbol{j}_{2}}^{k_{2}^{0}-d_{2}+1}\times\rho_{d_{3},\boldsymbol{j}_{3}}^{k_{3}^{0}}\right)-G_{0}\left(\rho_{d_{1},\boldsymbol{j}_{1}}^{k_{1}^{0}-d_{1}+1}\times\rho_{d_{2},\boldsymbol{j}_{2}}^{k_{2}^{0}-d_{2}+1}\times\rho_{d_{3},\boldsymbol{j}_{3}}^{k_{3}^{0}}\right)\right|<\frac{\epsilon}{2}\bigg)>0,
\end{eqnarray*}
which is implied if
\begin{eqnarray}
{\prob}&\bigg(&\nonumber\left|F\left(\rho_{d_{1},\boldsymbol{j}_{1}}^{k_{1}^{0}-d_{1}+1}\times\rho_{d_{2},\boldsymbol{j}_{2}}^{k_{2}^{0}-d_{2}+1}\times\rho_{d_{3},\boldsymbol{j}_{3}}^{k_{3}^{0}}\right)-G_{0}\left(\rho_{d_{1},\boldsymbol{j}_{1}}^{k_{1}^{0}-d_{1}+1}\times\rho_{d_{2},\boldsymbol{j}_{2}}^{k_{2}^{0}-d_{2}+1}\times\rho_{d_{3},\boldsymbol{j}_{3}}^{k_{3}^{0}}\right)\right|\bigg.\\
&\bigg.&\hspace{28mm}<\epsilon\left(2C(k_{1}^{0},k_{2}^{0},k_{3}^{0})\times \text{card}\left(I_{d_1}^{k_1^0}\right)\text{card}\left(I_{d_2}^{k_2^0}\right)(k_{3}^{0})^{d_{3}}\right)^{-1}\bigg)>0\nonumber
\end{eqnarray}
which is in turn implied if 
\begin{eqnarray}
{\prob}&\bigg(&\nonumber\left|F\left(\rho_{d_{1},\boldsymbol{j}_{1}}^{k_{1}^{0}-d_{1}+1}\times\rho_{d_{2},\boldsymbol{j}_{2}}^{k_{2}^{0}-d_{2}+1}\times\rho_{d_{3},\boldsymbol{j}_{3}}^{k_{3}^{0}}\right)-G_{0}\left(\rho_{d_{1},\boldsymbol{j}_{1}}^{k_{1}^{0}-d_{1}+1}\times\rho_{d_{2},\boldsymbol{j}_{2}}^{k_{2}^{0}-d_{2}+1}\times\rho_{d_{3},\boldsymbol{j}_{3}}^{k_{3}^{0}}\right)\right|\bigg.\\
&\bigg.&\hspace{52mm}<\epsilon\left(2C(k_{1}^{0},k_{2}^{0},k_{3}^{0})\times C_{d_1}^{k_1^0} C_{d_2}^{k_2^0}(k_{3}^{0})^{d_{3}}\right)^{-1}\bigg)>0,\label{eq:ferguson}
\end{eqnarray}
because
\begin{eqnarray}\label{eq:cardinality}
\text{card}\left(I_{d_i}^{k_i^0}\right)=C_{d_i}^{k_i^0},
\end{eqnarray} 
is the cardinality of the set of $\boldsymbol{j}_{i}$ for $i=1,2$.

As indicated by Proposition 3 in \cite{ferguson1973bayesian}, since $\rho_{d_{1},\boldsymbol{j}_{1}}^{k_{1}^{0}-d_{1}+1}\times\rho_{d_{2},\boldsymbol{j}_{2}}^{k_{2}^{0}-d_{2}+1}\times\rho_{d_{3},\boldsymbol{j}_{3}}^{k_{3}^{0}}$ is one of the measurable sets  (elements of a partition of $\tildeDelta_d$) and $F$ is a Dirichlet process, then inequality (\ref{eq:ferguson}) holds. Therefore, we have $\pi\left(Q_{\frac{\epsilon}{2}}({{b_{k_{1}^{0}k_{2}^{0}k_{3}^{0}}^{G_{0}}}})\right)>0$.

\subsection{Proof of Theorem \ref{thm:postweak}}\label{proof:postweak}
Before presenting the proof, we first introduce some notation. Let $\boldsymbol{k}=(k_1,k_2,k_{3})$, where $k_l\sim p_l(\cdot\mid \lambda_l)$, $l=1,2,3$ , $F\sim{\rm DP}(F_0,M_0)$, and $\boldsymbol{j}=(\boldsymbol{j}_1,\boldsymbol{j}_2,\boldsymbol{j}_3)$, where $\boldsymbol{j}_l$ be a random vector of dimension $d_l$, for $l=1,2,3$. Recall that $\mathcal{I}(\boldsymbol{k})=I_{d_1}^{k_1} \times I_{d_2}^{k_2} \times \{1,\ldots,k_3\}^{d_3}$ is defined in proof \ref{proof:mbp_cdf} with cardinality $C_{d_1}^{k_1} C_{d_2}^{k_2}(k_{3})^{d_{3}}$, and that the density function of the proposed MBP,
\begin{eqnarray*}
    b\left\{\bx;\boldsymbol{k},w_{\boldsymbol{k}}(F)\right\}&\coloneqq& b(\bx|\boldsymbol{k}, F),\\&=&\sum_{\boldsymbol{j}\in \mathcal{I}(\boldsymbol{k})}F\left({\rho_{d_1,\boldsymbol{j}_1}^{k_1-d_1+1}}\times{\rho_{d_2,\boldsymbol{j}_2}^{k_2-d_2+1}}\times\rho_{d_3,\boldsymbol{j}_3}^{k_3}\right){\rm dir}(\vx_1\mid\alpha(k_1-d_1+1,\boldsymbol{j}_1))\\
    &&\times{\rm dir}(\vx_2\mid\alpha(k_2-d_2+1,\boldsymbol{j}_2)) \prod_{l=1}^{d_3}{\rm beta}(x_{3,l}\mid j_{3,l},k_3-j_{3,l}+1),\\
    &=&\sum_{\boldsymbol{j}\in \mathcal{I}(\boldsymbol{k})} w_{\boldsymbol{j},\boldsymbol{k}}(F)
    {\rm dir}(\vx_1\mid\alpha(k_1-d_1+1,\boldsymbol{j}_1))\\
    &&\times {\rm dir}(\vx_2\mid\alpha(k_2-d_2+1,\boldsymbol{j}_2))\prod_{l=1}^{d_3}{\rm beta}(x_{3,l}\mid j_{3,l},k_3-j_{3,l}+1),
\end{eqnarray*}
where \[w_{\boldsymbol{j},\boldsymbol{k}}(F)=F\left({\rho_{d_1,\boldsymbol{j}_1}^{k_1-d_1+1}}\times{\rho_{d_2,\boldsymbol{j}_2}^{k_2-d_2+1}}\times\rho_{d_3,\boldsymbol{j}_3}^{k_3}\right)\]
is the probability mass associated with the partition of space $\tildeDelta_d$, and \[w_{\boldsymbol{k}}(F)=\mathrm{Vec}\left((w_{\boldsymbol{j},\boldsymbol{k}}(F))_{\boldsymbol{j}\in\mathcal{I}(\boldsymbol{k})}\right).\]

According to Theorem 6.1 in \cite{schwartz1965bayes}, to show that the posterior density of $G$ is weakly consistent at $g_0$, we only need to show that $\pi(KL_{\epsilon}({{g_0}}))>0$ for every $\epsilon>0$, where $KL_{\epsilon}({{g_0}})$ is a Kullback-Leibler neighborhood of $g_0$. To demonstrate that $g_0$ is in the Kullback-Leibler support of $\Pi$, we adopt a reasoning approach similar to that in Theorem 2 of \citet{petrone2002consistency}.

 First, we show that when $g_0$ is bounded away from $0$, i.e. $\inf_{\bx\in \tildeDelta_d}\{g_0(\bx)\}>0$, we have $\pi(KL_{\epsilon}({{g_0}}))>0$. To show this, we first claim that for any $\epsilon>0$, there exists $b\{\cdot;\boldsymbol{k}_0,w_{\boldsymbol{k}_0}(G_0)\}$ such that $KL(g_0,b\{\cdot;\boldsymbol{k}_0,w_{\boldsymbol{k}_0}(G_0)\})<\epsilon$, where $\boldsymbol{k}_0=(k_0,k_0,k_0)$ and $KL(g_0,b\{\cdot;\boldsymbol{k}_0,w_{\boldsymbol{k}_0}(G_0)\})$ is the Kullback–Leibler divergence between $g_0$ and $b\{\cdot;\boldsymbol{k}_0,w_{\boldsymbol{k}_0}(G_0)\}$. 
 
By Theorem \ref{thm:appro2f}, for any $G_0\in \mathscr{F}$ 
\[\lim_{\min\{k_1,k_2,k_3\}\to \infty}\sup_{\bx\in \tildeDelta_d} \left|b\{\bx;\boldsymbol{k},w_{\boldsymbol{k}}(G_0)\}-g_0(\bx)\right|=0.\] For fix $c>0$ sufficiently small, we can choose $k_0=k_0(c)$ such that, if $\min\{k_1,k_2,k_3\}\geq k_0$, $\sup_{\bx}|b\{\bx;\boldsymbol{k}, w_{\boldsymbol{k}}(G_0)\}-g_0(\bx)|<c$, then $b\{\cdot;\boldsymbol{k},w_{\boldsymbol{k}}(G_0)\}$ is bounded and bounded away from $0$. Therefore, there exists $M>0$ such that, for any $\bx\in \tildeDelta_d$ and $\min\{k_1,k_2,k_3\}\geq k_0$, 
\[\left|\log\left(\frac{g_0(\bx)}{b\{\bx;\boldsymbol{k},w_{\boldsymbol{k}}(G_0)\}}\right)\right|<M.\]

By dominated convergence theorem,
\begin{eqnarray*}
&\lim_{\min\{k_1,k_2,k_3\}\to\infty}\left(\int\log\left(\frac{g_0(\bx)}{b\{\bx;\boldsymbol{k},w_{\boldsymbol{k}}(G_0)\}}\right)g_0(\bx)d\bx\right)\\
&=\int\lim_{\min\{k_1,k_2,k_3\}\to\infty}\left\{\log\left(\frac{g_0(\bx)}{b\{\bx;\boldsymbol{k},w_{\boldsymbol{k}}(G_0)\}}\right)g_0(\bx)d\bx\right\}=0.
\end{eqnarray*}
Therefore, we have shown that for any $\epsilon>0$ there exists $k_0$ such that \[KL\left(g_0,b\{\cdot;\boldsymbol{k}_0,w_{\boldsymbol{k}_0}(G_0)\}\right)<\epsilon.\]
Then we show that there is $KL_{\epsilon}({{g_0}})$, a Kullback-Leibler neighborhood of $g_0$ (bounded away from 0) with positive prior probability. For $\delta>0$, let
\[N_{\delta}(g_0)=\left\{G \in \mathscr{F} :\max_{\boldsymbol{j}\in \mathcal{I}(\boldsymbol{k}_0)}|w_{\boldsymbol{j},\boldsymbol{k}_0}(G)-w_{\boldsymbol{j},\boldsymbol{k}_0}(G_0)|<\delta\right\}.\]
As $N_{\delta}(g_0)$ is a weak neighborhood, ${\prob} (G \in N_{\delta}(g_0))>0$. Recall that $b\{\cdot;\boldsymbol{k}_0,w_{\boldsymbol{k}_0}(G_0)\}$ is bounded and bounded away from $0$. Since 
\begin{eqnarray*}
C(k_{1},k_{2},k_{3})&\geq&{{\rm dir}(\vx_1\mid\alpha(k_{1}-d_{1}+1,\boldsymbol{j}_{1}))}{{\rm dir}(\vx_2\mid\alpha(k_{2}-d_{2}+1,\boldsymbol{j}_{2}))}\\&&\times\prod_{l=1}^{d_{3}}{{\rm beta}(x_{3,l}\mid j_{3,l},k_{3}-j_{3,l}+1)},
\end{eqnarray*} 
for all $\bx$ and all $\boldsymbol{j}\in \mathcal{I}(\boldsymbol{k})$, where 
\begin{eqnarray*}
C(k_{1},k_{2},k_{3})&=&\sup_{\vx_1}{\rm dir}(\vx_1\mid\alpha(k_{1}-d_{1}+1,\boldsymbol{j}_{1})) \sup_{\vx_2}{\rm dir}(\vx_2\mid\alpha(k_{2}-d_{2}+1,\boldsymbol{j}_{2}))\\ &&\times \sup_{\vx_3}\prod_{l=1}^{d_{3}}{{\rm beta}(x_{3,l}\mid j_{3,l},k_{3}-j_{3,l}+1)}.
\end{eqnarray*}
We have that 
\begin{eqnarray*}
&&\hspace{-2.5cm}\sup_{\bx\in \tildeDelta_d}|b\{\cdot;\boldsymbol{k}_0,w_{\boldsymbol{k}_0}(G)\}-b\{\cdot;\boldsymbol{k}_0,w_{\boldsymbol{k}_0}(G_0)\}|\\&\leq&\sup_{\bx\in \tildeDelta_d}\Bigg\{\sum_{\boldsymbol{j}\in \mathcal{I}(\boldsymbol{k}_0)}|w_{\boldsymbol{j},\boldsymbol{k}_0}(G)-w_{\boldsymbol{j},\boldsymbol{k}_0}(G_0)|{{\rm dir}(\vx_1\mid\alpha(k_{0}-d_{1}+1,\boldsymbol{j}_{1})) }\Bigg.\\&&\Bigg.\times{{\rm dir}(\vx_2\mid\alpha(k_{0}-d_{2}+1,\boldsymbol{j}_{2}))}\prod_{l=1}^{d_{3}}{{\rm beta}(x_{3,l}\mid j_{3,l},k_{0}-j_{3,l}+1)}\Bigg\},\\&\leq& C(k_{0},k_{0},k_{0})\times C_{d_1}^{k_0} C_{d_2}^{k_0}(k_{0})^{d_{3}}\times\max_{\boldsymbol{j}\in \mathcal{I}(\boldsymbol{k}_0)}|w_{\boldsymbol{j},\boldsymbol{k}_0}(G)-w_{\boldsymbol{j},\boldsymbol{k}_0}(G_0)|.
\end{eqnarray*}
By choosing $\delta$ sufficiently small, for any $G\in N_{\delta}(g_0)$, we can make $b\{\cdot;\boldsymbol{k}_0,w_{\boldsymbol{k}_0}(G)\}$ bounded and bounded away from $0$ also. It follows that, for any $\bx\in \tildeDelta_d$ and $G\in N_{\delta}(g_0)$, 
\[\left|\log\left(\frac{g_0(\bx)}{b\{\bx;\boldsymbol{k}_0,w_{\boldsymbol{k}_0}(G)\}}\right)\right|<M.\]
Therefore, by dominated convergence theorem, \[\lim_{w_{\boldsymbol{k}_0}(G)\to w_{\boldsymbol{k}_0}(G_0)} KL(g_0,b\{\cdot;\boldsymbol{k}_0,w_{\boldsymbol{k}_0}(G)\})=KL\left(g_0,b\{\cdot;\boldsymbol{k}_0,w_{\boldsymbol{k}_0}(G_0)\}]\right).\]
Hence, $KL(g_0,b\{\cdot;\boldsymbol{k}_0,w_{\boldsymbol{k}_0}(G)\})$ is continuous in $w_{\boldsymbol{k}_0}(G)$ at $w_{\boldsymbol{k}_0}(G_0)$. For any $\epsilon>0$, we can choose $\delta>0$ such that, $KL(g_0,b\{\cdot;\boldsymbol{k}_0,w_{\boldsymbol{k}_0}(G)\})<\epsilon$ for any $G\in N_{\delta}(g_0)$. Since ${\prob} (k_l=k_0)>0,l=1,2,3$, we have
\begin{eqnarray*}
\pi\left(KL_{\epsilon}({{g_0}})\right)&=&\pi \left( \big\{b\{\cdot;\boldsymbol{k}_0,w_{\boldsymbol{k}_0}(G)\}:KL(g_0,b\{\cdot;\boldsymbol{k}_0,w_{\boldsymbol{k}_0}(G)\})\leq \epsilon\big\} \right),\\&\geq& \pi\left(\big\{b\{\cdot;\boldsymbol{k}_0,w_{\boldsymbol{k}_0}(G)\}:G\in N_{\delta}(g_0)\big\}\right),\\
&=& p(\boldsymbol{k}_0){\prob}  (G \in N_{\delta}(g_0)), \\&=& {\prob} (k_1= k_0,k_2= k_0,k_3= k_0){\prob}  (G \in N_{\delta}(g_0))>0.
\end{eqnarray*}

Now, consider when $\inf_{\bx\in \tildeDelta_d}\{g_0(\bx)\}=0$. We make use of lemma 5.1 in \citet{ghosal1999consistent}: if $g_0$ and $g_1$ are densities such that, $g_0\leq C g_1$ for a positive constant $C$, then for any density $h$, 
\begin{eqnarray}
    KL(g_0,h)\leq (C+1)\log C+C\left\{KL(g_1,h)+\sqrt{KL(g_1,h)}\right\}.\label{eq:ghosal_ineq}
\end{eqnarray}
Choosing $a>0$ and defining \[g_1(\bx)=\frac{g_0(\bx)\vee a}{\int_{\tildeDelta_d}(g_0(\bx)\vee a)d\bx},\]
where $a\vee b$ is the maximum between $a$ and $b$, the density $g_1$ is bounded away from $0$ and $\infty$, therefore it is covered by the case above. A quick conclusion is that, for any $\epsilon>0$, we can choose $KL_{\epsilon}({{g_1}})$ with positive prior probability such that, $KL(g_1,b\{\cdot;\boldsymbol{k}_0,w_{\boldsymbol{k}_0}(G)\})<\epsilon$ for any $b\{\cdot;\boldsymbol{k}_0,w_{\boldsymbol{k}_0}(G)\}\in KL_{\epsilon}({{g_1}})$. Now $g_0\leq C g_1$, where $C=\int_{\tildeDelta_d}(g_0(\bx)\vee a) d\bx$, from inequality (\ref{eq:ghosal_ineq}), we have 
\begin{eqnarray*}
    &&\hspace{-2cm}KL(g_0,b\{\cdot;\boldsymbol{k}_0,w_{\boldsymbol{k}_0}(G)\})\\&\leq& (C+1)\log C+C\left\{KL(g_1,b\{\cdot;\boldsymbol{k}_0,w_{\boldsymbol{k}_0}(G)\})+\sqrt{KL(g_1,b\{\cdot;\boldsymbol{k}_0,w_{\boldsymbol{k}_0}(G)\})}\right\}.
\end{eqnarray*}
By choosing $a$ such that $(C+1)\log C<\frac{\epsilon}{2}$ and choosing $\delta$ in $N_{\delta}(g_1)$ sufficiently small, such that, \[C\left\{KL(g_1,b\{\cdot;\boldsymbol{k}_0,w_{\boldsymbol{k}_0}(G)\})+\sqrt{KL(g_1,b\{\cdot;\boldsymbol{k}_0,w_{\boldsymbol{k}_0}(G)\})}\right\}<\frac{\epsilon}{2},\] then we can have $KL(g_0,b\{\cdot;\boldsymbol{k}_0,w_{\boldsymbol{k}_0}(G)\})\leq\epsilon$. Therefore 
\begin{eqnarray*}
\pi\left(KL_{\epsilon}({{g_0}})\right)&=&\pi\left(\big\{b\{\cdot;\boldsymbol{k}_0,w_{\boldsymbol{k}_0}(G)\}:KL(g_0,b\{\cdot;\boldsymbol{k}_0,w_{\boldsymbol{k}_0}(G)\})\leq \epsilon\big\}\right),\\&\geq& \pi\left(\big\{b\{\cdot;\boldsymbol{k}_0,w_{\boldsymbol{k}_0}(G)\}:G\in N_{\delta}(g_1)\big\}\right),\\&>&0.
\end{eqnarray*}

\subsection{Proof of Theorem \ref{thm:poststrong}}\label{proof:poststrong}

Let $\mathcal{B}_{k_1,k_2,k_3}$ be the class of all proposed MBP densities of order $k_1,k_2,k_3$, notice that each MBP in $\mathcal{B}_{k_1,k_2,k_3}$ is defined using a partition of $\tildeDelta_d$, let $K=K(\boldsymbol{k})=K(k_1,k_2,k_3)$ denotes the size of that partition, from (\ref{eq:cardinality}), we know $K=C_{d_1}^{k_1} C_{d_2}^{k_2}k_{3}^{d_{3}}$. From lemma A.4 of \cite{ghosal2001entropies} and the proof of theorem 2.2 of \cite{ghosal2001convergence}, we find that:
\begin{eqnarray}
D(\epsilon,\mathcal{B}_{k_1,k_2,k_3},\|\cdot\|_1)\leq D(\epsilon,S_K,\|\cdot\|_1)\leq \left(\frac{C}{\epsilon}\right)^{K},
\end{eqnarray}\label{eq:pack_num_upbound}where $D(\epsilon,\mathcal{B}_{k_1,k_2,k_3},\|\cdot\|_1)$ is the $\epsilon$-packinging number of $\mathcal{B}_{k_1,k_2,k_3}$, the class of all convex combinations of $K$ densities, which is upper bounded by $\epsilon$-packinging number of the $K$-dimensional simplex, $D(\epsilon,S_K,\|\cdot\|_1)$, which is in turn upper bounded by $\left(\frac{C}{\epsilon}\right)^{K}$ for some absolute constant $C$.

Let $\mathcal{D}_n=\cup_{k_1=1}^{k_n^{(1)}}\cup_{k_2=1}^{k_n^{(2)}}\cup_{k_3=1}^{k_n^{(3)}}\mathcal{B}_{k_1,k_2,k_3}$, then \[\pi\left(\mathcal{D}_n^{\complement}\right)=\sum_{k_1 > k_n^{(1)}}\sum_{k_2 > k_n^{(2)}}\sum_{k_3 > k_n^{(3)}}p(k_1)p(k_2)p(k_3)<ce^{-nr},\] by condition (ii).

Also, we have 
\begin{eqnarray*}
D(\epsilon,\mathcal{D}_n,\|\cdot\|_1)&\leq&\sum_{k_1=1}^{k_n^{(1)}}\sum_{k_2=1}^{k_n^{(2)}}\sum_{k_3=1}^{k_n^{(3)}}D(\epsilon,\mathcal{B}_{k_1,k_2,k_3},\|\cdot\|_1)\\&\leq& k_n^{(1)}k_n^{(2)}k_n^{(3)}\left(\frac{C}{\epsilon}\right)^{K(\boldsymbol{k}_n)}\\&\leq& K(\boldsymbol{k}_n)\left(\frac{C}{\epsilon}\right)^{K(\boldsymbol{k}_n)}\\&\leq& \left(\frac{2C}{\epsilon}\right)^{K(\boldsymbol{k}_n)},
\end{eqnarray*}
where $K(\boldsymbol{k}_n)=K(k_n^{(1)},k_n^{(2)},k_n^{(3)})\geq k_n^{(1)}k_n^{(2)}k_n^{(3)}$ is obvious to see.

As for any $\epsilon>0$, $N(\epsilon,\mathcal{D}_n,\|\cdot\|_1)\leq D(\epsilon,\mathcal{D}_n,\|\cdot\|_1)$, we have,
\begin{eqnarray*}
N(\epsilon,\mathcal{D}_n,\|\cdot\|_1)\leq D(\epsilon,\mathcal{D}_n,\|\cdot\|_1)\leq \left(\frac{2C}{\epsilon}\right)^{K(\boldsymbol{k}_n)}.
\end{eqnarray*}
Thus, 
\begin{eqnarray*}
\log N(\epsilon,\mathcal{D}_n,\|\cdot\|_1)\leq K(\boldsymbol{k}_n)\log (\frac{2C}{\epsilon})=O(K(\boldsymbol{k}_n)).
\end{eqnarray*}
Since $C_{d_l}^{k_l} = O\left({k_l}^{\min\{k_l-d_l,d_l\}}\right)$ and $k_n^{(l)}\to\infty$, we have $C_{d_l}^{k_n^{(l)}} = O\left({k_n^{(l)}}^{d_l}\right)=o\left(n^{1/3}\right)$, $l=1,2$, then 
\begin{eqnarray*}
\log N(\epsilon,\mathcal{D}_n,\|\cdot\|_1)=O\left(K(\boldsymbol{k}_n)\right)=O\left(C_{d_1}^{k_n^{(1)}}\times C_{d_2}^{k_n^{(2)}}\times {k_n^{(3)}}^{d_{3}}\right)=o(n).
\end{eqnarray*} 
Since K-L support condition has already been proved in Theorem \ref{thm:postweak}, then, by Theorem 2 in \cite{ghosal1999posterior}, the posterior distribution of $g$ is $L_1$ consistent at $g_0$.

\subsection{Lemmas supporting the proof of posterior consistency for \break marginal and conditional densities}\label{lemma:supple_lemmas}

Similar to the neighborhoods used for the joint case, neighborhoods for collections of both marginal and conditional densities need to be defined. For the marginal, we introduce the $L_1$-marginal neighborhood of $g_0$, defined as follows,
\[T_{\epsilon,(2,3)}({{g_0}})=\left\{g\in \mathcal{D}:\int_{[0,1]^{d_3}}\int_{S_{d_2}}|g(\vx_2,\vx_3)-g_0(\vx_2,\vx_3)|d\vx_2d\vx_3 <\epsilon\right\}.\] For the conditional, we define the $g_0$-integrated $L_1$ distance $d_{g_0,L_1}$, where 
\begin{eqnarray*}
    &&\hspace{-17mm}d_{(1,G_0,L_1)}(g(\vx_1\mid\vx_2,\vx_3),g_0(\vx_1\mid\vx_2,\vx_3))\\&=&\int_{[0,1]^{d_3}}\int_{S_{d_2}} \int_{S_{d_1}}|g(\vx_1\mid\vx_2,\vx_3)-g_0(\vx_1\mid\vx_2,\vx_3)|d\vx_1g_0(\bx)d\vx_2d\vx_3,
\end{eqnarray*}
and consider the $g_0$-integrated $L_1$ conditional neighborhood of $g_0$ is given by \[T_{\epsilon,(2,3),L_1}({{g_0}})=\{g\in \mathcal{D}:d_{g_0,L_1}(g(\vx_1|\vx_2,\vx_3),g_0(\vx_1|\vx_2,\vx_3))<\epsilon\}.\]

We can now establish definitions of strong consistency for both marginal and conditional densities, similar to the joint case.  Assuming $\bx_i \overset{i.i.d.}{\sim} g_0, \, i=1, \ldots, n$, the posterior distribution of $g$ is said to be $L_1$-marginal consistent at $g_0(\vx_2, \vx_3)$ if, for any $\epsilon > 0$,
         \begin{eqnarray*}
             \pi{(T_{\epsilon,(2,3)}({{g_0}})}\mid \obs_1,\ldots,\obs_n) \underset{n \to \infty}{\longrightarrow} 1,\quad [g_0]\textit{a.s.}
         \end{eqnarray*}
Similarly, under the same assumptions, this posterior is said to be $g_0$-integrated $L_1$-consistent at 
$g_0(\vx_1|\vx_2,\vx_3)$ if for any $\epsilon>0$,
         \begin{eqnarray*}
             \pi{(T_{\epsilon,(2,3),L_1}({{g_0}})}\mid \obs_1,\ldots,\obs_n) \underset{n \to \infty}{\longrightarrow} 1,\quad [g_0]\textit{a.s.}
         \end{eqnarray*}

Since we have established that the proposed posterior distribution derived from the DMBPP prior is strong consistent in estimating the joint density $g_0$ (see Theorem \ref{thm:poststrong}), our results for the marginal and conditional distributions rely on the following two findings: 
\begin{itemize}
    \item[i)] If the joint density function $g(\vx_1, \vx_2,$ $ \vx_3)$ $L_1$-converges to a given joint density $g_0(\vx_1, \vx_2,$ $\vx_3)$, then the marginal density $g(\vx_2, \vx_3)$ also $L_1$-converges to the corresponding marginal density $g_0(\vx_2, \vx_3)$.
    \item[ii)] If the joint density function $g(\vx_1, \vx_2, \vx_3)$ $L_1$-converges to a given joint density $g_0(\vx_1, \vx_2,$ $\vx_3)$, then the conditional density $g(\vx_1 \mid \vx_2, \vx_3)$ converges to the corresponding conditional density $g_0(\vx_1 \mid \vx_2, \vx_3)$ in terms of the $g_0$-integrated $L_1$ distance, denoted as $d_{g_0,L_1}$.
\end{itemize}
These two results are formalized in Lemma \ref{lem:joint2mar} and Lemma \ref{lem:joint2con}, respectively. The proofs of these lemmas are provided in Sections \ref{proof:joint2mar} and \ref{proof:joint2con}.

\begin{lemma}{\bf{Closeness between joints indicates closeness between marginals.}}\label{lem:joint2mar}
    Suppose that for every $\epsilon>0$, $\int_{[0,1]^{d_3}}\int_{S_{d_2}}\int_{S_{d_1}}|g(\vx_1,\vx_2,\vx_3)-g_0(\vx_1,\vx_2,\vx_3)|d\vx_1d\vx_2d\vx_3<\epsilon$. Then, \begin{eqnarray*}
        \int_{[0,1]^{d_3}}\int_{S_{d_2}}|g(\vx_2,\vx_3)-g_0(\vx_2,\vx_3)|d\vx_2d\vx_3<\epsilon.
    \end{eqnarray*}
\end{lemma}

\begin{lemma}{\bf{Closeness between joints indicates closeness between conditionals.}}\label{lem:joint2con}
    Suppose that for every $\epsilon>0$,  $\int_{[0,1]^{d_3}}\int_{S_{d_2}}\int_{S_{d_1}}|g(\vx_1,\vx_2,\vx_3)-g_0(\vx_1,\vx_2,\vx_3)|d\vx_1d\vx_2d\vx_3<\epsilon$. Then, 
    \[d_{(1,G_0,L_1)}(g(\vx_1\mid\vx_2,\vx_3),g_0(\vx_1\mid\vx_2,\vx_3))
    <2\epsilon.\]
\end{lemma}

\subsection{Proof of Lemma \ref{lem:joint2mar}}\label{proof:joint2mar}
Notice that
\begin{eqnarray*}
&&\int_{[0,1]^{d_3}}\int_{S_{d_2}}|g(\vx_2,\vx_3)-g_0(\vx_2,\vx_3)|d\vx_2d\vx_3\\
&=&\int_{[0,1]^{d_3}}\int_{S_{d_2}}\left|\int_{S_{d_1}} g(\vx_1,\vx_2,\vx_3)d\vx_1-\int_{S_{d_1}} g_0(\vx_1,\vx_2,\vx_3)d\vx_1\right|d\vx_2d\vx_3,\\
&=&\int_{[0,1]^{d_3}}\int_{S_{d_2}}\left|\int_{S_{d_1}} \left(g(\vx_1,\vx_2,\vx_3)- g_0(\vx_1,\vx_2,\vx_3)\right)d\vx_1\right|d\vx_2d\vx_3,
\end{eqnarray*}
then by Jensen's inequality, 
\begin{eqnarray*}
&&\int_{[0,1]^{d_3}}\int_{S_{d_2}}\left|\int_{S_{d_1}} \left(g(\vx_1,\vx_2,\vx_3)- g_0(\vx_1,\vx_2,\vx_3)\right)d\vx_1\right|d\vx_2d\vx_3    \\&\leq&\int_{[0,1]^{d_3}}\int_{S_{d_2}}\int_{S_{d_1}}|g(\vx_1,\vx_2,\vx_3)-g_0(\vx_1,\vx_2,\vx_3)|d\vx_1d\vx_2d\vx_3<\epsilon.
\end{eqnarray*}

\subsection{Proof of Lemma \ref{lem:joint2con}}\label{proof:joint2con}
Notice that
\begin{eqnarray*}
    &&\int_{[0,1]^{d_3}}\int_{S_{d_2}} \int_{S_{d_1}}|g(\vx_1\mid\vx_2,\vx_3)-g_0(\vx_1\mid\vx_2,\vx_3)|d\vx_1g_0(\vx_2,\vx_3)d\vx_2d\vx_3\\
    &=&\int_{[0,1]^{d_3}}\int_{S_{d_2}}\int_{S_{d_1}}|g(\vx_1\mid\vx_2,\vx_3)g_0(\vx_2,\vx_3)\pm g(\vx_1,\vx_2,\vx_3)-g_0(\vx_1,\vx_2,\vx_3)|d\vx_1d\vx_2d\vx_3.
\end{eqnarray*}
Then, by triangle inequality,
\begin{eqnarray}
&&\int_{[0,1]^{d_3}}\int_{S_{d_2}}\int_{S_{d_1}}|g(\vx_1\mid\vx_2,\vx_3)g_0(\vx_2,\vx_3)\pm g(\vx_1,\vx_2,\vx_3)-g_0(\vx_1,\vx_2,\vx_3)|d\vx_1d\vx_2d\vx_3\nonumber\\&\leq&\int_{[0,1]^{d_3}}\int_{S_{d_2}}\int_{S_{d_1}}|g(\vx_1\mid\vx_2,\vx_3)g_0(\vx_2,\vx_3)-g(\vx_1,\vx_2,\vx_3)|d\vx_1d\vx_2d\vx_3\nonumber\\&+&\int_{[0,1]^{d_3}}\int_{S_{d_2}}\int_{S_{d_1}}|g(\vx_1,\vx_2,\vx_3)-g_0(\vx_1,\vx_2,\vx_3)|d\vx_1d\vx_2d\vx_3,\nonumber\\
&=&\int_{[0,1]^{d_3}}\int_{S_{d_2}} |g_0(\vx_2,\vx_3)-g(\vx_2,\vx_3)|d\vx_2d\vx_3\int_{S_{d_1}} g(\vx_1\mid\vx_2,\vx_3)d\vx_1\nonumber\\&+&\int_{[0,1]^{d_3}}\int_{S_{d_2}}\int_{S_{d_1}}|g(\vx_1,\vx_2,\vx_3)-g_0(\vx_1,\vx_2,\vx_3)|d\vx_1d\vx_2d\vx_3. \label{eq:conditional_ub}
\end{eqnarray}
By Lemma \ref{lem:joint2mar}, the first term of the summation in (\ref{eq:conditional_ub}) is upper bounded by $\epsilon$, and the second term of the summation in (\ref{eq:conditional_ub}) is upper bounded by $\epsilon$ by assumption. Thus, we conclude \[\int_{[0,1]^{d_3}}\int_{S_{d_2}} \int_{S_{d_1}}|g(\vx_1\mid\vx_2,\vx_3)-g_0(\vx_1\mid\vx_2,\vx_3)|d\vx_1g_0(\vx_2,\vx_3)d\vx_2d\vx_3<2\epsilon.\]

\subsection{Proof of Theorem \ref{thm:postmarcon}}\label{proof:postmarcon}
By Lemma \ref{lem:joint2mar}, we show that if $h\in T_{\epsilon}({g_0})$, then $h\in T_{\epsilon,(2,3)}({{g_0}})$. This indicate that $T_{\epsilon}({g_0})\subseteq T_{\epsilon,(2,3)}({{g_0}})$, which implies
\[\pi{(T_{\epsilon}({g_0})}\mid \obs_1,\ldots,\obs_n)\leq \pi{(T_{\epsilon,(2,3)}({{g_0}})}\mid \obs_1,\ldots,\obs_n)\leq 1\]
Under the condition of Theorem \ref{thm:poststrong}, we have $L_1$ consistency of joint density that, for any $\epsilon>0$,
\[\pi{(T_{\epsilon}({g_0})}\mid \obs_1,\ldots,\obs_n) \underset{n \to \infty}{\longrightarrow} 1,\quad [g_0]\textit{a.s}\]
Then we have $L_1$ marginal consistency of posterior that, \[\pi{(T_{\epsilon,(2,3)}({{g_0}})}\mid \obs_1,\ldots,\obs_n) \underset{n \to \infty}{\longrightarrow} 1,\quad [g_0]\textit{a.s}\]
Similar proof of $g_0$-integrated {$L_1$}-consistency of the conditional density follows.

By Lemma \ref{lem:joint2con}, we have that if $h\in T_{\epsilon/2}({g_0})$, then $h\in T_{\epsilon,(2,3),L_1}({{g_0}})$. This indicate that $T_{\epsilon/2}({g_0})\subseteq T_{\epsilon,(2,3),L_1}({{g_0}})$,  which implies
\[\pi{(T_{\epsilon/2}({g_0})}\mid \obs_1,\ldots,\obs_n)\leq \pi{(T_{\epsilon,(2,3),L_1}({{g_0}}))}\mid \obs_1,\ldots,\obs_n)\leq 1\]
Under the condition of Theorem \ref{thm:poststrong}, we have $L_1$ consistency of joint density that, for any $\epsilon>0$,
\[\pi{(T_{\epsilon/2}({g_0})}\mid \obs_1,\ldots,\obs_n) \underset{n \to \infty}{\longrightarrow} 1,\quad [g_0]\textit{a.s}\]
         
Then we have $g_0$-integrated $L_1$ conditional consistency of posterior that, \[\pi{(T_{\epsilon,(2,3),L_1}({{g_0}}))}\mid \obs_1,\ldots,\obs_n) \underset{n \to \infty}{\longrightarrow} 1,\quad [g_0]\textit{a.s}\]

\newpage
\section{Computational implementation of the model}\label{sec:Model_implement}

To implement the model, we use a finite-dimensional approximation of the Dirichlet process based on a truncated version of \cite{sethuraman1994constructive}'s stick-breaking representation. Under this finite-dimensional approximation, our model can be expressed as: 
\begin{eqnarray}
(\obs_i\mid \boldsymbol{\xi},\boldsymbol{\theta},\boldsymbol{k})&\overset{ind}{\sim}&K(\obs_i\mid \boldsymbol{\theta}_{\xi_{i}},\boldsymbol{k}),\quad i = 1,\ldots,n,\nonumber\\
(\xi_i\mid \boldsymbol{w})&\overset{i.i.d.}{\sim}&\sum_{j=1}^{N}w_j\delta_j(\cdot),\quad i = 1,\ldots,n,\nonumber\\
(\boldsymbol{w},\boldsymbol{\theta})&\sim& p(\boldsymbol{w})\times F_0^{N}(\boldsymbol{\theta}),\nonumber\\ \label{eq:truncated_model}
\boldsymbol{k}&\sim&p(\boldsymbol{k}),
\end{eqnarray}
where 
$\boldsymbol{k}=(k_1,k_2,k_3)$ is a random vector of independent polynomial degrees, and $k_m \sim {\rm{Poisson}}(\lambda_m)\mathbb{I}(k_m \geq d_m)$, for $m=1,2$, and $k_3 \sim {\rm{Poisson}}(\lambda_3)\mathbb{I}(k_3 \geq 1)$, therefore, $p(\boldsymbol{k})$ is a product of Poisson distributions.
$\boldsymbol{\xi}=(\xi_{1},\ldots,\xi_{n})$ denotes the cluster labels for each observation. The random vector $\boldsymbol{\theta}=(\boldsymbol{\theta}_1,\ldots,\boldsymbol{\theta}_N)$ denotes the atoms with 
$$
\boldsymbol{\theta}_{j} = (\theta_{1,1,j}, \, \ldots,  \,\theta_{1,d_1,j}, \, \theta_{2,1,j},  \, \ldots,  \,\theta_{2,d_2,j},  \, \theta_{3,1,j}, \, \ldots, \, \theta_{3,d_3,j}),\quad j=1,\ldots,N.
$$
The vector $\boldsymbol{\theta}_1,\ldots,\boldsymbol{\theta}_N$ are assumed to be independently and identically distributed according to $F_0$. The distribution $F_0$ on $\tildeDelta_d = S_{d_1} \times S_{d_2} \times [0,1]^{d_3}$ is defined as the product of two uniform Dirichlet distributions with dimensions $d_1$ and $d_2$, and $d_3$ uniform distributions on $[0,1]$. The vector $\boldsymbol{w}=(w_1,\ldots, w_N)$ denotes the weights of each atom, and $\pi(\boldsymbol{w})$ is the induced distribution through the stick-breaking representation
\[w_j = \begin{cases}
v_j \prod_{l<j} (1-v_l), &\quad j = 1,\ldots, N-1,\\
1-\sum_{l<N}w_l, &\quad j = N,
\end{cases}\]
with 
\begin{eqnarray*}
v_j|M_0 &\overset{i.i.d.}{\sim}& {\rm Beta}(1,M_0),\quad j = 1,\ldots,N-1, \\
    M_0 &\sim & {\rm Gamma}(1,1).
\end{eqnarray*}
Finally, the kernel
\begin{eqnarray*}
K(\bx \mid \boldsymbol{\theta}_{j},\boldsymbol{k}) & = & {\rm dir}\left(\vx_1\mid\alpha(k_1-d_1+1,\lceil k_1 \boldsymbol{\theta}_{1,j} \rceil)\right){\rm dir}(\vx_2\mid\alpha(k_2-d_2+1, \lceil k_2 \boldsymbol{\theta}_{2,j} \rceil)) \\ & & \times
    \prod_{l=1}^{d_3}{\rm beta}(x_{3,l}\mid \lceil k_3 \theta_{3,l,j}\rceil,k_3-\lceil k_3 \theta_{3,l,j} \rceil+1),   
\end{eqnarray*}
where $\lceil \cdot \rceil$ denotes the ceiling function and 
$$
\lceil k_m \boldsymbol{\theta}_{m,j} \rceil = 
(\lceil k_m \theta_{m,1,j} \rceil, \ldots, \lceil k_m \theta_{m,d_m,j} \rceil),\quad m =1,2.
$$

The computational implementation of model (\ref{eq:truncated_model}) relies on the Blocked Gibbs Algorithm proposed by \cite{ishwaran2001gibbs}. This algorithm, available in NIMBLE, is used with a truncation number of $N=25$. 

\newpage
\section{Additional simulation results} \label{sec:Simulation_fig}

The true joint density Scenario I is given by
\begin{eqnarray*}
    g_0(\vx_1,x_2)
        &=&0.3{\rm dir}(\vx_1\mid (2.1,10,3))
        {\rm{beta}}(x_2\mid 1,10) + 0.5{\rm dir}(\vx_1\mid (10,3.1,10))
        \\&&{\rm{beta}}(x_2\mid 5,5)
        + 0.2{\rm dir}(\vx_1\mid (10,3.1,10)){\rm{beta}}(x_2\mid 10,1),\\&& \vx_1\in S_2,\quad x_2\in [0,1]
    \end{eqnarray*}

The true joint density Scenario II is given by
\begin{eqnarray*}
    g_0(\vx_1,\vx_2,x_3,x_4)&=&
    0.3 {\rm dir}(\vx_1\mid (8, 14, 2, 3)){\rm dir}(\vx_2\mid (2.1,10,3)){\rm{beta}}(x_3\mid 1,5){\rm beta}(x_4\mid 1,10)
    \\&+& 0.2 {\rm dir}(\vx_1\mid (18, 2, 4, 5)){\rm dir}(\vx_2\mid (10,3.1,10)){\rm{beta}}(x_3\mid 1,5){\rm{beta}}(x_4\mid 5,5)
    \\&+& 0.2
    {\rm dir}(\vx_1\mid (9, 18, 2, 4)){\rm dir}(\vx_2\mid (10,3.1,10)){\rm{beta}}(x_3\mid 2,3){\rm{beta}}(x_4\mid 5,5)
    \\&+& 0.1
    {\rm dir}(\vx_1\mid (14, 2, 8, 3)){\rm dir}(\vx_2\mid (10,3.1,10)){\rm{beta}}(x_3\mid 10,8){\rm{beta}}(x_4\mid 5,5)
    \\&+& 0.1
    {\rm dir}(\vx_1\mid (2, 28, 4, 2)){\rm dir}(\vx_2\mid (10,3.1,10)){\rm{beta}}(x_3\mid 15,10){\rm{beta}}(x_4\mid 10,1)
    \\&+&0.1
    {\rm dir}(\vx_1\mid (22, 4, 8, 1)){\rm dir}(\vx_2\mid (10,3.1,10)){\rm{beta}}(x_3\mid 20,1){\rm{beta}}(x_4\mid 10,1),\\&& 
        \vx_1\in S_3,\quad \vx_2\in S_2,\quad x_3\in [0,1],\quad x_4\in [0,1]
    \end{eqnarray*}
\begin{figure}[H] 
\captionsetup[subfloat]{labelformat=empty}
\subfloat[]{
\includegraphics[page=2]{rescaled/simu_fig/simu_plots_marginal_S2_Univariates.pdf}}
\subfloat[]{
\includegraphics[page=4]{rescaled/simu_fig/simu_plots_marginal_S2_Univariates.pdf}}
\subfloat[]{
\includegraphics[page=6]{rescaled/simu_fig/simu_plots_marginal_S2_Univariates.pdf}}
\subfloat[]{
\includegraphics[page=8]{rescaled/simu_fig/simu_plots_marginal_S2_Univariates.pdf}}\\[-10pt]
\subfloat[]{
\includegraphics[page=1]{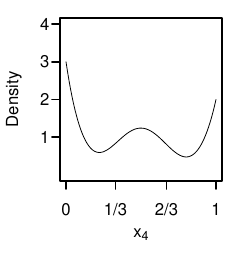}}
\subfloat[]{
\includegraphics[page=3]{rescaled/simu_fig/simu_plots_marginal_S2_Univariates.pdf}}
\subfloat[]{
\includegraphics[page=5]{rescaled/simu_fig/simu_plots_marginal_S2_Univariates.pdf}}
\subfloat[]{
\includegraphics[page=7]{rescaled/simu_fig/simu_plots_marginal_S2_Univariates.pdf}}
\caption{Simulated data – Scenario II. Posterior inference for the marginal distributions. The dotted line
represents the true marginal density, the continuous line represents the mean, across Monte Carlo simulations,
of the posterior mean of the marginal density, and the gray area represents the point-wise 95\% confidence
band. The first column display the true univariate
marginal densities, while the second, third and forth columns show the results for $n = 250$, $500$, and $1000$, respectively. The first and 
second row show the results for $x_3$ and $x_4$, respectively.}
\label{fig:s2mar1}
\end{figure}



\begin{figure}[H]
\centering
\captionsetup[subfloat]{labelformat=empty}
\subfloat[]{
\includegraphics[page=2]{rescaled/simu_fig/simu_plots_marginal_S2_Simplex.pdf}}
\subfloat[]{
\includegraphics[page=6]{rescaled/simu_fig/simu_plots_marginal_S2_Simplex.pdf}}
\subfloat[]{
\includegraphics[page=10]{rescaled/simu_fig/simu_plots_marginal_S2_Simplex.pdf}}
\subfloat[]{
\includegraphics[page=14]{rescaled/simu_fig/simu_plots_marginal_S2_Simplex.pdf}}\\[-10pt]
\subfloat[]{
\includegraphics[page=3]{rescaled/simu_fig/simu_plots_marginal_S2_Simplex.pdf}}
\subfloat[]{
\includegraphics[page=7]{rescaled/simu_fig/simu_plots_marginal_S2_Simplex.pdf}}
\subfloat[]{
\includegraphics[page=11]{rescaled/simu_fig/simu_plots_marginal_S2_Simplex.pdf}}
\subfloat[]{
\includegraphics[page=15]{rescaled/simu_fig/simu_plots_marginal_S2_Simplex.pdf}}\\[-10pt]
\subfloat[]{
\includegraphics[page=4]{rescaled/simu_fig/simu_plots_marginal_S2_Simplex.pdf}}
\subfloat[]{
\includegraphics[page=8]{rescaled/simu_fig/simu_plots_marginal_S2_Simplex.pdf}}
\subfloat[]{
\includegraphics[page=12]{rescaled/simu_fig/simu_plots_marginal_S2_Simplex.pdf}}
\subfloat[]{
\includegraphics[page=16]{rescaled/simu_fig/simu_plots_marginal_S2_Simplex.pdf}}\\[-10pt]
\subfloat[]{
\includegraphics[page=1]{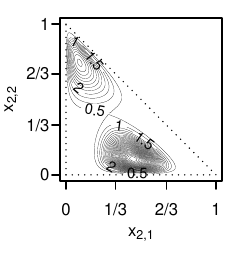}}
\subfloat[]{
\includegraphics[page=5]{rescaled/simu_fig/simu_plots_marginal_S2_Simplex.pdf}}
\subfloat[]{
\includegraphics[page=9]{rescaled/simu_fig/simu_plots_marginal_S2_Simplex.pdf}}
\subfloat[]{
\includegraphics[page=13]{rescaled/simu_fig/simu_plots_marginal_S2_Simplex.pdf}}
\caption{Simulated data – Scenario II. Posterior inference for the bivariate marginal distributions for $(x_{1,1}, x_{1,2}, x_{1,3})$ and $(x_{2,1}, x_{2,2})$. The first column display the contour plots of the true bivariate
marginal densities, while the second, third, and fourth columns show the contour plots of the mean, across
Monte Carlo simulations, of the posterior mean of the bivariate density for $n = 250$, $500$, and $1000$, respectively.}
\label{fig:s2mar2}
\end{figure}




\begin{table}[H]
    \centering
    \caption{In-sample data points used for the conditional density estimation in Scenario I.}
    \label{tab:S1data} \begin{tabular}{cccc}
 \hline\hline
     Data point & \multicolumn{2}{c}{$\bx_1$} & $x_2$ \\
 \hline
$\vec p_1$     & 0.276 &    0.400  &  0.032  \\
 $\vec p_2$     & 0.098 &    0.532 &  0.164  \\
 $\vec p_3$      & 0.546 &    0.038 &  0.279 \\
 $\vec p_4$      & 0.339 &    0.366 &  0.961 \\
 \hline \hline
\end{tabular}
\end{table}



\begin{table}[H]
    \centering
    \caption{In-sample data points used for the conditional density estimation in Scenario II.}
    \label{tab:S2data}    
\begin{tabular}{cccccc}
 \hline\hline
 Data point & $x_{1,1}$ & $x_{1,2}$ & $x_{1,3}$ & $x_3$ & $x_4$\\
 \hline
 $\vec p_1$   &    0.262 &    0.634 &    0.045 & 0.322 &  0.001 \\
 $\vec p_2$   &    0.191 &    0.668 &    0.044 & 0.313 &  0.182 \\
 $\vec p_3$   &    0.229 &    0.431 &    0.059 & 0.208 &  0.284 \\
 $\vec p_4$   &    0.091 &    0.824 &    0.070 & 0.631 &  0.924 \\
 \hline\hline
\end{tabular}
\end{table}

\begin{figure}[H]
\centering
\captionsetup[subfloat]{labelformat=empty}
\subfloat[$\bx_1 \mid x_2 = 0.032$]{
\includegraphics[page=1]{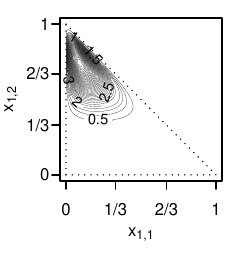}}
\subfloat[$n=250$]{
\includegraphics[page=5]{rescaled/simu_fig/simu_plots_conditional_S1_Simplex.pdf}}
\subfloat[$n=500$]{
\includegraphics[page=9]{rescaled/simu_fig/simu_plots_conditional_S1_Simplex.pdf}}
\subfloat[$n=1000$]{
\includegraphics[page=13]{rescaled/simu_fig/simu_plots_conditional_S1_Simplex.pdf}}\\
\subfloat[$\bx_1 \mid x_2 = 0.164$]{
\includegraphics[page=2]{rescaled/simu_fig/simu_plots_conditional_S1_Simplex.pdf}}
\subfloat[$n=250$]{
\includegraphics[page=6]{rescaled/simu_fig/simu_plots_conditional_S1_Simplex.pdf}}
\subfloat[$n=500$]{
\includegraphics[page=10]{rescaled/simu_fig/simu_plots_conditional_S1_Simplex.pdf}}
\subfloat[$n=1000$]{
\includegraphics[page=14]{rescaled/simu_fig/simu_plots_conditional_S1_Simplex.pdf}}\\
\subfloat[$\bx_1 \mid x_2 = 0.279$]{
\includegraphics[page=3]{rescaled/simu_fig/simu_plots_conditional_S1_Simplex.pdf}}
\subfloat[$n=250$]{
\includegraphics[page=7]{rescaled/simu_fig/simu_plots_conditional_S1_Simplex.pdf}}
\subfloat[$n=500$]{
\includegraphics[page=11]{rescaled/simu_fig/simu_plots_conditional_S1_Simplex.pdf}}
\subfloat[$n=1000$]{
\includegraphics[page=15]{rescaled/simu_fig/simu_plots_conditional_S1_Simplex.pdf}}\\
\subfloat[$\bx_1 \mid x_2 = 0.961$]{
\includegraphics[page=4]{rescaled/simu_fig/simu_plots_conditional_S1_Simplex.pdf}}
\subfloat[$n=250$]{
\includegraphics[page=8]{rescaled/simu_fig/simu_plots_conditional_S1_Simplex.pdf}}
\subfloat[$n=500$]{
\includegraphics[page=12]{rescaled/simu_fig/simu_plots_conditional_S1_Simplex.pdf}}
\subfloat[$n=1000$]{
\includegraphics[page=16]{rescaled/simu_fig/simu_plots_conditional_S1_Simplex.pdf}}
\caption{Simulated data – Scenario I. Posterior inference for the bi-variate conditional distributions. The first column display the contour plots of the true bi-variate
conditional densities, while the second, third, and fourth columns show the contour plots of the mean, across
Monte Carlo simulations, of the posterior mean of the bi-variate conditional density for $n = 250$, $500$, and $1000$, respectively.
The first, second, and third row show the results for $\bx_1$, for the different in-sample values of $x_2$, respectively.}
\label{fig:s1con2}
\end{figure}    

\begin{figure}[H]
\centering
\captionsetup[subfloat]{labelformat=empty}
\subfloat[$x_2 \mid \vx_1 = (0.276,0.400)$]{\includegraphics[page=1]{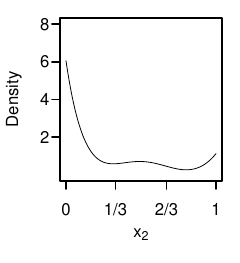}}
\subfloat[$n=250$]{
\includegraphics[page=5]{rescaled/simu_fig/simu_plots_conditional_S1_Univariates.pdf}}
\subfloat[$n=500$]{
\includegraphics[page=9]{rescaled/simu_fig/simu_plots_conditional_S1_Univariates.pdf}}
\subfloat[$n=1000$]{
\includegraphics[page=13]{rescaled/simu_fig/simu_plots_conditional_S1_Univariates.pdf}}\\
\subfloat[$x_2 \mid \vx_1 = (0.098,0.532)$]{\includegraphics[page=2]{rescaled/simu_fig/simu_plots_conditional_S1_Univariates.pdf}}
\subfloat[$n=250$]{
\includegraphics[page=6]{rescaled/simu_fig/simu_plots_conditional_S1_Univariates.pdf}}
\subfloat[$n=500$]{
\includegraphics[page=10]{rescaled/simu_fig/simu_plots_conditional_S1_Univariates.pdf}}
\subfloat[$n=1000$]{
\includegraphics[page=14]{rescaled/simu_fig/simu_plots_conditional_S1_Univariates.pdf}}\\
\subfloat[$x_2 \mid \vx_1 = (0.546,0.038)$]{\includegraphics[page=3]{rescaled/simu_fig/simu_plots_conditional_S1_Univariates.pdf}}
\subfloat[$n=250$]{
\includegraphics[page=7]{rescaled/simu_fig/simu_plots_conditional_S1_Univariates.pdf}}
\subfloat[$n=500$]{
\includegraphics[page=11]{rescaled/simu_fig/simu_plots_conditional_S1_Univariates.pdf}}
\subfloat[$n=1000$]{
\includegraphics[page=15]{rescaled/simu_fig/simu_plots_conditional_S1_Univariates.pdf}}\\
\subfloat[$x_2 \mid \vx_1 = (0.339,0.366)$]{\includegraphics[page=4]{rescaled/simu_fig/simu_plots_conditional_S1_Univariates.pdf}}
\subfloat[$n=250$]{
\includegraphics[page=8]{rescaled/simu_fig/simu_plots_conditional_S1_Univariates.pdf}}
\subfloat[$n=500$]{
\includegraphics[page=12]{rescaled/simu_fig/simu_plots_conditional_S1_Univariates.pdf}}
\subfloat[$n=1000$]{
\includegraphics[page=16]{rescaled/simu_fig/simu_plots_conditional_S1_Univariates.pdf}}
\caption{Simulated data – Scenario I. Posterior inference on conditional density of $x_2 \mid \bx_1$. The first, second, and  third columns show the plot of the mean, across Monte Carlo simulations,
of the posterior mean of the conditional density , and the gray area represents the point-wise 95\% confidence
band, for $n = 250$, $500$, and $1000$, respectively. The
first, second, third, and fourth row show the in-sample values of $\bx_1$.
The dotted line represents the true conditional density and the continuous line represents the posterior mean for the corresponding conditional density.}
\label{fig:s1con1}
\end{figure}     

\newpage
\section{Additional results for NHANES data} \label{sec:NHANES_fig}

\begin{figure}[H]
\centering
\includegraphics[page=1,trim={0 0 0 0},clip]{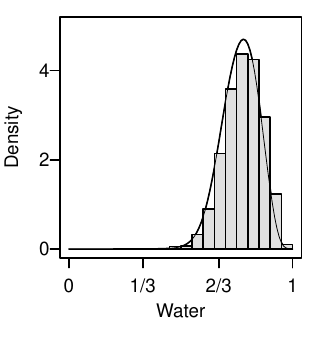}
\includegraphics[page=2,trim={0 0 0 0},clip]{rescaled/real_fig/alldata_multi_with_inits_Marginal.pdf}\\
\includegraphics[page=3,trim={0 0 0 0},clip]{rescaled/real_fig/alldata_multi_with_inits_Marginal.pdf}
\includegraphics[page=4,trim={0 0 0 0},clip]{rescaled/real_fig/alldata_multi_with_inits_Marginal.pdf}
\caption{Posterior predictive marginal distributions (continuous line) for water-intake, healthy nutrient, unhealthy nutrient intake, and fiber, with histograms representing observed data points.}
\label{fig:NHANESnutr1}
\end{figure}

\begin{figure}[H]
\centering
\includegraphics[page=5,trim={0 0 0 0},clip]{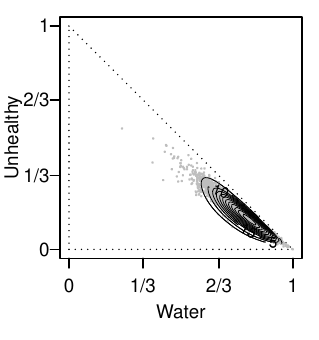}
\includegraphics[page=6,trim={0 0 0 0},clip]{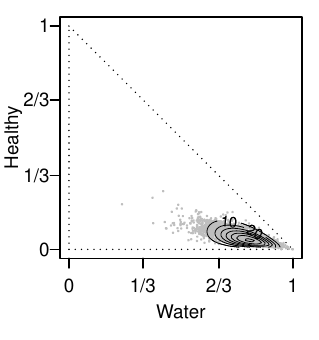}
\includegraphics[page=7,trim={0 0 0 0},clip]{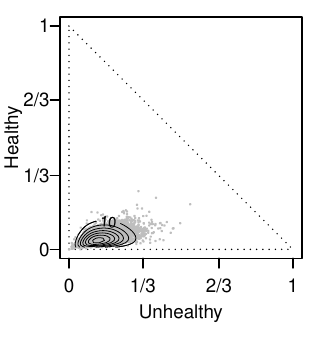}\\
\includegraphics[page=8,trim={0 0 0 0},clip]{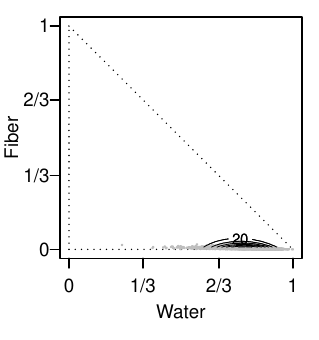}
\includegraphics[page=9,trim={0 0 0 0},clip]{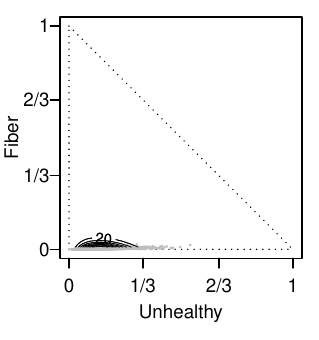}
\includegraphics[page=10,trim={0 0 0 0},clip]{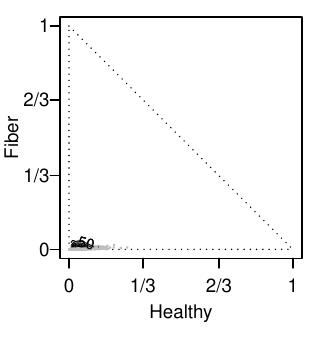}
\caption{Contour plots showing the posterior mean of the bi-variate  marginal density for combinations of nutrient intake components, with gray dots indicating observed data points.}
\label{fig:NHANESnutr2}
\end{figure}


\begin{figure}[H]
\centering
\includegraphics[page=17,trim={0 0 0 0},clip]{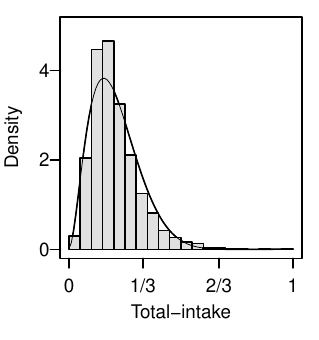}
\includegraphics[page=18,trim={0 0 0 0},clip]{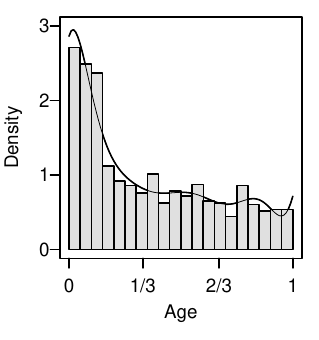}
\includegraphics[page=19,trim={0 0 0 0},clip]{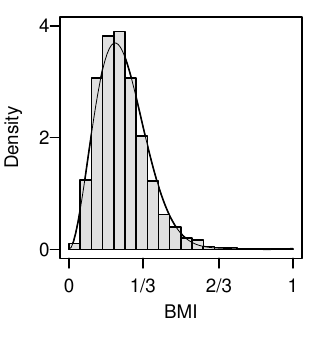}
\caption{
Posterior mean of the marginal
densities, along with the histogram, for total intake, age, and BMI.}
\label{fig:NHANESabt}
\end{figure}


\begin{figure}[H]
\captionsetup[subfloat]{labelformat=empty}
\centering
\subfloat[Observed data]{
\includegraphics[width=0.64\textwidth,page=1,trim={5 5 3 5},clip]{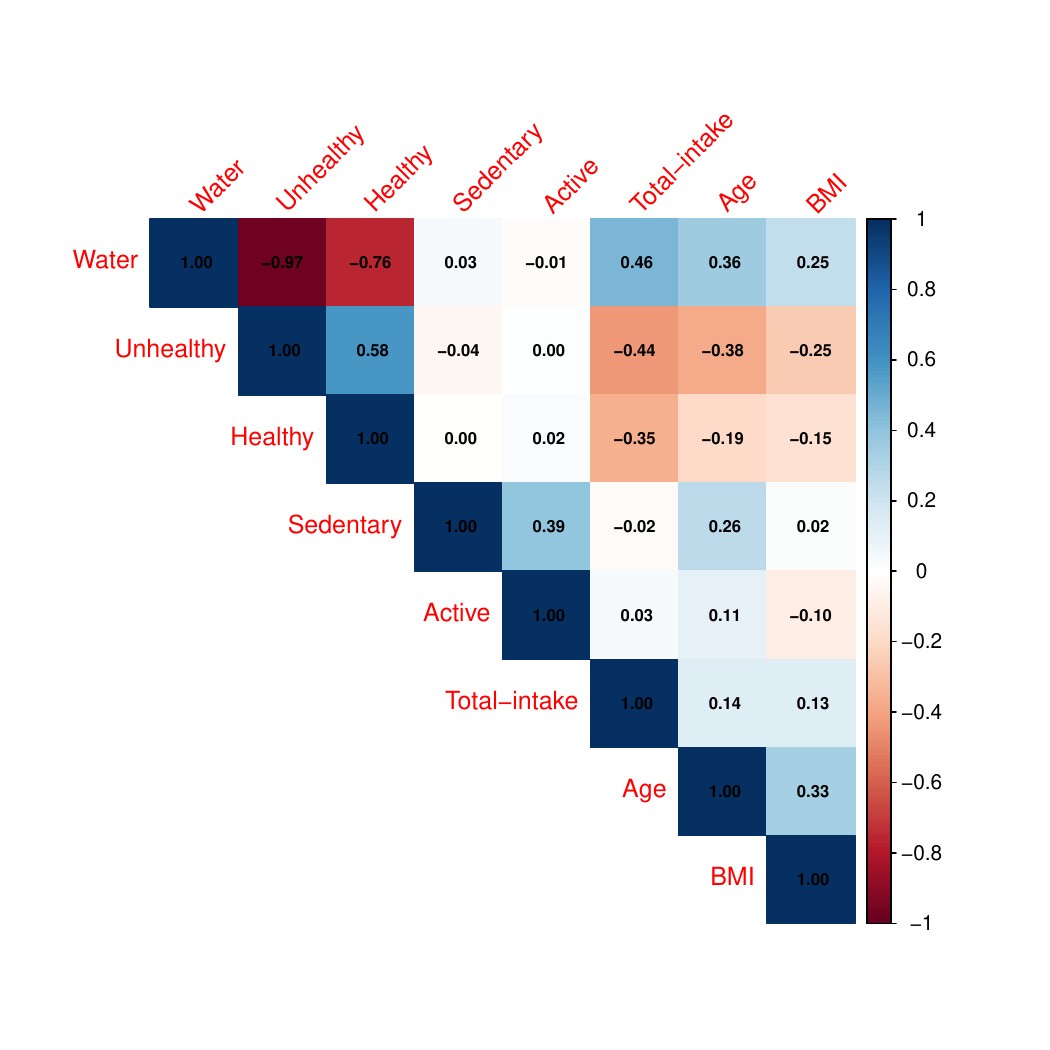}}

\subfloat[Posterior mean]{
\includegraphics[width=0.64\textwidth,page=2,trim={5 5 3 5},clip]{rescaled/real_fig/Corrplot_pred_10000_large_sample.pdf}}
\caption{Empirical Person's correlation matrix (left panel) and posterior  mean and 95\% credible interval $(l,u)$ for Pearson' correlation matrix (right panel) .}
\label{fig:cormat_extra}
\end{figure}

\begin{figure}[H]
\centering
\captionsetup[subfloat]{labelformat=empty,position=top, justification=centering}  
\subfloat[(a) \\ Low-age \\ Low-BMI]{
\includegraphics[page=25,trim={0 0 0 5.6},clip]{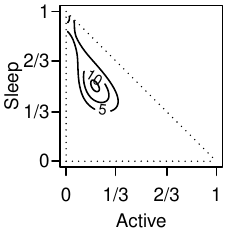}}
 \subfloat[(b) \\ Low-age \\ High-BMI]{
\includegraphics[page=26,trim={0 0 0 5.6},clip]{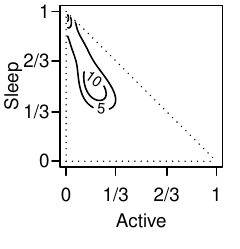}}
 \subfloat[(c) \\ High-age \\ Low-BMI]{
\includegraphics[page=27,trim={0 0 0 5.6},clip]{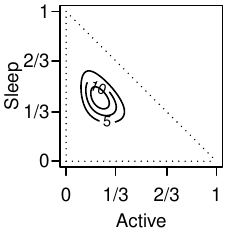}}
 \subfloat[(d) \\ High-age \\ High-BMI]{
\includegraphics[page=28,trim={0 0 0 5.6},clip]{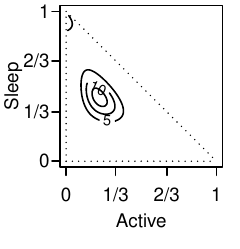}}\\
\subfloat[(e) \\ Low-total-intake \\ Low-water-intake]{
\includegraphics[page=29,trim={0 0 0 5.6},clip]{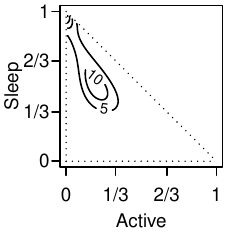}}
\subfloat[(f) \\ 
Low-total-intake \\ High-wate-intake]{
\includegraphics[page=30,trim={0 0 0 5.6},clip]{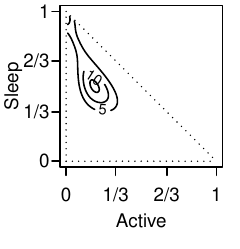}}
\subfloat[(g) \\ High-total-intake \\ Low-water-intake]{
\includegraphics[page=31,trim={0 0 0 5.6},clip]{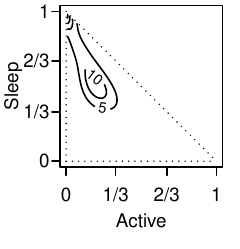}}
\subfloat[(h) \\ High-total-intake \\ High-water-intake]{
\includegraphics[page=32,trim={0 0 0 5.6},clip]{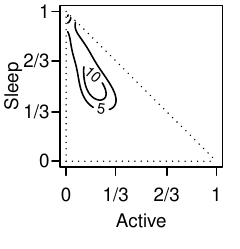}}\\
\subfloat[(i) \\ Low-age \\ Low-water-intake]{
\includegraphics[page=33,trim={0 0 0 5.6},clip]{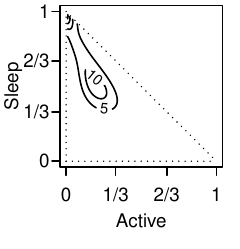}}
\subfloat[(j) \\ Low-age \\ High-water-intake]{
\includegraphics[page=34,trim={0 0 0 5.6},clip]{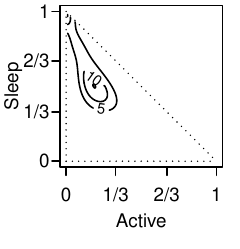}}
\subfloat[(k) \\ High-age \\ Low-water-intake]{
\includegraphics[page=35,trim={0 0 0 5.6},clip]{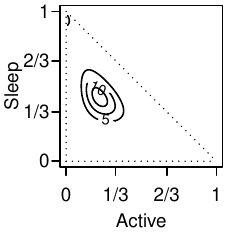}}
\subfloat[(l) \\ High-age \\ High-water-intake]{
\includegraphics[page=36,trim={0 0 0 5.6},clip]{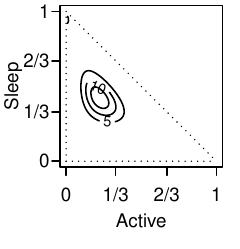}}

\caption{MHANES data -- Conditional density for active and sleep behaviors, given different values of age, BMI, total intake, and nutrient intake. 
Here low-age = 16.5 years, high-age = 62 years, low-BMI = 19.46, high-BMI = 35.54, low-total-intake = 1912.82 grams, high-total-intake = 4059.21 grams, low-water-intake 64\%, and high-water-intake = 82\%.}
\label{fig:NHANESconcontour2}
\end{figure}   

\begin{figure}[H]
\centering
\captionsetup[subfloat]{labelformat=empty,position=top, justification=centering}  
\subfloat[(a) \\ Low-age \\ Low-BMI]{
\includegraphics[page=13,trim={0 0 0 5.6},clip]{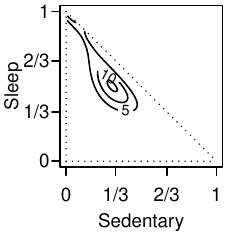}}
 \subfloat[(b) \\ Low-age \\ High-BMI]{
\includegraphics[page=14,trim={0 0 0 5.6},clip]{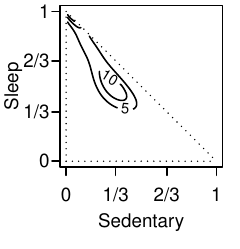}}
 \subfloat[(c) \\ High-age \\ Low-BMI]{
\includegraphics[page=15,trim={0 0 0 5.6},clip]{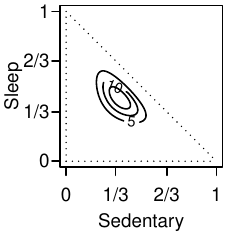}}
 \subfloat[(d) \\ High-age \\ High-BMI]{
\includegraphics[page=16,trim={0 0 0 5.6},clip]{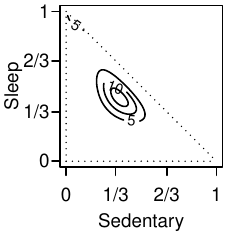}}\\
\subfloat[(e) \\ Low-total-intake \\ Low-water-intake]{
\includegraphics[page=17,trim={0 0 0 5.6},clip]{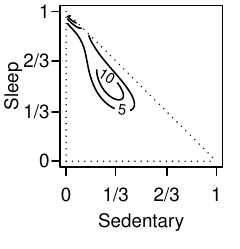}}
\subfloat[(f) \\ 
Low-total-intake \\ High-wate-intake]{
\includegraphics[page=18,trim={0 0 0 5.6},clip]{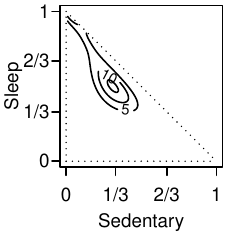}}
\subfloat[(g) \\ High-total-intake \\ Low-water-intake]{
\includegraphics[page=19,trim={0 0 0 5.6},clip]{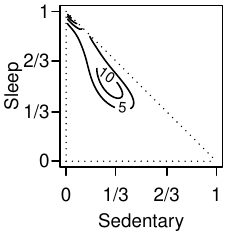}}
\subfloat[(h) \\ High-total-intake \\ High-water-intake]{
\includegraphics[page=20,trim={0 0 0 5.6},clip]{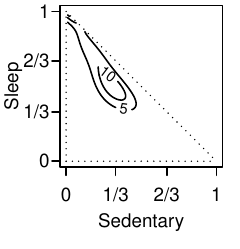}}\\
\subfloat[(i) \\ Low-age \\ Low-water-intake]{
\includegraphics[page=21,trim={0 0 0 5.6},clip]{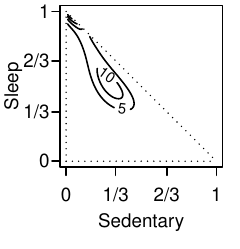}}
\subfloat[(j) \\ Low-age \\ High-water-intake]{
\includegraphics[page=22,trim={0 0 0 5.6},clip]{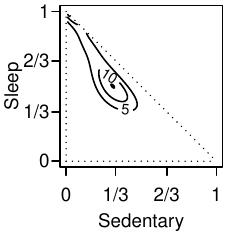}}
\subfloat[(k) \\ High-age \\ Low-water-intake]{
\includegraphics[page=23,trim={0 0 0 5.6},clip]{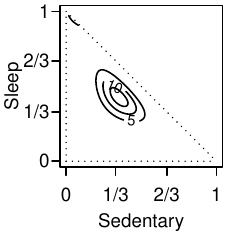}}
\subfloat[(l) \\ High-age \\ High-water-intake]{
\includegraphics[page=24,trim={0 0 0 5.6},clip]{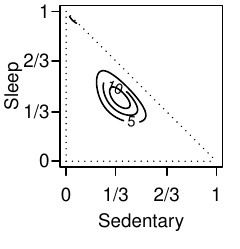}}
\caption{MHANES data -- Conditional density for sedentary and sleep behaviors, given different values of age, BMI, total intake, and nutrient intake. 
Here low-age = 16.5 years, high-age = 62 years, low-BMI = 19.46, high-BMI = 35.54, low-total-intake = 1912.82 grams, high-total-intake = 4059.21 grams, low-water-intake = 64\%, and high-water-intake = 82\%.}
\label{fig:NHANESconcontour3}
\end{figure}

\end{document}